\documentclass{article}

\usepackage{enumitem}

\usepackage{microtype}
\usepackage{graphicx}
\usepackage{subcaption}
\usepackage[normalem]{ulem}
\usepackage{booktabs} %
\usepackage{amsmath}
\usepackage[bottom]{footmisc}

\PassOptionsToPackage{hyphens}{url}\usepackage{hyperref}   %

\newcommand{\studyA}{Study-1}
\newcommand{\studyB}{Study-2}

\newif\ifcomments
\commentsfalse

\ifcomments
\newcommand{\todo}[1]{{\color{red}{TODO: #1}}}
\newcommand{\cs}[1]{{\color{blue}{CS: #1 :SC}}}
\newcommand{\correct}[1]{{\color{red}{#1}}}
\newcommand{\ds}[1]{{\color{magenta}{DS: #1}}}
\newcommand{\dcm}[1]{{\color{blue}{DCM: #1}}}
\newcommand{\michelle}[1]{{\color{cyan}{MLM: #1}}}
\newcommand{\mct}[1]{{\color{magenta}{MCT: #1}}}
\newcommand{\jpd}[1]{{\color{magenta}{JPD: #1}}}
\newcommand{\newdcm}[1]{{\color{blue}{NEW-DCM: #1}}}
\newcommand{\dsnew}[1]{{\color{magenta}{DS: #1}}}

\else
\newcommand{\todo}[1]{}
\newcommand{\cs}[1]{}
\newcommand{\correct}[1]{{\color{red}{#1}}}
\newcommand{\ds}[1]{}
\newcommand{\dcm}[1]{}
\newcommand{\michelle}[1]{}
\newcommand{\mct}[1]{}
\newcommand{\jpd}[1]{}
\newcommand{\newdcm}[1]{#1}
\newcommand{\dsnew}[1]{#1}
\fi

\newcommand{\Appref}[1]{Appendix~\ref{#1}}
\newcommand{\checkbox}{\text{\fboxsep=-.15pt\fbox{\rule{0pt}{1.5ex}\rule{1.5ex}{0pt}}}}

\usepackage[accepted]{icml2020}

\icmltitlerunning{Measuring Non-Expert Comprehension of Machine Learning Fairness Metrics}

\begin{document}

\twocolumn[
\icmltitle{Measuring Non-Expert Comprehension of Machine Learning Fairness Metrics}

\begin{icmlauthorlist}
\icmlauthor{Debjani Saha}{umd}
\icmlauthor{Candice Schumann}{umd}
\icmlauthor{Duncan C. McElfresh}{umd}
\icmlauthor{John P. Dickerson}{umd}
\icmlauthor{Michelle L. Mazurek}{umd}
\icmlauthor{Michael Carl Tschantz}{icsi}
\end{icmlauthorlist}

\icmlaffiliation{umd}{University of Maryland, College Park, MD}
\icmlaffiliation{icsi}{ICSI, Berkeley, CA}

\icmlcorrespondingauthor{Michelle L. Mazurek}{mmazurek@cs.umd.edu}

\icmlkeywords{Machine Learning, ICML}

\vskip 0.3in
]

\printAffiliationsAndNotice{}  %

\begin{abstract}
Bias in machine learning has manifested injustice in several areas, such as medicine, hiring, and criminal justice. In response, computer scientists have developed myriad definitions of \emph{fairness} to correct this bias in fielded algorithms. While some definitions are based on established legal and ethical norms, others are largely mathematical. It is unclear whether the general public agrees with these fairness definitions, and perhaps more importantly, whether they \emph{understand} these definitions.  We take initial steps toward bridging this gap between ML researchers and the public, by addressing the question: \emph{does a lay audience understand a basic definition of ML fairness?} We develop a metric to measure comprehension of three such definitions--demographic parity, equal opportunity, and equalized odds. We evaluate this metric using an online survey, and investigate the relationship between comprehension and sentiment, demographics, and the definition itself.
\end{abstract}

\section{Introduction}
Research into algorithmic fairness has grown in both importance and volume over the past few years, driven in part by the emergence of a grassroots Fairness, Accountability, Transparency, and Ethics (FATE) in Machine Learning (ML) community.  Different metrics and approaches to algorithmic fairness have been proposed, many of which are based on prior legal and philosophical concepts, such as disparate impact and disparate treatment~\cite{feldman2015certifying,chouldechova2017fair,binns2017fairness}. However, definitions of ML fairness do not always fit well within pre-existing legal and moral frameworks. The rapid expansion of this field makes it difficult for professionals to keep up, let alone the general public.
Furthermore, misinformation about notions of fairness can have significant legal implications.\footnote{\url{https://www.cato.org/blog/misleading-veritas-accusation-google-bias-could-result-bad-law}}

Computer scientists have largely focused on developing mathematical notions of fairness and incorporating them into ML systems. A much smaller collection of studies have measured public perception of bias and (un)fairness in algorithmic decision-making. 
\newdcm{
However, as both the academic community and society in general continue to discuss issues of ML fairness, it remains unclear whether non-experts--who will be \emph{impacted} by ML-guided decisions--understand various mathematical definitions of fairness sufficiently to provide opinions and critiques. 
We emphasize that these technologies are likely to have greater impact on marginalized populations, and those with lower levels of education, as in the case of hiring and criminal justice~\cite{barocas2016big,frey2017future}.
For this reason, we focus on a non-expert audience and a context (hiring) that most people would find relatively familiar. 
}

\noindent\textbf{Our Contributions.}
We take a step toward addressing this issue by studying peoples' comprehension and perceptions of three definitions of ML fairness: \emph{demographic parity}, \emph{equal opportunity,} and \emph{equalized odds} \cite{Hardt16:Equality}.
Specifically, we address the following research questions:
\vspace{-9pt}
\begin{itemize}[itemsep=0cm,leftmargin=1cm]
    \item[\textbf{RQ1}] When provided with an explanation intended for a non-technical audience, do non-experts comprehend each definition and its implications?
    \item[\textbf{RQ2}] What factors play a role in comprehension?
    \item[\textbf{RQ3}] How are comprehension and sentiment related?
    \item[\textbf{RQ4}] How do the different definitions compare in terms of comprehension? 
\end{itemize}
\vspace{-9pt}

We developed two online surveys to address these research questions. We presented participants with a simplified decision-making scenario and an accompanying \emph{fairness rule} expressed in the scenario's context. We asked questions related to the participants' comprehension of and sentiment toward this rule. Tallying the number of correct responses to the comprehension questions gives us a \emph{comprehension score} for each participant.
In \studyA{}, we found that this comprehension score is a consistent and reliable indicator of understanding demographic parity. %

Then, in \studyB{}, we used a similar approach to compare comprehension among all three definitions of interest. We find that (1) education is a significant predictor of rule understanding, (2) the counterintuitive definition of Equal Opportunity with False Negative Rate was significantly harder to understand than other definitions, and (3) participants with low comprehension scores tended to express less negative sentiment toward the fairness rule.
\newdcm{%
This underlines the importance of considering stakeholders before deploying a ``fair'' ML system, because some stakeholders may not understand or agree with an ML-specific notion of fairness. 
Our goal is to help to designers and adopters of fairness approaches understand whether they are communicating with stakeholders effectively. 
}

\section{Related Work}\label{sec:related}

In response to many instances of bias in fielded artificial intelligence (AI) and machine learning (ML) systems,  ML fairness has received significant attention from the computer-science community.
Notable examples include gender bias in job-related ads~\cite{datta2015automated}, racial bias in evaluating names on resumes~\cite{caliskan2017semantics}, and racial bias in predicting criminal recidivism~\cite{angwin2016machine}.
To correct  biased behavior, researchers have proposed several mathematical and algorithmic notions of fairness.

Most algorithmic fairness definitions found in literature are motivated by the philosophical notion of individual fairness (e.g., see~\cite{Rawls71a}), and legal definitions of disparate impact/treatment (e.g., see~\cite{barocas2016big}). 
Several ML-specific definitions of fairness have been proposed which claim to uphold these philosophical and legal concepts. 
These definitions of ``ML fairness'' fall loosely into two categories (for a review, see~\cite{chouldechova2018frontiers}). \emph{Statistical Parity} posits that in a \emph{fair} outcome, individuals from different protected groups have the same chance of receiving a positive (or negative) outcome.
Similarly, \emph{Predictive Parity}~\cite{Hardt16:Equality} asserts that the predictive accuracy should be similar across different protected groups--often measured by the false positive rate (FPR) or false negative rate (FNR) in binary classification settings. 
Myriad other definitions have been proposed, based on concepts such as calibration~\cite{pleiss2017fairness} and causality~\cite{kusner2017counterfactual}. 
Of course, all of these definitions make limiting assumptions; no concept of fairness is perfect~\cite{Hardt16:Equality}. The question remains, \emph{which} of these fairness definitions are appropriate, and in \emph{what context?}
There are two important components to answering this question: \emph{communicating} these fairness definitions to a general audience, and \emph{measuring their perception} of these definitions in context.

Communicating ML-related concepts is an active and growing research area.
In particular, \emph{interpretable ML} focuses on communicating the decision-making process and results of ML-based decisions to a general audience~\cite{lipton2018mythos}. 
Many tools have been developed to make ML models more interpretable, and many demonstrably improve understanding of ML-based decisions~\cite{ribeiro2016should,Huysmans2011}.
These models often rely on concepts from probability and statistics---teaching these concepts has long been an active area of research.
\citet{batanero2016research} provide an overview of teaching probability and how students learn probability; our surveys use their method of communicating probability, which relies on proportions.
We draw on several other concepts from this literature for our study design; for example avoiding numerical and statistical representations~\cite{gigerenzer2003simple,gigerenzer2007helping}, which can be confusing to a general audience.
Instead we provide relatable examples, accompanied by examples and graphics~\cite{hogarth2015providing}.

Effectively communicating ML concepts is necessary to achieve our second goal of understanding peoples' perceptions of these concepts.
One particularly active research area focuses on how people perceive bias in algorithmic systems.
For example, \citet{woodruff2018qualitative} investigated perceptions of algorithmic bias among marginalized populations, using a focus group-style workshop;~\citet{grgic2018human} study the underlying factors causing perceptions of bias, highlighting the importance of selecting appropriate features in algorithmic decision-making; \citet{plane2017exploring} look at perceptions of discrimination of online advertising;~\newdcm{\citet{harrison2020empirical} studies perceptions of fairness in stylized machine learning models;} \dsnew{\citet{srivastava2019mathematical} note that perceived appropriateness of an ML notion of fairness may depend on the domain in which the decision-making system is deployed, but suggest that simpler notions may best capture lay perceptions of fairness.}

A related body of work studied how people perceive algorithmic decision-makers.
\citet{lee2018understanding} studies perceptions of fairness, trust, and emotional response of algorithmic decision-makers --- as compared to human decision-makers. 
Similar work studies perception of fairness in the context of splitting goods or tasks, and in loan decisions~\cite{Lee2017,Lee2019,saxena2020fairness}.
\citet{binns2018s} studies how different explanation styles impact perceptions of algorithmic decision-makers.

This substantial body of prior research provided inspiration and guidance for our work.
Prior work has studied both the effective communication of, and perceptions of, ML-related concepts.
We hypothesize that these concepts are in fact related; to that end, we design experiments to simultaneously study peoples' \emph{comprehension} of and \emph{perceptions} of common ML fairness definitions.

\section{Methods}\label{sec:methods}

To study perceptions of ML fairness, we conducted two online surveys where participants were presented with a hypothetical decision-making scenario. Participants were then presented with a ``rule'' for enforcing fairness. We then asked each participant several questions on their comprehension and perceptions of this fairness rule. 
We first conducted \studyA{} to validate our methodology; we then conducted the larger and broader \studyB{} to address our main research questions.
Both studies were approved by the University of Maryland Institutional Review Board (IRB).

\subsection{\studyA{}}\label{sec:studyA}

In \studyA{} we tested three different decision-making scenarios based on real-world decision problems: hiring, giving employee awards, and judging a student art project. 
However, we observed no difference in participant responses between these scenarios; for this reason, 
\dsnew{we focus exclusively on hiring in 
\studyB{} (see \ref{sec:studyB}).}
Please see Appendix~\ref{app:survey} for a description of the \studyA{} scenarios, and \Appref{app:scenario_analysis} for relevant survey results.
In \studyA{}, we chose (what we believe is) the simplest definition of ML fairness, namely, demographic parity. In short, this rule requires that the fraction of one group who receives a \emph{positive} outcome (e.g., an award or job offer) is equal for both groups. 

\subsubsection{Survey Design}\label{methods:design}

Here we provide a high-level discussion of the survey design; the full text of each survey can be found in Appendix~\ref{app:survey}. 
The participant first receives a consent form (see Appendix~\ref{app:consent}). If consent is obtained, the participant sees a short paragraph explaining the decision-making scenario. To make demographic parity accessible to a non-technical audience, and to avoid bias related to algorithmic decision-making, we frame this notion of fairness as a \emph{rule} that the decision-maker must follow to be fair. 
In the hiring scenario, we framed this decision rule as follows:
\emph{The fraction of applicants who receive job offers that are female should equal the fraction of applicants that are female. Similarly, the fraction of applicants who receive job offers that are male should equal the fraction of applicants that are male.}

We then ask two questions concerning participant evaluation of the scenario, nine comprehension questions about the fairness rule, two self-report questions on participant understanding and use of the rule, and four free response questions on comprehension and sentiment.
For example, one comprehension question is:
\emph{Is the following statement TRUE OR FALSE: This hiring rule always allows the hiring manager to send offers exclusively to the most qualified applicants}.
Finally, we collect demographic information (age, gender, race/ethnicity, education level, and expertise in a number of  relevant fields).

We conducted in-person cognitive interviews ~\cite{harrell2009data} to pilot our survey, leading to several improvements in the question design. Most notably, because some cognitive interview participants appeared to use their own personal notions of fairness rather than our provided rule, we added questions to assess this compliance issue. 

\subsubsection{Recruitment and Participants} \label{subsubsec:methods:study1:recruitment}

We recruited participants using the online service Cint \cite{cint}, which allowed us to loosely approximate the 2017 U.S. Census distributions \cite{census07} for ethnicity and education level, allowing for broad representation. %
We required that participants be 18 years of age or older, and fluent in English. Participants were compensated using Cint's rewards system; according to a Cint representative: ``[Participants] can choose to receive their rewards in cash sent to their bank accounts (e.g. via PayPal), online shopping opportunities with one of multiple online merchants, or donations to a charity."

\dsnew{Data was collected during August 2019.} In total 147 participants were included in the \studyA{} analysis, including 75 men (51.0\%), 71 women (48.3\%), and 1 (0.7\%) preferring not to answer. The average age was 46 years (SD = 16). Ethnicity and educational attainment are summarized in Table~\ref{tab:demo}. %
On average, participants completed the survey in 14 minutes.

Table~\ref{tab:demo} summarizes the ethnicity and education level of participants in both \studyA{} and \studyB{}.

\begin{table}[ht]
\centering
\caption{\label{tab:demo} Participant demographics across ethnicity and education level, compared to the 2017 U.S. Census. AI = American Indian, AN = Alaska Native, NH = Native Hawaiian, PI = Pacific Islander, AA = African American. Note that in \studyB{}, two participants did not report their education level.}
\vspace{5pt}
{\small
\begin{tabular}{@{}lrrr@{}}
    \toprule
    & \multicolumn{3}{c}{\textbf{Percent of Sample}} \\
    & \textbf{Census} & \textbf{\studyA{}} & \textbf{\studyB{}} \\
    \midrule
    \textbf{Ethnicity} & & \\
    AI or AN & 0.7 & 0.7 & 0.9\\
    Asian or NH or PI & 5.7 & 1.4 & 2.3 \\
    Black or AA & 12.3 & 10.2 & 15.8 \\
    Hispanic or Latinx & 18.1 & 12.2 & 7.7\\
    Other & 2.6 & 2.7 & 1.4 \\
    White & 60.6 & 72.8 & 71.9 \\
    
  \addlinespace[1.5 ex]
    \textbf{Education Level} & & \\
    Less than HS & 12.1 & 6.1 & 6.9 \\
    HS or equivalent & 27.7 & 29.9 & 24.9 \\
    Some post-secondary & 30.8 & 30.6 & 24.9 \\
    Bachelor's and above & 29.4 & 33.3 & 42.7 \\
    \bottomrule
\end{tabular}%
}
\vspace{-5pt}
\end{table}

\subsection{\studyB}\label{sec:studyB}
\studyB{} follows a very similar structure to \studyA{} with a few changes. First, we decided to use only the hiring (HR) decision scenario (See \Appref{app:scenario_analysis} for more in-depth discussion). %
Second, we expanded to three definitions of fairness: \emph{demographic parity} (DP), \emph{equal opportunity} (EP), and \emph{equalized odds} (EO) ~\cite{Hardt16:Equality}. Within EP, we tested both False Negative Rate (FNR) and False Positive Rate (FPR), resulting in a total of four conditions. 

\subsubsection{Survey Design}
Here we provide a high-level discussion of the differences between \studyB{} and \studyA{}; the full text of each survey can be found in Appendix~\ref{app:survey}. We used a between-subjects design with random assignment among the four conditions (DP, FNR, FPR, EO). Again, we frame each notion of fairness as a \emph{hiring rule} that the decision-maker must follow to be fair. For example, in FPR we define the award rule as follows: 
\emph{The fraction of unqualified male candidates who receive job offers should equal the fraction of unqualified female candidates who receive job offers.}

\begin{figure}[h]
    \centering
    \includegraphics[height=1in]{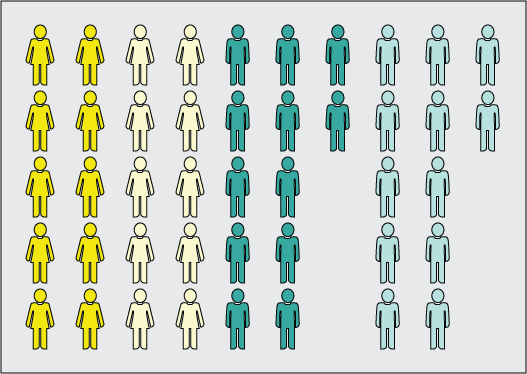}
    
    \vspace{10pt}
    \includegraphics[height=1in]{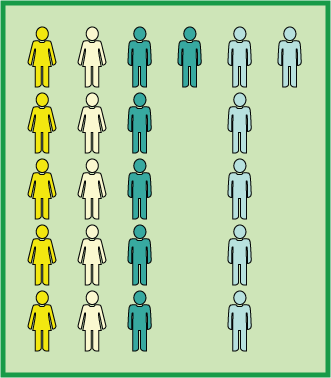}
    \space\space
    \includegraphics[height=1in]{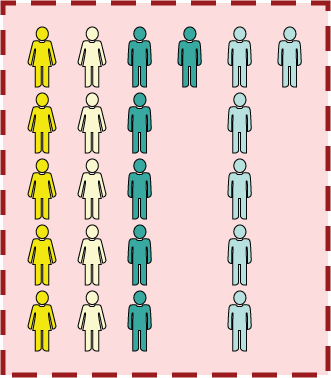}
    \caption{A graphical example to describe a fair hiring outcome for EO. Yellow people represent females while green people represent males. The darker colors represent qualified individuals while the lighter colors represent unqualified individuals. The gray box represents the original pool of applicants. The green box represent individuals that received job offers while the red box with a dashed border represents individuals that did \emph{not} receive job offers.}
    \label{fig:example_people}
\end{figure}

For this version, we added graphical examples to further clarify our explanations (see Fig.~\ref{fig:example_people} for an example). 
We used the all the same %
questions as in \studyA{} but added  two additional Likert-scale questions %
assessing participant sentiment: one asked whether they liked the rule, and the other asked whether they agreed with the rule. One free response question (asking how participants personally would go about the hiring process to ensure it was fair), which did not consistently provide useful responses in \studyA{}, was removed from the \studyB{} survey in an effort to keep the expected completion time similar. %

\subsubsection{Recruitment and Participants}
We again used the Cint service to recruit participants. \dsnew{Compensation for participation was handled in the same manner as described in \S\ref{subsubsec:methods:study1:recruitment}.} Because our initial sample (intended to target education, ethnicity, gender and age distributions approximating the U.S. census) skewed more highly educated than we had hoped, we added a second round of recruitment one week
later primarily targeting participants without bachelor's degrees. Hereafter, we report on both samples together.

\dsnew{Data was collected during January and February 2020.} In total 349 participants were included in the \studyB{} analysis, including 142 men (40.7\%), 203 women (58.2\%), 1 other (0.3\%), and 3 (0.9\%) preferring not to answer. The average age was 45 years (SD = 15). Ethnicity and educational attainment are summarized in Table~\ref{tab:demo}. %
On average, participants completed the survey in 16 minutes. %

\subsection{Data Analysis}

Free response questions were qualitatively coded for statistical testing. In \studyA{}, one question was coded by a single researcher for simple correctness (see \Appref{results:1:rq1}), and the other was independently coded by three researchers (resolved to 100\%) to capture sentiment information (see \Appref{results:1:rq3}). In \studyB{}, both questions were independently coded by 2-3 researchers (resolved to 100\%). Participants who provided nonsensical answers, answers not in English, or other non-responsive answers to free response questions were excluded from all analysis.

The following methods were used for all statistical analyses unless otherwise specified. Correlations with nonparamentric ordinal data were assessed using Spearman's rho. Omnibus comparisons on nonparametric ordinal data were performed with a Kruskal--Wallis (K-W) test, and relevant post-hoc comparisons with Mann--Whitney U (M-WU) tests. Post-hoc $p$-values were adjusted for multiple comparisons using Bonferroni correction. $\chi^2$ tests were used for comparisons of nominal data.
Boxplots show median and first and third quartiles; whiskers extend to $1.5 * \text{IQR}$ (interquartile range), with outliers indicated by points. \dsnew{The full analysis script for both studies can be found on github. \footnote{\url{https://github.com/saharaja/ICML2020-fairness}}}

\subsection{Limitations}
As with all surveys, our study has certain limitations. We recruited a demographically broad population, but web panels are generally more tech-savvy than the broader population \cite{redmiles2019well}. We consider this acceptable for a first effort. Some participants may 
be satisficing rather than answering carefully. We mitigate this by 
disqualifying participants with off-topic or non-responsive free-text responses. Further, this limitation can be expected to be consistent across conditions, enabling reasonable comparison. Finally, better or clearer explanations of the fairness definitions we explored are certainly possible; we believe our explanations were sufficient to allow us to investigate our research questions, especially because they were designed to be consistent across conditions.

\section{Results}\label{sec:results}

In this section we first discuss the preliminary findings from \studyA{} (see \S\ref{results:a}). These findings were used as hypotheses for further exploration and testing in \studyB{}; we discuss those results second (see \S\ref{results:b}).

\subsection{\studyA{}} \label{results:a}
We analyze survey responses for \studyA{} and make several observations. We first validate our comprehension score as a measure of participant understanding; we then generate hypotheses for further exploration in \studyB{}.

\subsubsection{Our Survey Effectively Captures Rule Comprehension} \label{results:a:validity}

We find that we can measure comprehension of the fairness rule. The comprehension score was calculated as the total correct responses out of a possible 9. All questions were weighted equally. The relevant questions included 2 multiple choice, 4 true/false, and 3 yes/no questions. The average score was 6.2 (SD=2.3).

We validate our comprehension score using two methods: internal validity testing, and correlation against two self-report and one free response question included in our survey (see \Appref{results:1:rq1} for further details).

\vspace{-5pt}

\paragraph{Internal Validity}
Cronbach's $\alpha$ and item-total correlation were used to assess internal validity of the comprehension score. Both measures met established thresholds \cite{nunnally1978,everitt2010}: Cronbach's $\alpha = 0.71$, and item-total correlation for 8 of the 9 items (all but Q5) $> 0.3$. %

\vspace{-5pt}

\paragraph{Question Correlation}

We find that self-reported rule understanding and use are reflected in comprehension score. First, we compared comprehension score to self-reported rule understanding (Q13): ``I am confident I know how to apply the award rule described above,'' rated on a five-point Likert scale from strongly agree (1) to strongly disagree (5). The median response was ``agree'' ($\text{Q1}=1$, $\text{Q3}=3$). Higher comprehension scores tended to be associated with greater confidence in understanding (Spearman's $\rho = 0.39$, $p<0.001$), supporting the notion that  comprehension score is a valid measure of rule comprehension.

Next, we compared comprehension score to a self-report question about the participant's use of the rule (Q14), with the following options: (a) ``I applied the provided award rule only,'' (b) ``I used my own ideas of what the correct award decision should be rather than the provided award rule,'' or (c) ``I used a combination of the provided award rule and my own ideas of what the correct award decision should be.'' We find that participants who claimed to use only the rule scored significantly higher (mean 7.09) than those who used their own notions (4.90) or a combination (4.68) (post-hoc M-WU,
$p<0.001$ for both tests; corrected $\alpha = 0.05/3 = 0.017$). This further corroborates our comprehension score.

Finally, we asked participants to explain the rule in their own words (Q12). Each response was then qualitatively coded as one of five categories -- \textbf{Correct}: describes rule correctly; \textbf{Partially correct}: description has some errors or is somewhat vague; \textbf{Neither}: vague description of purpose of the rule rather than how it works, or pure opinion; \textbf{Incorrect}: incorrect or irrelevant; and \textbf{None}: no answer, or expresses confusion. Participants whose responses were either correct (mean comprehension score = 7.71) or partially correct (7.03) performed significantly better on our survey than those responding with neither (5.13) or incorrect (4.24) (post-hoc M-WU, $p<0.001$ for these four comparisons, corrected $\alpha = 0.05/10 = 0.005$). These findings further validate our comprehension score. Additional details of these results and the associated statistical tests can be found in \Appref{results:1:rq1}.

\subsubsection{Hypotheses Generated} \label{results:a:hypotheses}

We analyzed the data from \studyA{} in an exploratory fashion intended to generate hypotheses that could be tested in \studyB{}. 
We highlight here three key hypotheses that emerged from the data.

\paragraph{Education Influences Comprehension}
We used poisson regression models to explore whether various demographic factors were associated with differences in comprehension. We found that a model including education as a regressor had greater explanatory power than a model without (see \Appref{results:1:rq2} for further details).

\paragraph{Disagreement with the Rule is Associated with Higher Comprehension Scores}

We asked participants for their opinion on the presented rule in a free response question (Q15). These responses were qualitatively coded to capture participant sentiment toward the rule in one of five categories -- \textbf{Agree}: generally positive sentiment towards rule; \textbf{Depends}: describes both pros and cons of the given rule; \textbf{Disagree}: generally negative sentiment towards rule; \textbf{Not understood}: expresses confusion about rule; \textbf{None}: no answer, or lacks opinion on appropriateness of the rule. Participants who expressed disagreement with the rule performed better (mean comprehension score = 7.02) than those who expressed agreement (5.50), did not understand the rule (4.44), or provided no response (5.09) to the question (post-hoc M-WU, $p<0.005$ for these three comparisons;
corrected $\alpha=0.05/10=0.005$). \Appref{results:1:rq3} provides further details.

\begin{figure*}[h]
    \centering
\begin{subfigure}[t]{0.30\textwidth}
    \includegraphics[width=1\linewidth]{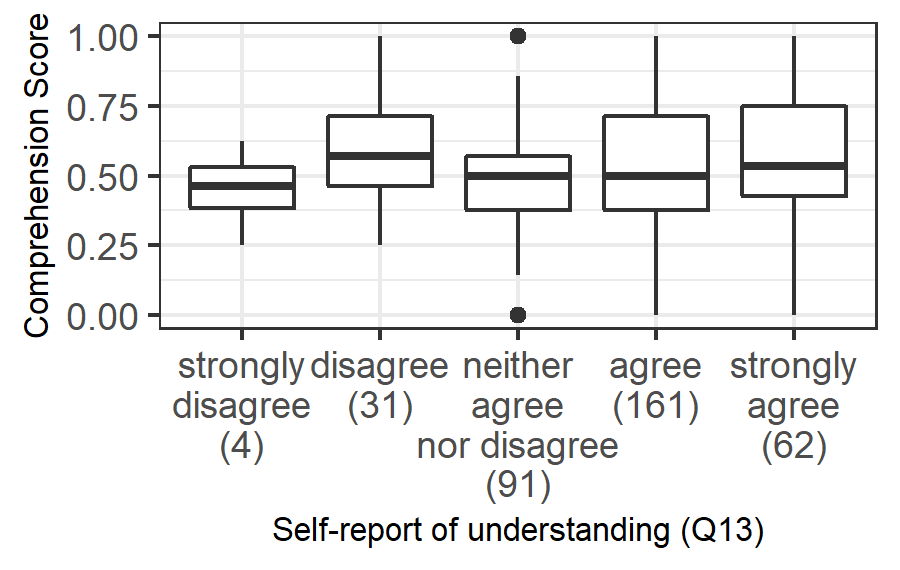}
    \caption{Grouped by response to Q13}
    \label{fig:studyB_q13}
    \vspace{-5pt}
\end{subfigure}
\quad
\begin{subfigure}[t]{0.30\textwidth}
    \includegraphics[width=1\linewidth]{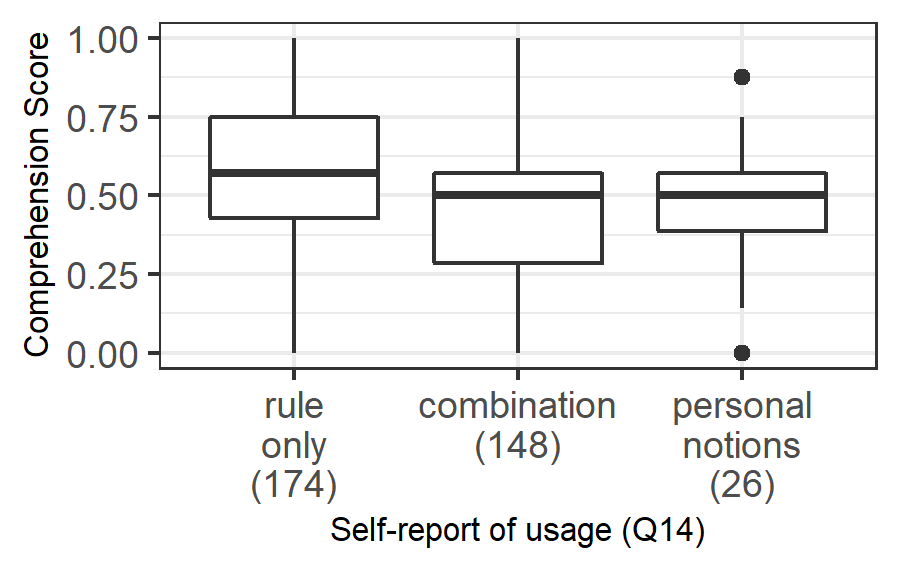}
    \caption{Grouped by response to Q14.}
    \label{fig:studyB_q14}
    \vspace{-5pt}
\end{subfigure}
\quad
\begin{subfigure}[t]{0.30\textwidth}
    \includegraphics[width=1\linewidth]{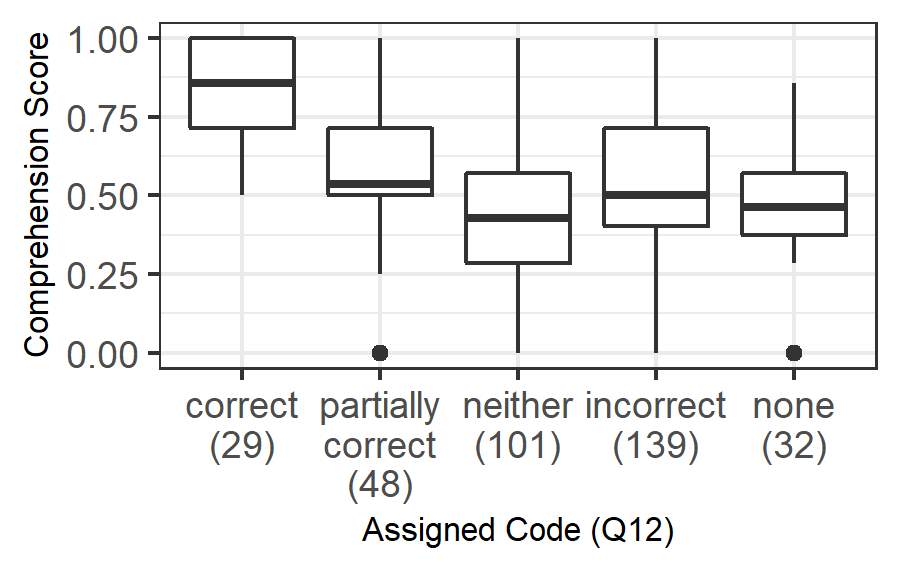}
    \caption{Grouped by coded response to Q12.}
    \label{fig:studyB_q12}
    \vspace{-5pt}
\end{subfigure}
\caption{Comprehension scores grouped by questions. In (a), self-reported understanding of the rule was not related to comprehension score. X-axis is reversed for figure and correlation test. In (b), rule compliance (leftmost on the x-axis) was associated with higher comprehension scores. One participant who did not provide a response was excluded from this figure and the relevant analysis. Finally, in (c), participants who provided either correct or partially correct responses tended to perform better.}
\end{figure*}

\paragraph{Non-Compliance is Associated with Lack of Understanding} \label{results:a:non-comp}

We were interested in understanding why some participants failed to adhere to the rule, as measured by their self-report of rule usage in Q14. We labeled those who responded with either having used their own personal notions of fairness ($n=29)$ or some combination of their personal notions and the rule ($n=28$) as ``non-compliant" (NC), with the remaining $n=89$ labeled as ``compliant" (C). One participant who did not provide a response was excluded from this analysis, conducted using $\chi^2$ tests.

Non-compliant participants were less likely to self-report high understanding of the rule in Q13 (see Fig. \ref{fig:q13q14}). Moreover, non-compliance also appears to be associated with a reduced ability to correctly explain the rule in Q12 (see Fig. \ref{fig:q12q14}). This fits with the overall strong relationship we observed among comprehension scores, self-reported understanding, ability to explain the rule, and compliance. 

Further, negative participant sentiment towards the rule (Q15) also appears to be associated with greater compliance (see Fig. \ref{fig:q15q14}). %
Thus, non-compliant participants appear to behave this way because they do not \emph{understand} the rule, rather than because they do not \emph{like} it. Refer to \Appref{results:a:non-comp} for further details.

\subsection{\studyB{}} \label{results:b}
We first confirm the validity of our 
comprehension score, then compare comprehension across 
definitions and examine the hypotheses generated in \studyA{}.

\subsubsection{Score Validation} \label{results:b:validation}

We validated our metric using the same approach used in \studyA{}, i.e., assessing both internal validity and correlation with self-report and free-response questions. We report the results of this assessment here.

\paragraph{Internal Validity}

We again used Cronbach's $\alpha$ and item-total correlation to assess internal validity of the comprehension score. An initial assessment using all 349 responses yielded Cronbach's $\alpha = 0.38$, and item-total correlation $> 0.3$ for only four of the nine comprehension questions. Since both measures performed below established thresholds \cite{nunnally1978,everitt2010}, we investigated further and repeated these measurements individually for each fairness-definition condition (DP, FNR, FPR, EO). This procedure showed stark differences in Cronbach's $\alpha$ based on definition: DP = 0.64, FNR = 0.39, FPR = 0.49, EO = 0.62. Item-total correlations followed a similar pattern:  best in DP, worst in FNR. Based on these differences, we iteratively removed problematic questions from the score on a per-definition basis until all remaining questions achieved an item-total correlation of $> 0.3$ \cite{everitt2010}.
By removing poorly performing questions, we 
increase our confidence that the measured comprehension scores are meaningful for further analysis. Table \ref{tab:dropped_qs} specifies which questions were retained for analysis in each definition.
\vspace{-10pt}
\begin{table}[ht]
\centering
\caption{\label{tab:dropped_qs} Questions that were used for downstream analysis after iterative removal of questions with poor item-total correlation.}
\vspace{5pt}
{\small
\begin{tabular}{@{}lrrrrrrrrr@{}}
    \toprule
     & \multicolumn{9}{c}{\textbf{Questions}}\\
    \midrule
     & Q3 & Q4 & Q5 & Q6 & Q7 & Q8 & Q9 & Q10 & Q11\\
    \midrule
    DP & X & X &  &  & X & X & X & X & X \\
    FNR & X & X & X &  &  & X &  &  &   \\
    FPR & X & X & X & X &  & X &  & X & X \\
    EO & X & X & X &  & X & X & X & X & X \\
    \bottomrule
\end{tabular}%
}
\end{table}

Because questions were dropped on a per-definition basis, the maximum of the resulting scores varied from 4-8 depending on the definition, rather than being a uniform 9. We normalized this treating comprehension score as a percentage of the maximum for each condition rather than a raw score. %
We report this \textit{adjusted score} in the remainder of \S\ref{results:b}. The average score was 0.53 (SD=0.22).

\paragraph{Question Correlation} \label{qcorr}

As in \studyA{}, we compare comprehension scores with responses to self-report and free response questions included in our survey.

First, we compared comprehension score to self-reported rule understanding (Q13), as described in \S\ref{results:a:validity}.
The median response was ``agree'' ($\text{Q1}=2$, $\text{Q3}=3$). We assess the correlation between these responses and comprehension score using Spearman's rho (appropriate for ordinal data). Unlike in \studyA{}, there was no relationship between self-reported understanding and comprehension score (Fig.~\ref{fig:studyB_q13}).

Next, we compared comprehension score to a self-report question about the participant's use of the rule (Q14), as described in \S\ref{results:a:validity}.
A K-W test revealed a relationship between self-reported rule usage and comprehension score ($p<0.001$). %
We find that participants who claimed to use only the rule tended to score higher (mean comprehension score = 0.58) than those who used a combination of the rule and their own notions of fairness (0.47, $p<0.01$). No other differences were found (post-hoc M-WU;  %
corrected $\alpha = 0.05/3 = 0.017$). This suggests that participants are answering at least somewhat honestly: when they try to apply the rule, comprehension scores improve (see Fig. \ref{fig:studyB_q14}).

Finally, we asked participants to explain the rule in their own words (Q12). Each response was then qualitatively coded as one of five categories, as described in \S\ref{results:a:validity}.
These results can be seen in Fig.~\ref{fig:studyB_q12}. A K-W test revealed a relationship between comprehension score and coded responses to Q12 %
($p<0.001$). Correct (mean comprehension score = 0.83) responses were associated with higher comprehension scores than partially correct (0.58), neither (0.44), incorrect (0.52), and none (0.48) responses %
($p<0.001$ for all); partially correct responses were also associated with higher comprehension scores than neither responses 
($p<0.001$); and incorrect responses were associated with higher comprehension scores that neither responses 
($p<0.005$). No other differences were found (post-hoc M-WU; corrected $\alpha = 0.05/10 = 0.005$). These findings support our claim that our comprehension score is a valid measure of fairness-rule comprehension.

\subsubsection{Education and Definition are Related to Comprehension Score} \label{results:b:edu}

One hypothesis generated by \studyA{} was that comprehension score is positively correlated with education level.
We investigated this hypothesis further in \studyB{} using linear regression models followed by model selection. 
\dsnew{We believe this exploratory approach to be appropriate despite the previously formulated hypothesis, given the introduction of a new variable in \studyB{}, i.e. fairness definition.} %

Eleven models were tested, regressing different combinations of demographics (ethnicity, gender, education, and age) and condition (fairness definition). Models were compared using Akaike information criterion (AIC), a standard method of evaluating model quality and performing model selection \cite{akaike1974}. Comparison by AIC revealed that the model using just education (edu) and fairness definition (def) as regressors was the model of best fit. In this model, having a Bachelor's degree or above resulted in a score increase of 0.14, and the FNR condition caused a score decrease of -0.11 ($p < 0.004$ for both; corrected $\alpha = 0.05/11 = 0.0045$). A regression table of the best fit model can be found in Table \ref{tab:GLM}.

\begin{table}[h]
\centering
\caption{\label{tab:GLM} Regression table for the best fit model, with two covariates: education (baseline: no HS) and definition (baseline: DP). %
Est. = estimate, CI = confidence interval.}
\vspace{5pt}
{\small
\begin{tabular}{@{}lrrc@{}}
    \toprule
    \textbf{Covariate} & \textbf{Est.} & \textbf{95\% CI} & \textbf{$p$} \\
    \midrule
    \emph{Education} \\
    HS & 0.00 & [-0.10, 0.10] & 0.989 \\ %
    Post-secondary, no BS & 0.09 & [-0.01, 0.18] & 0.078\\ %
    Bachelor's and above & 0.14 & [0.04, 0.23] & $<0.004$ \\ %
  \addlinespace[1.5 ex]
  \emph{Definition} \\
    EO & -0.08 & [-0.14, 0.01] & 0.020 \\ %
    FPR & -0.05 & [-0.11, 0.01] & 0.124 \\ %
    FNR & -0.11 & [-0.18, -0.05] & $<0.001$ \\ %
    \bottomrule
\end{tabular}%
}
\vspace{-7pt}
\end{table}

\begin{figure}[th]
    \centering
    \includegraphics[width=0.8\columnwidth]{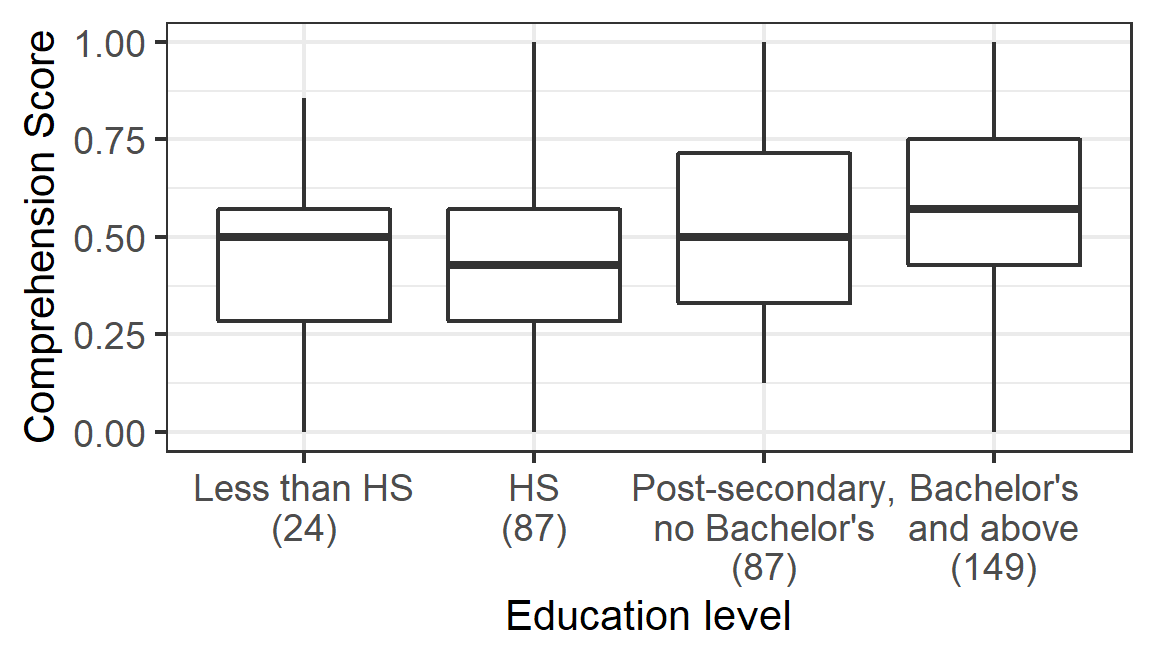}
    \vspace{-15pt}
    \caption{Comprehension score grouped by education level. Higher education was associated with higher comprehension scores. Note that two participants who did not report their education level were removed from this figure and the relevant analysis.}
    \label{fig:studyB_edu}
    \vspace{-5pt}
\end{figure}

\begin{figure}[th]
    \centering
    \includegraphics[width=0.8\columnwidth]{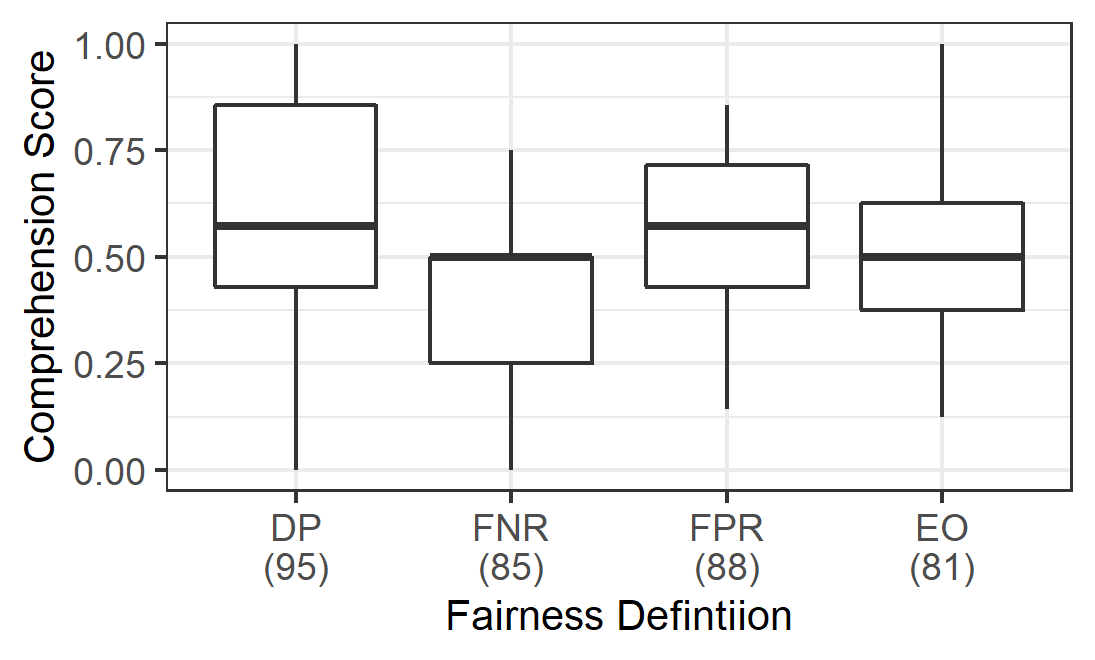}
    \vspace{-15pt}
    \caption{Comprehension score grouped by fairness definition. The FNR condition was associated with lower comprehension sore.}
    \label{fig:studyB_scores}
    \vspace{-10pt}
\end{figure}

AIC results of each of the eleven models, along with the relevant regressors, can be seen in Table \ref{tab:AIC} in \Appref{app:b:model_selection}. Comprehension score as a function of education and fairness definition can be seen in Figs. \ref{fig:studyB_edu} and \ref{fig:studyB_scores}.

\subsubsection{Greater Negative Sentiment Toward the Rule is Associated with Higher Comprehension Scores} \label{results:b:sentiment}

In \studyA{}, we found a relationship between participant sentiment towards the rule and comprehension score. To better interrogate this phenomenon, in \studyB{} we added two more questions to the survey to directly address the issue of sentiment, rather than relying on a free-response question. One (Q15) asks, ``To what extent do you agree with the following statement: I like the hiring rule?", and is evaluated on a five-point Likert scale from ``strongly agree" (1) to ``strongly disagree" (5). The other (Q16) asks, ``To what extent do you agree with the following statement: I agree with the hiring rule?", and is also evaluated on a five-point Likert scale from ``strongly agree" (1) to ``strongly disagree" (5).

\begin{figure}[t]
    \centering
    \includegraphics[width=0.8\columnwidth]{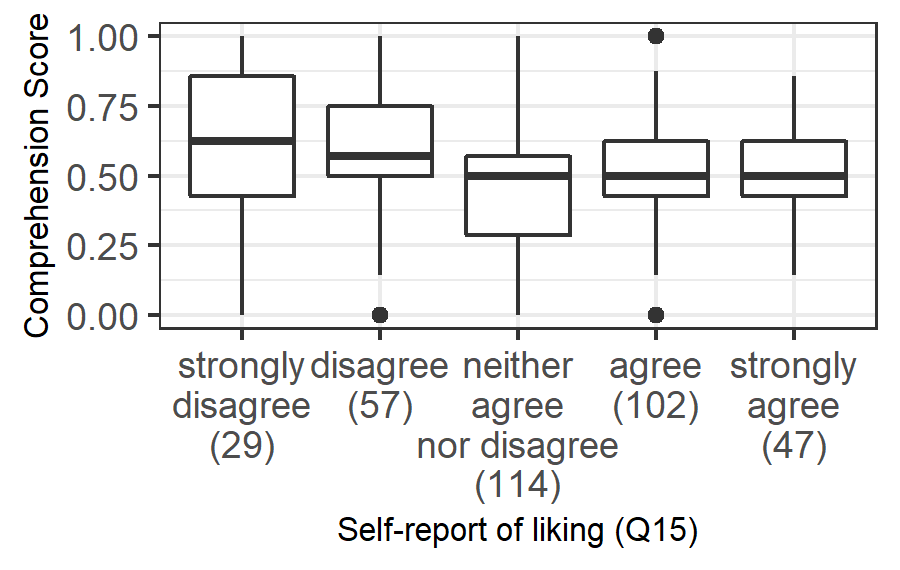}
    \vspace{-15pt}
    \caption{Comprehension score grouped by response to Q15. Dislike of the rule was associated with higher comprehension scores. X-axis is reversed for figure and correlation test.}
    \label{fig:studyB_q15}
    \vspace{10pt}
\end{figure}

\begin{figure}[t]
    \centering
    \includegraphics[width=0.8\columnwidth]{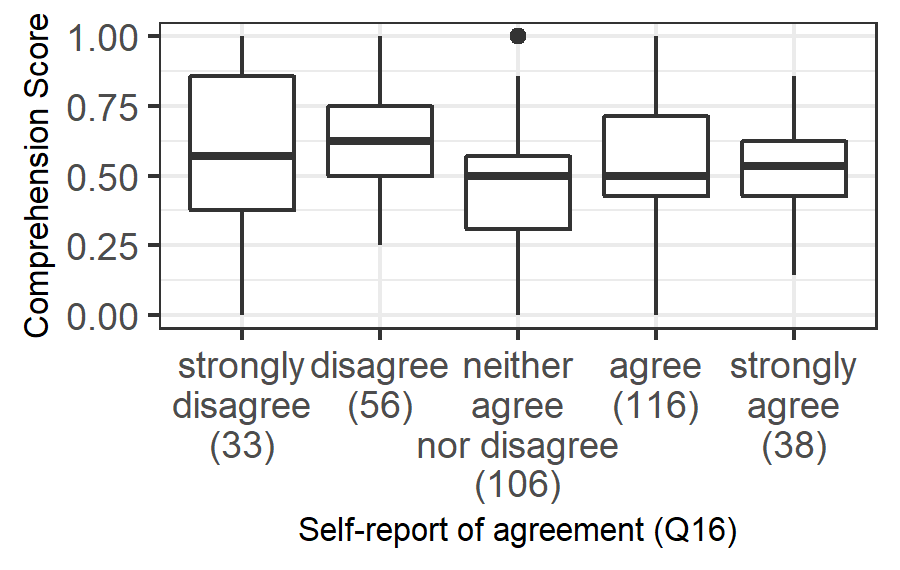}
    \vspace{-15pt}
    \caption{Comprehension score grouped by response to Q16. Disagreement with the rule was associated with higher comprehension score. X-axis is reversed for figure and correlation test.}
    \label{fig:studyB_q16}
    \vspace{-10pt}
\end{figure}

Using Spearman's rho, we assessed the correlation between responses to these two questions and comprehension score. A minor correlation was found between liking the rule and comprehension score, i.e. those who disliked the rule were more likely to have higher comprehension scores ($\rho = -0.11, p < 0.05$; see Fig.~\ref{fig:studyB_q15}).
A slight correlation was also found between agreeing with the rule and comprehension score, i.e. disagreement was associated with higher comprehension scores ($\rho = -0.11, p < 0.05$; see Fig.~\ref{fig:studyB_q16}).

\subsubsection{Non-Compliance is Associated with Lack of Understanding} \label{results:b:non-comp}

A final hypothesis generated in \studyA{} involves non-compliance: i.e., why do participants who report \textit{not} using the rule to answer the comprehension questions behave this way? 
In \studyA{}, we found that this was due to the fact that non-compliant participants were less able to \textit{understand} the rule, rather than because they did not \textit{like} it. 
We also observed this in our results from \studyB{}:
compliant participants exhibited higher self-reported understanding of the rule ($p < 0.001$, Fig. \ref{fig:studyB_nc_q13q14}), were more likely to correctly explain the rule ($p < 0.001$, Fig. \ref{fig:studyB_nc_q12q14}), and were more likely to dislike the rule ($p < 0.05$, Fig. \ref{fig:studyB_nc_q15q14}). We observed no relationship between compliance and agreement with the rule (Fig. \ref{fig:studyB_nc_q16q14}). Refer to \Appref{app:b:compliance} for more details.

\section{Discussion} \label{sec:discussion}

Bias in machine learning is a growing threat to justice; to date, ML bias has been documented in both commercial and government applications, in sectors such as medicine, criminal justice, and employment. In response, ML researchers have proposed various notions of \emph{fairness} to correct these biases. Most ML fairness definitions are purely mathematical, and require some knowledge of machine learning. While they are intended to benefit the general public, it is unclear whether the general public agrees with --- or even understands --- these notions of ML fairness.

We take an initial step to bridge this gap by asking \emph{do people understand the notions of fairness put forth by ML researchers?} To answer this question we develop a short questionnaire to assess understanding of three particular notions of ML fairness (demographic parity, equal opportunity, and equalized odds). We find that our comprehension score (with some adjustments for each definition) appears to be a consistent and reliable indicator of understanding the fairness metrics. 
The comprehension score demonstrated in this work lays a foundation for many future studies exploring other fairness definitions.

We do find, however, that comprehension is lower for equal opportunity, false negative rate than other definitions. 
In general, comprehension scores for equal opportunity (both FNR and FPR) were less internally consistent than other fairness rules, suggesting  participant responses were also more ``noisy'' for equal opportunity.
This is somewhat intuitive: equal opportunity is difficult to understand, as it only involves one type of error (FNR or FPR) rather than both.
Furthermore, FNR participants had the lowest comprehension scores \emph{and} the lowest consistency of all conditions.
We believe this finding also matches intuition: FNR is a strange notion in the context of hiring, as it concerns only those qualified applicants who were \emph{not} hired or offered jobs.
Indeed, in free-response questions several participants mentioned that they do not understand why qualified candidates are \emph{not} hired.
We believe many participants fixated on this strange setting, impacting their comprehension scores.
This finding is potentially problematic, as equal opportunity definitions are increasingly used in practice. Indeed, major fairness tools such as Google What-If tool \cite{wexler2019if} and the IBM AI Fairness 360 \cite{bellamy2019ai} specifically focus on equal opportunity. Further work should be put into making descriptions of nuanced fairness metrics more accessible.

Our analysis also identified other issues that should be considered when thinking about mathematical notions of fairness.
First, we find that education is a strong
predictor of comprehension. This is especially troubling, as the negative impacts of biased ML are expected to disproportionately impact the most marginalized~\cite{barocas2016big} and displace employment opportunities for those with the least education~\cite{frey2017future}. Lack of understanding may hamper these groups' ability to effectively advocate for themselves. Designing more accessible explanations of fairness should be a top research priority. 

Second, we find that those with the weakest comprehension of fairness metrics also express the least negative sentiment toward them. When fairness is a concern, there are always trade-offs---between accuracy and equity, or between different stakeholders, and so on. Balancing these trade-offs is an uncomfortable dilemma often lacking an objectively correct solution. It is possible that those who comprehend this dilemma \emph{also} recognize the precarious trade-off struck by any mathematical definition of fairness, and are therefore dissatisfied with it. From another perspective, this finding is more insidious. If those with the weakest understanding of AI bias are also least likely to protest, then major problems in algorithmic fairness may remain uncorrected.

\section*{Acknowledgments}
\jpd{Apparently these don't count toward the page limit for the body of the paper, ``It has now been decided that the acknowledgments section might appear after the 9th page (along with references). The paper checker has been updated accordingly.''}
Dickerson, McElfresh, and Schumann were supported in part by NSF CAREER Award IIS-1846237, DARPA GARD Award \#HR112020007, DARPA SI3-CMD Award \#S4761, DoD WHS Award \#HQ003420F0035, NIH R01 Award NLM-013039-01, and a Google Faculty Research Award.
We gratefully acknowledge funding support from the NSF (Grants 1844462 and 1844518).
The opinions in this paper are those of the authors and do not necessarily reflect the opinions of any funding sponsor or the United States Government.

\bibliography{refs}

\begin{thebibliography}{39}
\providecommand{\natexlab}[1]{#1}
\providecommand{\url}[1]{\texttt{#1}}
\expandafter\ifx\csname urlstyle\endcsname\relax
  \providecommand{\doi}[1]{doi: #1}\else
  \providecommand{\doi}{doi: \begingroup \urlstyle{rm}\Url}\fi

\bibitem[Akaike(1974)]{akaike1974}
Akaike, H.
\newblock A new look at the statistical model identification.
\newblock In \emph{IEEE Transactions on Automatic Control}, volume~19, pp.\
  716–723, 1974.
\newblock \doi{10.1109/TAC.1974.1100705}.

\bibitem[Angwin et~al.(2016)Angwin, Larson, Mattu, and
  Kirchner]{angwin2016machine}
Angwin, J., Larson, J., Mattu, S., and Kirchner, L.
\newblock Machine bias.
\newblock \emph{ProPublica, May}, 23:\penalty0 2016, 2016.

\bibitem[Barocas \& Selbst(2016)Barocas and Selbst]{barocas2016big}
Barocas, S. and Selbst, A.~D.
\newblock Big data's disparate impact.
\newblock \emph{Calif. L. Rev.}, 104:\penalty0 671, 2016.

\bibitem[Batanero et~al.(2016)Batanero, Chernoff, Engel, Lee, and
  S{\'a}nchez]{batanero2016research}
Batanero, C., Chernoff, E.~J., Engel, J., Lee, H.~S., and S{\'a}nchez, E.
\newblock Research on teaching and learning probability.
\newblock In \emph{Research on teaching and learning probability}, pp.\  1--33.
  Springer, Cham, 2016.

\bibitem[Bellamy et~al.(2019)Bellamy, Dey, Hind, Hoffman, Houde, Kannan, Lohia,
  Martino, Mehta, Mojsilovi{\'c}, et~al.]{bellamy2019ai}
Bellamy, R.~K., Dey, K., Hind, M., Hoffman, S.~C., Houde, S., Kannan, K.,
  Lohia, P., Martino, J., Mehta, S., Mojsilovi{\'c}, A., et~al.
\newblock Ai fairness 360: An extensible toolkit for detecting and mitigating
  algorithmic bias.
\newblock \emph{IBM Journal of Research and Development}, 63\penalty0
  (4/5):\penalty0 4--1, 2019.

\bibitem[Binns(2017)]{binns2017fairness}
Binns, R.
\newblock Fairness in machine learning: Lessons from political philosophy.
\newblock \emph{Proceedings of Machine Learning Research}, 81:\penalty0 1--11,
  2017.

\bibitem[Binns et~al.(2018)Binns, Van~Kleek, Veale, Lyngs, Zhao, and
  Shadbolt]{binns2018s}
Binns, R., Van~Kleek, M., Veale, M., Lyngs, U., Zhao, J., and Shadbolt, N.
\newblock `{I}t's reducing a human being to a percentage': Perceptions of
  justice in algorithmic decisions.
\newblock In \emph{Proceedings of the 2018 CHI Conference on Human Factors in
  Computing Systems}, pp.\  377. ACM, 2018.

\bibitem[Bureau(2017)]{census07}
Bureau, U.~C.
\newblock 2017 us census demographics, 2017.
\newblock URL \url{https://data.census.gov/cedsci}.

\bibitem[Caliskan et~al.(2017)Caliskan, Bryson, and
  Narayanan]{caliskan2017semantics}
Caliskan, A., Bryson, J.~J., and Narayanan, A.
\newblock Semantics derived automatically from language corpora contain
  human-like biases.
\newblock \emph{Science}, 356\penalty0 (6334):\penalty0 183--186, 2017.

\bibitem[Chouldechova(2017)]{chouldechova2017fair}
Chouldechova, A.
\newblock Fair prediction with disparate impact: A study of bias in recidivism
  prediction instruments.
\newblock \emph{Big data}, 5\penalty0 (2):\penalty0 153--163, 2017.

\bibitem[Chouldechova \& Roth(2018)Chouldechova and
  Roth]{chouldechova2018frontiers}
Chouldechova, A. and Roth, A.
\newblock The frontiers of fairness in machine learning.
\newblock \emph{arXiv preprint arXiv:1810.08810}, 2018.

\bibitem[Cint()]{cint}
Cint.
\newblock Cint.
\newblock URL \url{https://www.cint.com/}.

\bibitem[Datta et~al.(2015)Datta, Tschantz, and Datta]{datta2015automated}
Datta, A., Tschantz, M.~C., and Datta, A.
\newblock Automated experiments on ad privacy settings.
\newblock \emph{Proceedings on Privacy Enhancing Technologies}, 2015\penalty0
  (1):\penalty0 92--112, 2015.

\bibitem[Everitt \& Skrondal(2010)Everitt and Skrondal]{everitt2010}
Everitt, B. and Skrondal, A.
\newblock \emph{The Cambridge Dictionary of Statistics}.
\newblock Cambridge University Press, 4th edition, 2010.

\bibitem[Feldman et~al.(2015)Feldman, Friedler, Moeller, Scheidegger, and
  Venkatasubramanian]{feldman2015certifying}
Feldman, M., Friedler, S.~A., Moeller, J., Scheidegger, C., and
  Venkatasubramanian, S.
\newblock Certifying and removing disparate impact.
\newblock In \emph{Proceedings of the 21th ACM SIGKDD International Conference
  on Knowledge Discovery and Data Mining}, pp.\  259--268. ACM, 2015.

\bibitem[Frey \& Osborne(2017)Frey and Osborne]{frey2017future}
Frey, C.~B. and Osborne, M.~A.
\newblock The future of employment: How susceptible are jobs to
  computerisation?
\newblock \emph{Technological forecasting and social change}, 114:\penalty0
  254--280, 2017.

\bibitem[Gigerenzer \& Edwards(2003)Gigerenzer and
  Edwards]{gigerenzer2003simple}
Gigerenzer, G. and Edwards, A.
\newblock Simple tools for understanding risks: from innumeracy to insight.
\newblock \emph{Bmj}, 327\penalty0 (7417):\penalty0 741--744, 2003.

\bibitem[Gigerenzer et~al.(2007)Gigerenzer, Gaissmaier, Kurz-Milcke, Schwartz,
  and Woloshin]{gigerenzer2007helping}
Gigerenzer, G., Gaissmaier, W., Kurz-Milcke, E., Schwartz, L.~M., and Woloshin,
  S.
\newblock Helping doctors and patients make sense of health statistics.
\newblock \emph{Psychological science in the public interest}, 8\penalty0
  (2):\penalty0 53--96, 2007.

\bibitem[Grgic-Hlaca et~al.(2018)Grgic-Hlaca, Redmiles, Gummadi, and
  Weller]{grgic2018human}
Grgic-Hlaca, N., Redmiles, E.~M., Gummadi, K.~P., and Weller, A.
\newblock Human perceptions of fairness in algorithmic decision making: A case
  study of criminal risk prediction.
\newblock In \emph{Proceedings of the 2018 World Wide Web Conference}, pp.\
  903--912. International World Wide Web Conferences Steering Committee, 2018.

\bibitem[Hardt et~al.(2016)Hardt, Price, and Srebro]{Hardt16:Equality}
Hardt, M., Price, E., and Srebro, N.
\newblock Equality of opportunity in supervised learning.
\newblock In \emph{NeurIPS}, pp.\  3315--3323, 2016.

\bibitem[Harrell \& Bradley(2009)Harrell and Bradley]{harrell2009data}
Harrell, M.~C. and Bradley, M.~A.
\newblock Data collection methods. semi-structured interviews and focus groups.
\newblock Technical report, Rand National Defense Research Inst santa monica
  ca, 2009.

\bibitem[Harrison et~al.(2020)Harrison, Hanson, Jacinto, Ramirez, and
  Ur]{harrison2020empirical}
Harrison, G., Hanson, J., Jacinto, C., Ramirez, J., and Ur, B.
\newblock An empirical study on the perceived fairness of realistic, imperfect
  machine learning models.
\newblock In \emph{Proceedings of the 2020 Conference on Fairness,
  Accountability, and Transparency}, pp.\  392--402, 2020.

\bibitem[Hogarth \& Soyer(2015)Hogarth and Soyer]{hogarth2015providing}
Hogarth, R.~M. and Soyer, E.
\newblock Providing information for decision making: Contrasting description
  and simulation.
\newblock \emph{Journal of Applied Research in Memory and Cognition},
  4\penalty0 (3):\penalty0 221--228, 2015.

\bibitem[Huysmans et~al.(2011)Huysmans, Dejaeger, Mues, Vanthienen, and
  Baesens]{Huysmans2011}
Huysmans, J., Dejaeger, K., Mues, C., Vanthienen, J., and Baesens, B.
\newblock An empirical evaluation of the comprehensibility of decision table,
  tree and rule based predictive models.
\newblock \emph{Decis. Support Syst.}, 51\penalty0 (1):\penalty0 141--154,
  April 2011.
\newblock ISSN 0167-9236.
\newblock \doi{10.1016/j.dss.2010.12.003}.
\newblock URL \url{http://dx.doi.org/10.1016/j.dss.2010.12.003}.

\bibitem[Kusner et~al.(2017)Kusner, Loftus, Russell, and
  Silva]{kusner2017counterfactual}
Kusner, M.~J., Loftus, J., Russell, C., and Silva, R.
\newblock Counterfactual fairness.
\newblock In \emph{Advances in Neural Information Processing Systems}, pp.\
  4066--4076, 2017.

\bibitem[Lee(2018)]{lee2018understanding}
Lee, M.~K.
\newblock Understanding perception of algorithmic decisions: Fairness, trust,
  and emotion in response to algorithmic management.
\newblock \emph{Big Data \& Society}, 5\penalty0 (1):\penalty0
  2053951718756684, 2018.

\bibitem[Lee \& Baykal(2017)Lee and Baykal]{Lee2017}
Lee, M.~K. and Baykal, S.
\newblock Algorithmic mediation in group decisions: Fairness perceptions of
  algorithmically mediated vs. discussion-based social division.
\newblock In \emph{Proceedings of the 2017 ACM Conference on Computer Supported
  Cooperative Work and Social Computing}, CSCW '17, pp.\  1035--1048, New York,
  NY, USA, 2017. ACM.
\newblock ISBN 978-1-4503-4335-0.
\newblock \doi{10.1145/2998181.2998230}.
\newblock URL \url{http://doi.acm.org/10.1145/2998181.2998230}.

\bibitem[Lee et~al.(2019)Lee, Jain, Cha, Ojha, and Kusbit]{Lee2019}
Lee, M.~K., Jain, A., Cha, H.~J., Ojha, S., and Kusbit, D.
\newblock Procedural justice in algorithmic fairness: Leveraging transparency
  and outcome control for fair algorithmic mediation.
\newblock In \emph{Proc. ACM Hum.-Comput. Interact., 3, CSCW}, pp.\  Article
  182, New York, NY, USA, 2019. ACM.
\newblock URL \url{https://doi.org/10.1145/3359284}.

\bibitem[Lipton(2018)]{lipton2018mythos}
Lipton, Z.~C.
\newblock The mythos of model interpretability.
\newblock \emph{Communications of the ACM}, 61\penalty0 (10):\penalty0 36--43,
  2018.

\bibitem[Nunnally(1978)]{nunnally1978}
Nunnally, J.
\newblock \emph{Psychometric Theory}.
\newblock McGraw-Hill, 2nd edition, 1978.

\bibitem[Plane et~al.(2017)Plane, Redmiles, Mazurek, and
  Tschantz]{plane2017exploring}
Plane, A.~C., Redmiles, E.~M., Mazurek, M.~L., and Tschantz, M.~C.
\newblock Exploring user perceptions of discrimination in online targeted
  advertising.
\newblock In \emph{26th {USENIX} Security Symposium ({USENIX} Security 17)},
  pp.\  935--951, 2017.

\bibitem[Pleiss et~al.(2017)Pleiss, Raghavan, Wu, Kleinberg, and
  Weinberger]{pleiss2017fairness}
Pleiss, G., Raghavan, M., Wu, F., Kleinberg, J., and Weinberger, K.~Q.
\newblock On fairness and calibration.
\newblock In \emph{Advances in Neural Information Processing Systems}, pp.\
  5680--5689, 2017.

\bibitem[Rawls(1971)]{Rawls71a}
Rawls, J.
\newblock \emph{A Theory of Justice}.
\newblock Harvard University Press, 1971.

\bibitem[Redmiles et~al.(2019)Redmiles, Kross, and Mazurek]{redmiles2019well}
Redmiles, E.~M., Kross, S., and Mazurek, M.~L.
\newblock How well do my results generalize? comparing security and privacy
  survey results from mturk, web, and telephone samples.
\newblock In \emph{2019 IEEE Symposium on Security and Privacy (SP)}, pp.\
  1326--1343. IEEE, 2019.

\bibitem[Ribeiro et~al.(2016)Ribeiro, Singh, and Guestrin]{ribeiro2016should}
Ribeiro, M.~T., Singh, S., and Guestrin, C.
\newblock Why should i trust you?: Explaining the predictions of any
  classifier.
\newblock In \emph{Proceedings of the 22nd ACM SIGKDD international conference
  on knowledge discovery and data mining}, pp.\  1135--1144. ACM, 2016.

\bibitem[Saxena et~al.(2020)Saxena, Huang, DeFilippis, Radanovic, Parkes, and
  Liu]{saxena2020fairness}
Saxena, N.~A., Huang, K., DeFilippis, E., Radanovic, G., Parkes, D.~C., and
  Liu, Y.
\newblock How do fairness definitions fare? {T}esting public attitudes towards
  three algorithmic definitions of fairness in loan allocations.
\newblock \emph{Artificial Intelligence}, 283:\penalty0 103238, 2020.

\bibitem[Srivastava et~al.(2019)Srivastava, Heidari, and
  Krause]{srivastava2019mathematical}
Srivastava, M., Heidari, H., and Krause, A.
\newblock Mathematical notions vs. human perception of fairness: {A}
  descriptive approach to fairness for machine learning.
\newblock \emph{CoRR}, abs/1902.04783, 2019.
\newblock URL \url{http://arxiv.org/abs/1902.04783}.

\bibitem[Wexler et~al.(2019)Wexler, Pushkarna, Bolukbasi, Wattenberg,
  Vi{\'e}gas, and Wilson]{wexler2019if}
Wexler, J., Pushkarna, M., Bolukbasi, T., Wattenberg, M., Vi{\'e}gas, F., and
  Wilson, J.
\newblock The what-if tool: Interactive probing of machine learning models.
\newblock \emph{IEEE transactions on visualization and computer graphics},
  26\penalty0 (1):\penalty0 56--65, 2019.

\bibitem[Woodruff et~al.(2018)Woodruff, Fox, Rousso-Schindler, and
  Warshaw]{woodruff2018qualitative}
Woodruff, A., Fox, S.~E., Rousso-Schindler, S., and Warshaw, J.
\newblock A qualitative exploration of perceptions of algorithmic fairness.
\newblock In \emph{Proceedings of the 2018 CHI Conference on Human Factors in
  Computing Systems}, pp.\  656. ACM, 2018.

\end{thebibliography}
\bibliographystyle{icml2020}

\appendix
\clearpage

\section{Methods}
\subsection{Cognitive Interviews} \label{methods:design:cog}
We recruited $9$ participants from the DC Metropolitan area using Craigslist. We required participants to be over 18 years of age and fluent in English. Participants ranged between the ages of 20 and 66.
These interviews took place on the University of Maryland campus and lasted about $1$ hour. All participants signed a written consent form prior to the interview, and were paid \$30 for their time.

During these interviews, participants completed a preliminary version of the survey used in \studyA{}. 
After each survey question, we asked the participants several interview questions related to their comprehension of and feelings toward the survey. We found that some participants tended to use their own personal notions of fairness when answering comprehension questions rather than using the definition we provided. We were concerned that this would limit our ability to effectively measure comprehension. To address this problem, we rewrote several parts of our survey and added two new questions (Q14 and Q15).

\subsection{Non-Expert Verification}

We designed this study to assess \emph{non-expert} understanding and opinions of ML fairness metrics. To this end, we asked respondents to self-rate their level of expertise in a variety of fields, including ML, at the end of the survey (see \Appref{app:demographics}). A number of participants did report having ``expert" level experience in ML ($n = 2$ out of 147 in \studyA{}, and $n = 15$ out of 349 in \studyB{}). We considered removing these participants from the analyses, but ultimately did not because there was no relationship between self-reported ML expertise and comprehension score (Spearman's rho, for both studies).

\section{\studyA{}: Detailed Results} \label{app:studyA_results}

\subsection{Our Survey Effectively Captures Rule Comprehension} \label{results:1:rq1}

\begin{figure}[h]
    \centering
    \includegraphics[width=0.8\columnwidth]{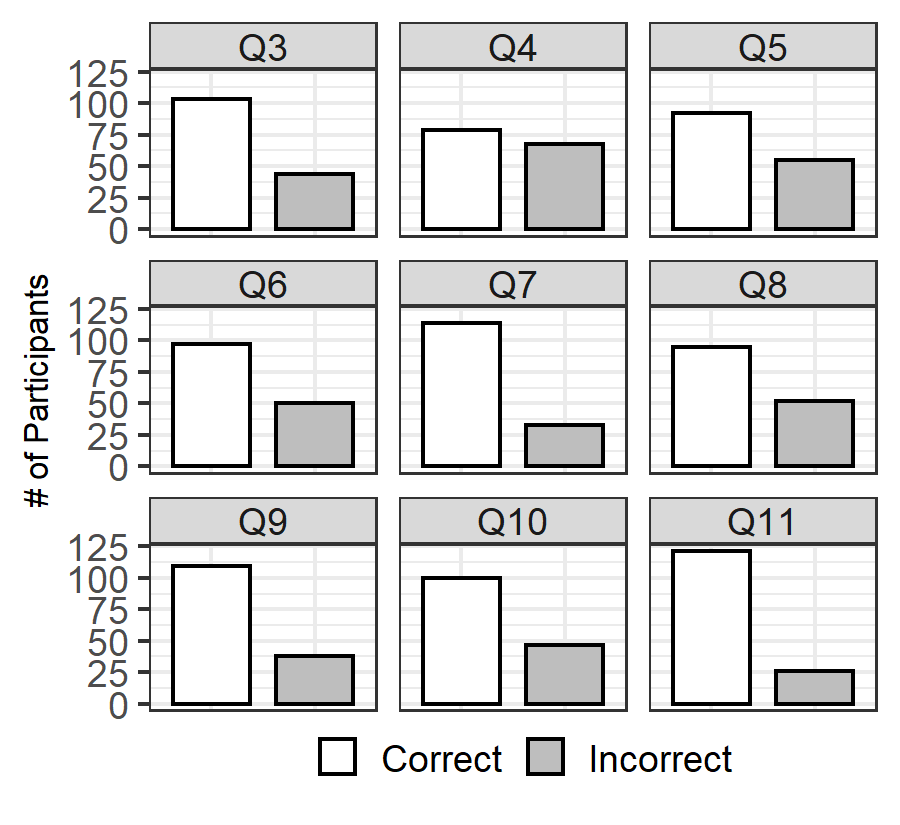}
    \vspace{-10pt}
    \caption{Number of participants answering each question correctly. Each panel contains all 147 participants. 
    }
    \label{fig:question_breakdown}
    \vspace{-5pt}
\end{figure}

We find that our survey is internally consistent, and effectively measures participant comprehension of demographic parity. The former we evaluated using Cronbach's $\alpha$ and item-total correlation (discussed in \S\ref{results:a:validity}), and the latter using two self-report measures and one free response question.
See Fig.~\ref{fig:question_breakdown} for participant performance per question.

\subsubsection{Self-reported rule understanding and use are reflected in comprehension score}
\begin{figure}[ht]
    \centering
    \includegraphics[width=0.8\columnwidth]{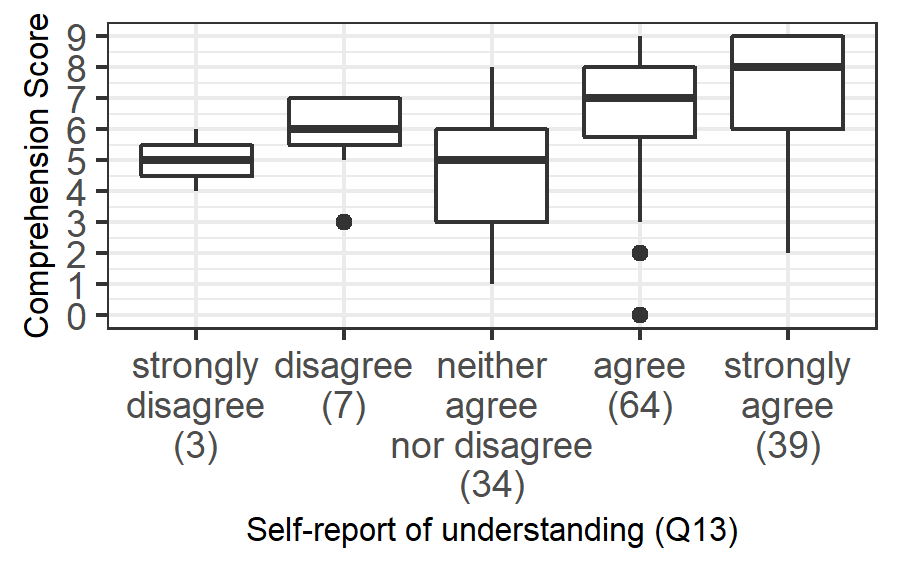}
    \vspace{-10pt}
    \caption{Comprehension score grouped by response to Q13. Self-reported understanding of the rule was associated with higher comprehension scores. X-axis is reversed for figure and correlation test.}
    \label{fig:q13}
\end{figure}

First, we compared comprehension score to self-reported rule understanding (Q13). Higher comprehension scores were associated with greater confidence in understanding (Spearman's rho), suggesting that participants were accurately assessing their ability to apply the rule (see Fig. \ref{fig:q13}).

\begin{figure}[ht]
    \centering
    \includegraphics[width=0.8\columnwidth]{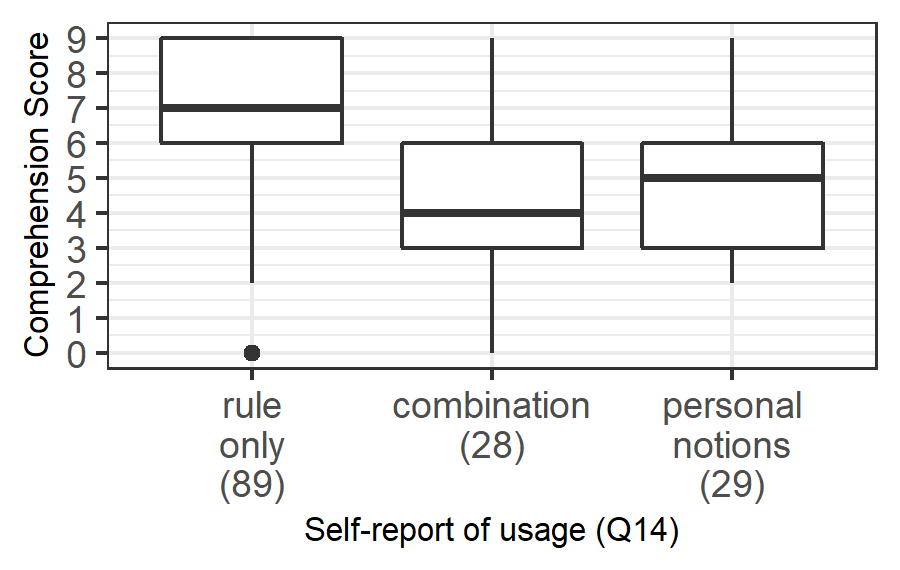}
    \vspace{-10pt}
    \caption{Comprehension score grouped by response to Q14. Rule compliance (leftmost on the x-axis) was associated with higher comprehension scores. One participant who did not provide a response was excluded from the figure and relevant analysis.}
    \label{fig:q14}
\end{figure}

Next, we compared comprehension score to a self-report question about the participant's use of the rule (Q14)
Participants who claimed to use only the rule tended to score higher than those who used their own notions of fairness or a combination thereof (K-W test, and post-hoc M-WU), suggesting that participants are answering somewhat honestly: when they try to apply the rule, comprehension scores improve (see Fig.~\ref{fig:q14}).

\subsubsection{Participants with higher comprehension scores are better able to explain the rule}
\begin{figure}[ht]
    \centering
    \includegraphics[width=0.8\columnwidth]{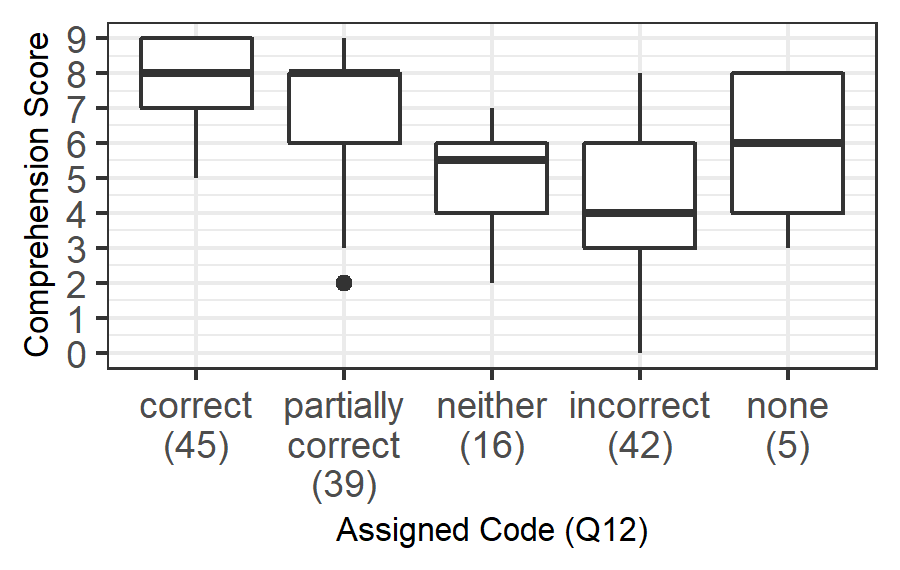}
    \vspace{-10pt}
    \caption{Comprehension score grouped by code assigned to Q12 response. Participants who provided either correct or partially correct responses tended to perform better.}
    \label{fig:q12}
    \vspace{-10pt}
\end{figure}

To further validate our comprehension score, we asked participants to explain the rule in their own words (Q12). %
Responses were qualitatively coded as one of five categories: \textbf{correct}, \textbf{partially correct}, \textbf{neither}, \textbf{incorrect}, or \textbf{none} (as discussed in \S\ref{results:a:validity}). The results of this coding can be seen can be seen in Fig. \ref{fig:q12}. Participants providing correct explanations of the rule attained higher comprehension scores (k-W test, and post-hoc M-WU), further corroborating our claim that our comprehension score is a valid measure of fairness rule comprehension.

\subsection{Education Influences Comprehension} \label{results:1:rq2}

During the cognitive interview phase, we observed a possible trend of comprehension scores being lower for older participants and those with less educational attainment. If true, this would suggest that fairness explanations should be carefully validated to ensure they can be used with diverse populations. We investigated this hypothesis, in an exploratory fashion, using poisson regression models.

Three models were tested. The first regressed score against all four demographic categories as predictors (gender, age, ethnicity, and education), the second omitted education, and the third tested only education. Models were compared using Akaike information criterion (AIC), a standard method of evaluating model quality and performing model selection \cite{akaike1974}. Comparison by AIC revealed that model 1 (all four categories) was a better predictor for comprehension score than models 2 or 3 (AIC = 643.3, 651.2, and 660.5, respectively; difference = 0.0, 7.9, and 17.1). 
In model 1, only education showed correlation with comprehension score (effect size = $1.40$, $p<0.05$). %
Further work is needed to confirm this exploratory result.

\begin{figure}[ht]
    \centering
    \includegraphics[width=0.8\columnwidth]{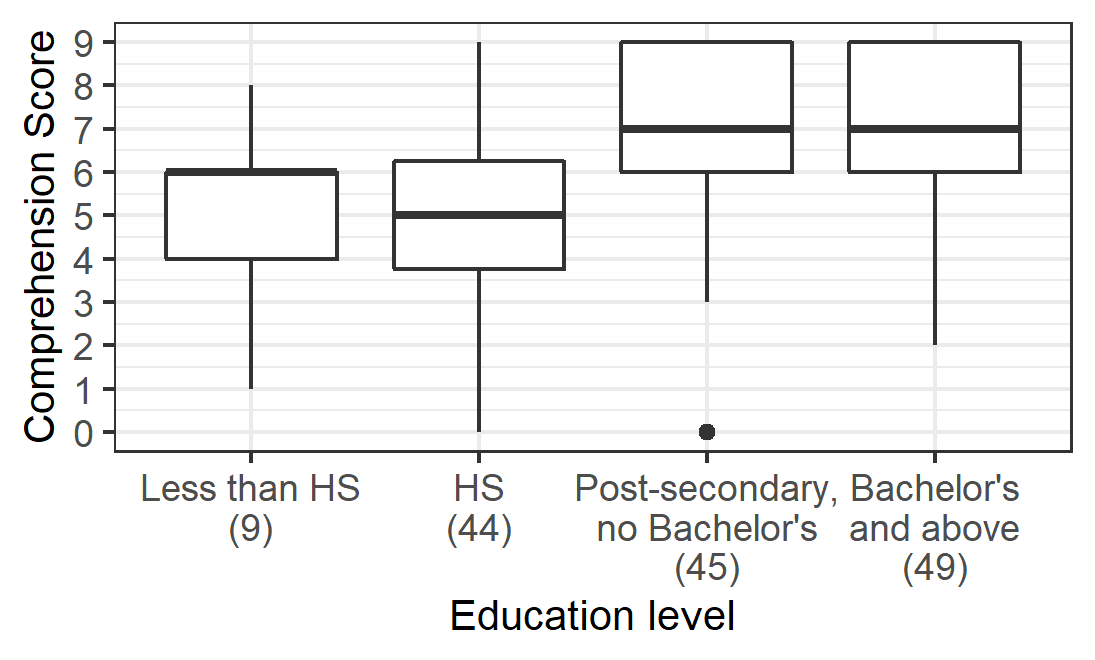}
    \vspace{-10pt}
    \caption{Comprehension score grouped by education level. Higher education level was associated with higher comprehension scores.}
    \label{fig:edu}
    \vspace{-10pt}
\end{figure}

\subsection{Disagreement with the Rule is Associated with Higher Comprehension Scores} \label{results:1:rq3}

Participants were asked for their opinion on the presented rule in another free response question (Q15). These responses were then qualitatively coded to capture participant sentiment towards the rule as one of five categories: \textbf{agree}, \textbf{depends}, \textbf{disagree}, \textbf{not understood}, or \textbf{none} (as discussed in \S\ref{results:a:hypotheses}).

\begin{figure}[ht]
    \centering
    \includegraphics[width=0.8\columnwidth]{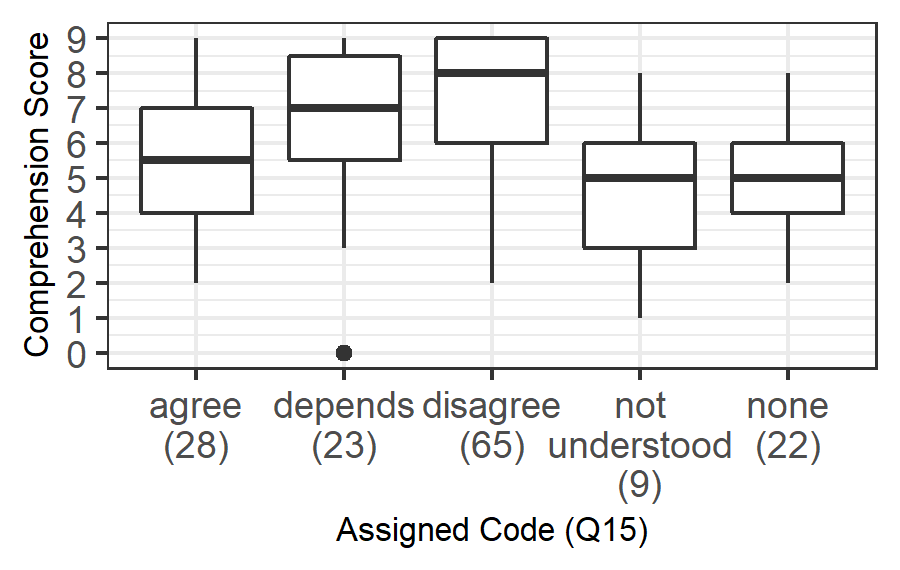}
    \caption{Comprehension score grouped by code assigned to Q15 response. Participants who exhibited negative sentiment toward the rule responses tended to perform better.}
    \label{fig:q15}
    \vspace{-10pt}
\end{figure}
This question was added based on the cognitive interviews (see \Appref{methods:design:cog}), where perception seemed to influence compliance.
The results of coding Q15 can be seen in Fig. \ref{fig:q15}. Participants who expressed disagreement with the rule performed better than those who expressed agreement, did not understand the rule, or provided no response to the question (K-W test, post-hoc M-WU). Note that this result should not be interpreted as an overall finding on the appropriateness of demographic parity. Instead we anticipate the perceptions of appropriateness of any fairness definition will be highly context-dependent.

\subsection{Non-Compliance is Associated with Lack of Understanding} \label{results:a:non-comp}

We were interested in understanding why some participants failed to adhere to the rule, as measured by their self-report of rule usage in Q14. 
After labeling participants as either ``non-compliant" (NC, $n=57)$ or ``compliant" (C, $n=89$), we conducted a series of $\chi^2$ tests to investigate this phenomenon.

\begin{figure}[h]
    \centering
    \includegraphics[width=0.8\columnwidth]{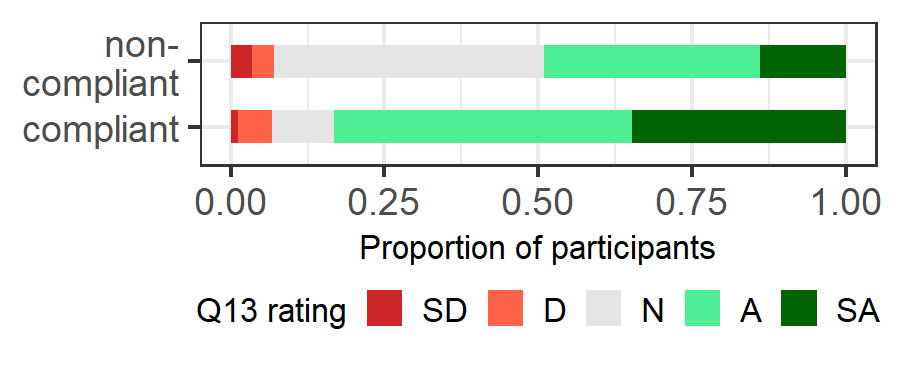}
    \vspace{-15pt}
    \caption{Self-report of understanding (Q13) split by compliance (Q14). NC participants tend to report less confidence in their ability to apply the rule. SD = strongly disagree, D = disagree, N = neither agree nor disagree, A = agree, SA = strongly agree.}
    \label{fig:q13q14}
    \vspace{-5pt}
\end{figure}

\begin{figure}[h]
    \centering
    \includegraphics[width=0.8\columnwidth]{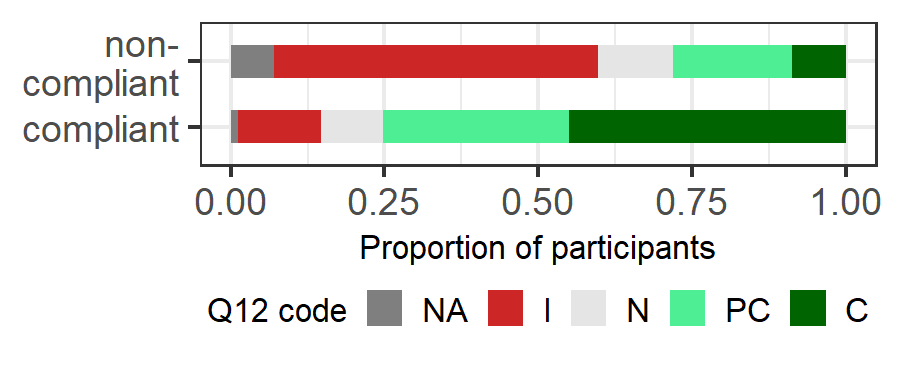}
    \vspace{-15pt}
    \caption{Correctness of rule explanation (Q12) split by compliance (Q14). NC participants tend to be less able to explain the presented rule in their own words. NA = none, I = incorrect, N = neither, PC = partially correct, C = correct.}
    \label{fig:q12q14}
    \vspace{-5pt}
\end{figure}

\begin{figure}[h]
    \centering
    \includegraphics[width=0.8\columnwidth]{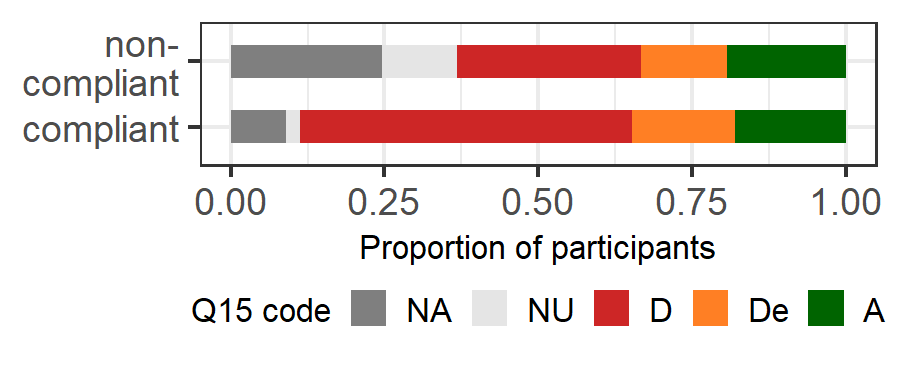}
    \vspace{-15pt}
    \caption{Participant agreement with rule (Q15) split by compliance (Q14). NC participants tend to harbor less negative sentiment towards the rule. NA = none, NU = not understood, D = disagree, De = depends, A = agree.}
    \label{fig:q15q14}
    \vspace{-5pt}
\end{figure}

Non-compliant participants were less likely to self-report high understanding of the rule in Q13 (see Fig. \ref{fig:q13q14}). %
Moreover, non-compliance also appears to be associated with a reduced ability to correctly explain the rule in Q12 (see Fig. \ref{fig:q12q14}). %
Further, negative participant sentiment towards the rule (Q15) also appears to be associated with greater compliance (see Fig. \ref{fig:q15q14}). %
Thus, non-compliant participants appear to behave this way because they do not \emph{understand} the rule, rather than because they do not \emph{like} it.

\subsection{Decision Scenarios} \label{app:scenario_analysis}

For \studyA{} we designed three decision-making scenarios to test whether the perceived importance or realism of a particular scenario influenced comprehension score. They are as follows:
\begin{itemize} \itemsep=0cm
    \item \textbf{Art Project (AP):} distributing awards for art projects to primary school students,
    \item \textbf{Employee Awards (EA):} distributing employee awards at a sales company, and
    \item \textbf{Hiring (HR):} distributing job offers to applicants.
\end{itemize}
In each scenario the students/employees/applicants are partitioned into two groups (parents' occupation for the first scenario, and binary gender for the other two scenarios).
We use a between-subjects design: participants are randomly partitioned into three conditions, one for each scenario (AP, EA, or HR).
For each condition we define the \emph{fairness rule} in the context of the decision-making scenario (see Appendix~\ref{app:survey} for the full surveys).

Next we describe our main conclusion related to the different decision-making scenarios in \studyA: the scenario does not influence comprehension score.

\subsubsection{Scenario does not Influence Comprehension Scores (RQ4)} \label{results:rq4}

We were concerned that less important and/or realistic scenarios would cause participants to take the survey less seriously, and therefore perform more poorly.
To test this,
participants were randomly assigned to a scenario, resulting in the following distribution: AP = 41, EA = 49, HR = 57.

\begin{figure}[ht]
    \centering
    \includegraphics[width=0.8\columnwidth]{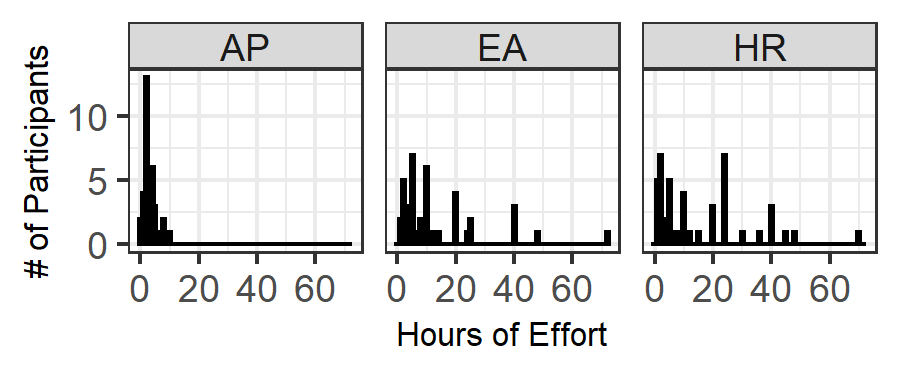}
    \vspace{-10pt}
    \caption{Importance of a scenario by proxy of hours of effort necessary to make a decision in each scenario. AP merited less hours of effort than both EA and HR.}
    \label{fig:q2}
\end{figure}

A K-W test revealed no differences between scenarios in terms of comprehension score (mean comprehension scores: AP = 6.0, EA = 6.74, HR = 5.86%
). However, differences did exist between scenarios in terms of importance (assessed in Q2), measured in hours of effort deemed necessary to make the relevant decision (K-W, %
$p<0.001$). Post-hoc M-WU revealed that participants believed making a decision in the AP scenario merited fewer hours of effort (mean = 3.15hrs) than in the EA (13.52hrs, $p<0.001$)
or HR (15.23hrs, $p<0.001$)
scenarios (corrected $\alpha=0.05/3=0.017$). See Fig. \ref{fig:q2} for distributions of responses.

Of note, it is possible that perceived realism, assessed in Q1 on a five-point Likert scale, was also influenced by scenario (K-W, $p=0.051$), but we may need larger sample sizes to confirm this. Regardless, while the nature of a scenario does influence participant perception in terms of importance and (possibly) realism, it does not appear to influence comprehension (at least for the scenarios we chose). For this reason, we chose to test a single scenario (HR) in \studyB{}. %

\section{\studyB{}: Detailed Results}

\subsection{Model Selection} \label{app:b:model_selection}

In \S\ref{results:b:edu} we assessed eleven linear regression models for predicting comprehension scores. The best fit model, determined by model selection via AIC, included only education (edu) and fairness definition (def) as regressors. The results of model selection are below in Table \ref{tab:AIC}.

\begin{table}[bht]
\centering
\caption{\label{tab:AIC} Models tested in \S\ref{results:b:edu}, sorted by best to least fit. The first model in the table (edu + def) is the model of best fit. dAIC = difference from model with lowest AIC value.}
\vspace{7pt}
{\small
\begin{tabular}{@{}lrr@{}}
    \toprule
    \textbf{Model regressors} & \textbf{AIC} & \textbf{dAIC} \\
    \midrule
    edu + def & -80.4 & 0 \\
    edu & -72.8 & 7.6 \\
    gender + edu & -70.3 & 10.1 \\
    age + edu & -63.7 & 16.7 \\
    gender + age + edu & -61.1 & 19.2 \\
    gender + age + eth + edu + def & -61.1 & 19.2 \\
    def & -60.8 & 19.6 \\
    gender + age + eth + edu & -55.5 & 24.9 \\
    gender + age + def & -46.4 & 34 \\
    gender + age + eth + def & -41.6 & 38.8 \\
    gender + age + eth & -37.2 & 43.2 \\
    \bottomrule
\end{tabular}%
}
\vspace{-10pt}
\end{table}

\subsection{Non-Compliance} \label{app:b:compliance}

In \S\ref{results:b:non-comp} we sought to further investigate the findings of \studyA{} with regards to compliance (Q14). To do so, we labeled those who responded (in \studyB{}) with either having used their own personal notions of fairness ($n=26)$ or some combination of their personal notions and the rule ($n=148$) as ``non-compliant" (NC), with the remaining $n=174$ labeled as ``compliant" (C). One participant who did not provide a response was excluded from this analysis, conducted using KW and $\chi^2$ tests.

\begin{figure}[h]
    \centering
    \includegraphics[width=0.8\columnwidth]{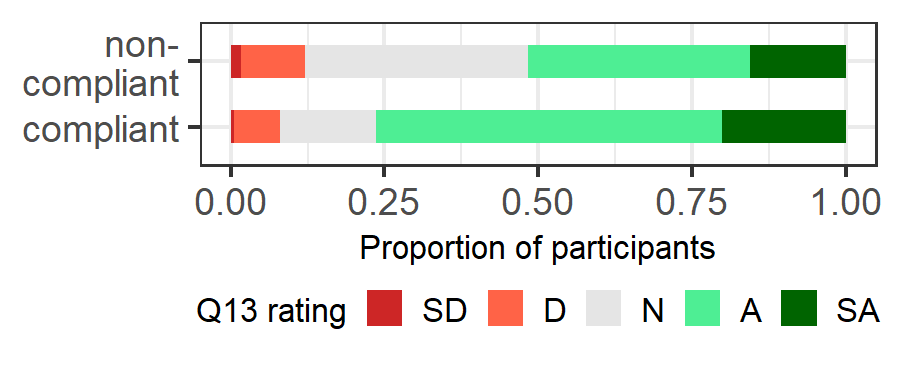}
    \vspace{-15pt}
    \caption{Self-report of understanding (Q13) split by compliance (Q14). NC participants tend to report less confidence in their ability to apply the rule. SD = strongly disagree, D = disagree, N = neither agree nor disagree, A = agree, SA = strongly agree.}
    \label{fig:studyB_nc_q13q14}
    \vspace{-5pt}
\end{figure}

\begin{figure}[h]
    \centering
    \includegraphics[width=0.8\columnwidth]{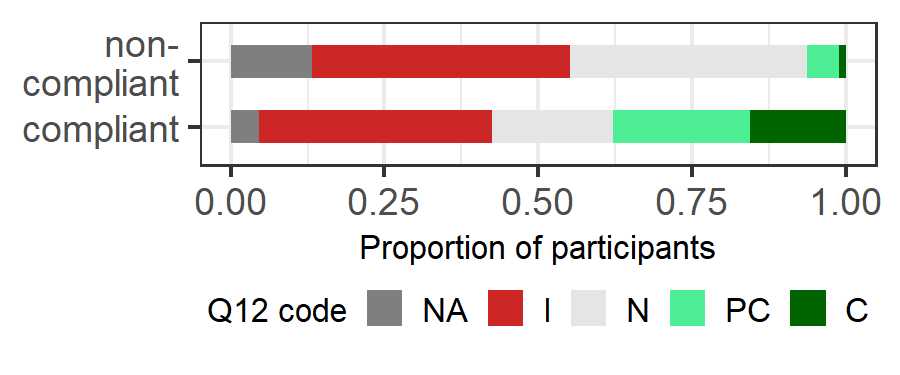}
    \vspace{-15pt}
    \caption{Correctness of rule explanation (Q12) split by compliance (Q14). NC participants tend to be less able to explain the presented rule in their own words. NA = none, I = incorrect, N = neither, PC = partially correct, C = correct.}
    \label{fig:studyB_nc_q12q14}
    \vspace{-5pt}
\end{figure}

Non-compliant participants were less likely to self-report high understanding of the rule in Q13 %
(KW test, $p<0.001$, see Fig. \ref{fig:studyB_nc_q13q14}). Moreover, non-compliance also appears to be associated with a reduced ability to correctly explain the rule in Q12 %
($\chi^2$ test, $p<0.001$, see Fig. \ref{fig:studyB_nc_q12q14}). This fits with the overall strong relationship we observed among comprehension scores, %
ability to explain the rule, and compliance. 

\begin{figure}[h]
    \centering
    \includegraphics[width=0.8\columnwidth]{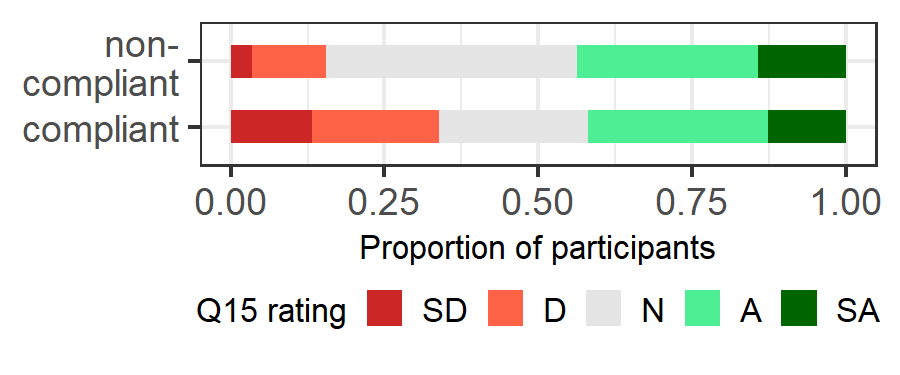}
    \vspace{-15pt}
    \caption{Participant liking for rule (Q15) split by compliance (Q14). NC participants tend to dislike the rule less than C participants. SD = strongly disagree, D = disagree, N = neither agree nor disagree, A = agree, SA = strongly agree.}
    \label{fig:studyB_nc_q15q14}
    \vspace{-5pt}
\end{figure}

\begin{figure}[h]
    \centering
    \includegraphics[width=0.8\columnwidth]{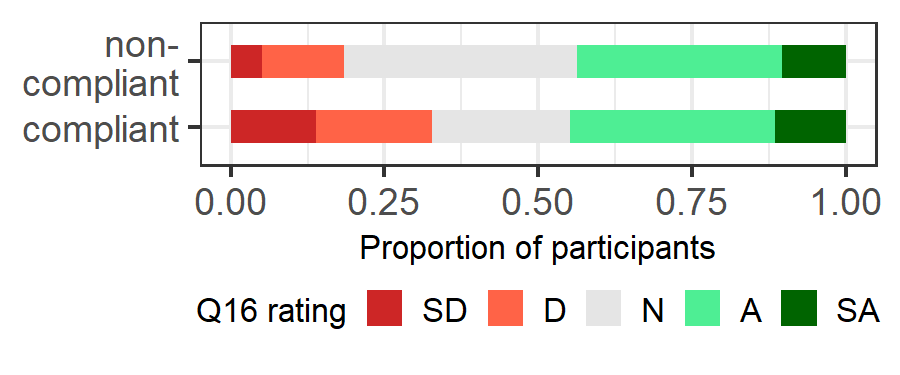}
    \vspace{-15pt}
    \caption{Participant agreement with rule (Q16) split by compliance (Q14). No differences were found between NC and C participants. SD = strongly disagree, D = disagree, N = neither agree nor disagree, A = agree, SA = strongly agree.}
    \label{fig:studyB_nc_q16q14}
    \vspace{-5pt}
\end{figure}

Further, greater dislike towards the rule (Q15) also appears to be associated with greater compliance %
(KW test, $p<0.05$, see Fig. \ref{fig:studyB_nc_q15q14}). %
However, there was no relationship between disagreement towards the rule (Q16) and compliance (see Fig. \ref{fig:studyB_nc_q16q14}).

These results largely corroborate the notion that non-compliant participants appear to behave this way because they do not \emph{understand} the rule, rather than because they do not \emph{like} it. %

\section{Surveys} \label{app:survey}

\subsection{\studyA{} Survey} \label{app:surveyA}

Each of the surveys are split into four main sections. The first section is the consent form which can be found in Appendix~\ref{app:consent}. The second section describes the scenario and asks questions about the given scenario (\S\ref{app:scenarios}). The third section describes the fairness metric, defined as the rule, used (in this case it is demographic parity) and asks specific questions about the metric (\S\ref{app:rules}). Finally the last section asks for demographic information (\S\ref{app:demographics}).

\subsubsection{Scenario descriptions and questions}\label{app:scenarios}
The following is shown to each participant:

It is very important that you read each question carefully and think about your answers. The success of our research relies on our respondents being thoughtful and taking this task seriously.
\begin{itemize}
    \item[\checkbox] I have read the above instructions carefully.
\end{itemize}

We then introduce one of three different decision making scenarios, described below, followed by two questions. Words that vary across scenario in the questions are shown as $<$art project, employee awards, hiring$>$.

\paragraph{Art project}
A fourth grade teacher is reviewing 20 student art projects. They will award lollipops to the top 4 students who put the most effort into their projects. The teacher knows that some of the students have artists as parents, who might have helped their children with their art project. The teacher’s goal is to give out lollipops only based on the amount of effort that the student \emph{themselves} put into their projects.

The teacher uses the following criteria to decide who should get a lollipop:
\begin{itemize}
    \item Elaborateness of each project.
    \item Creativity of each project.
\end{itemize}

About 50\% of the students have artists as parents, and 50\% do not.  

In the past, students with artists as parents typically put more effort into their projects. 

In this group of students there is a wide range of project quality (as measured by elaborateness and creativity). However, this range of quality is about the same between students with artists as parents and those without.

The teacher wants to make sure that they award lollipops in a fair way, no matter whether the students’ parents are artists or not.

\paragraph{Employee awards}
A manager at a sales company is deciding which of their 100 employees should receive each of 10 mid-year awards. The manager’s goal is to give awards to employees who \emph{will} have high net sales at the end of the year. 

The manager uses the following criteria to decide who should get an award:
\begin{itemize}
    \item Recent performance reviews
    \item Mid-year net sales
    \item Number of years on the job
\end{itemize}

About 50\% of the employees are men, and 50\% are women.  

In the past, men have achieved higher end-of-year net sales than women.

In this group of employees, there is a wide range of qualifications (as measured by performance reviews, mid-year net sales, and number of years on the job). However, this range of qualifications is about the same between male and female employees.

The manager wants to make sure that this awards process is fair to the employees, no matter their gender. 

\paragraph{Hiring}
A hiring manager at a new sales company is reviewing 100 new job applications. Each applicant has submitted a resume, and has had an interview. The manager will send job offers to 10 out of the 100 applicants. Their goal is to make offers to applicants who will have high net sales after a year on the job.  

The manager will use the following to decide which applicants should receive job offers:
\begin{itemize}
    \item Interview scores
    \item Quality of recommendation letters
    \item Number of years of prior experience in the field
\end{itemize}

About 50\% of the applicants are men, and 50\% are women.  

In the past, men have achieved higher net sales than women, after one year on the job. 

In this applicant pool there is a wide range of applicant quality (as measured by interview scores, recommendation letters, and years of prior experience in the field). However, the range of quality is about the same for both male and female applicants.

The hiring manager wants to make sure that this hiring process is fair to applicants, no matter their gender. 

\paragraph{Questions}
\begin{enumerate}
    \item To what extent do you agree with the following statement: a scenario similar to the one described above might occur in real life. 
    \begin{itemize}
        \item Strongly agree
        \item Agree
        \item Neither agree nor disagree
        \item Disagree
        \item Strongly Disagree
    \end{itemize}
    \item How much effort should the $<$teacher, manager, hiring manager$>$ put in to make sure this decision is fair? [short answer - number of hours]
\end{enumerate}

\subsubsection{Rule descriptions and questions}\label{app:rules}
Unless otherwise noted the rule description is shown above each of the questions for reference. Correct answers are noted in \correct{red}.

\paragraph{Art project}
The teacher uses the following award rule to distribute lollipops: \emph{The fraction of students who receive lollipops that have artist parents should equal the fraction of students in the class that have artist parents. Similarly, the fraction of students who receive lollipops that do not have artist parents should equal the fraction of students in the class that do not have artist parents.}

Example 1: If 10 out of the 20 students in the class have artist parents, then 2 out of the 4 lollipops would be awarded to students with artist parents (and the remaining 2 would be awarded to students without artist parents).

Example 2: If 5 out of the 20 students in the class have artist parents, then 1 out of the 4 lollipops would be awarded to students with artist parents (and the remaining 3 would be awarded to students without artist parents).

In the next section, we will ask you some questions about the information you have just read. Please note that this is not a test of your abilities. We want to measure the quality of the description you read, not your ability to take tests or answer questions.

\textbf{Please note that we ask you to apply and use ONLY the above award rule when answering the following questions. You will have an opportunity to state your opinions and feelings on the rule later in the survey.}

\begin{enumerate}
    \setcounter{enumi}{2}
    \item Suppose a different teacher is considering awarding lollipops to the whole 4th grade. There are 100 students with artist parents, and 200 students without artist parents. The teacher decides to award 10 lollipops to students with artist parents. \textbf{Assuming the teacher is required to use the award rule above}, how many students without artist parents need to receive lollipops?
    \begin{enumerate}
        \item 10
        \item \correct{20}
        \item 40
        \item 50
    \end{enumerate}
    \item \textbf{Assuming the teacher is required to use the award rule above}, in which of these cases can a teacher award more lollipops to students without artist parents than to students with artist parents?
    \begin{enumerate}
        \item When the students without artist parents have higher-quality projects (i.e., more elaborate and more creative) than those with artist parents.
        \item \correct{When there are more students without artist parents than those with artist parents.}
        \item When students without artist parents have more creative projects than those with artist parents.
        \item This cannot happen under the award rule.
    \end{enumerate}
    \item \textbf{Assuming the teacher is required to use the award rule above}, is the following statement \correct{TRUE} OR FALSE: Even if a student with artist parents has a project that is of the same quality (i.e., equally elaborate and equally creative) as another project by a student without artist parents, they can be treated differently (ie., only one of the students might get a lollipop).
    \item \textbf{Assuming the teacher is required to use the award rule above}, is the following statement TRUE OR \correct{FALSE}: If all students without artist parents have low-quality projects (i.e., low elaborateness and low creativity), but the teacher awards lollipops to some of them, then any lollipops awarded to students with artist parents must be awarded to those who have low-quality projects.
    \item \textbf{Assuming the teacher is required to use the award rule above}, is the following statement \correct{TRUE} OR FALSE: Suppose the teacher is distributing 10 lollipops amongst a pool of students that includes students with and without artist parents. Even if all students with artist parents have low-quality (i.e., low elaborateness and low creativity) projects, some of them must still receive lollipops.
    \item \textbf{Assuming the teacher is required to use the award rule above}, is the following statement TRUE OR \correct{FALSE}: This award rule always allows the teacher to award lollipops exclusively to the students who have the highest quality (i.e., most elaborate and most creative) projects.
\end{enumerate}

In the two examples above there are 20 students. Consider a different scenario, with \textbf{6 students -- 4 with artist parents and 2 without, as illustrated below}. The next three questions each give a potential outcome for all six students (i.e., which of the 6 students receive awards). Please indicate which of the outcomes follow \textbf{the award rule above}.

\vspace{10pt}
\includegraphics[height=1in]{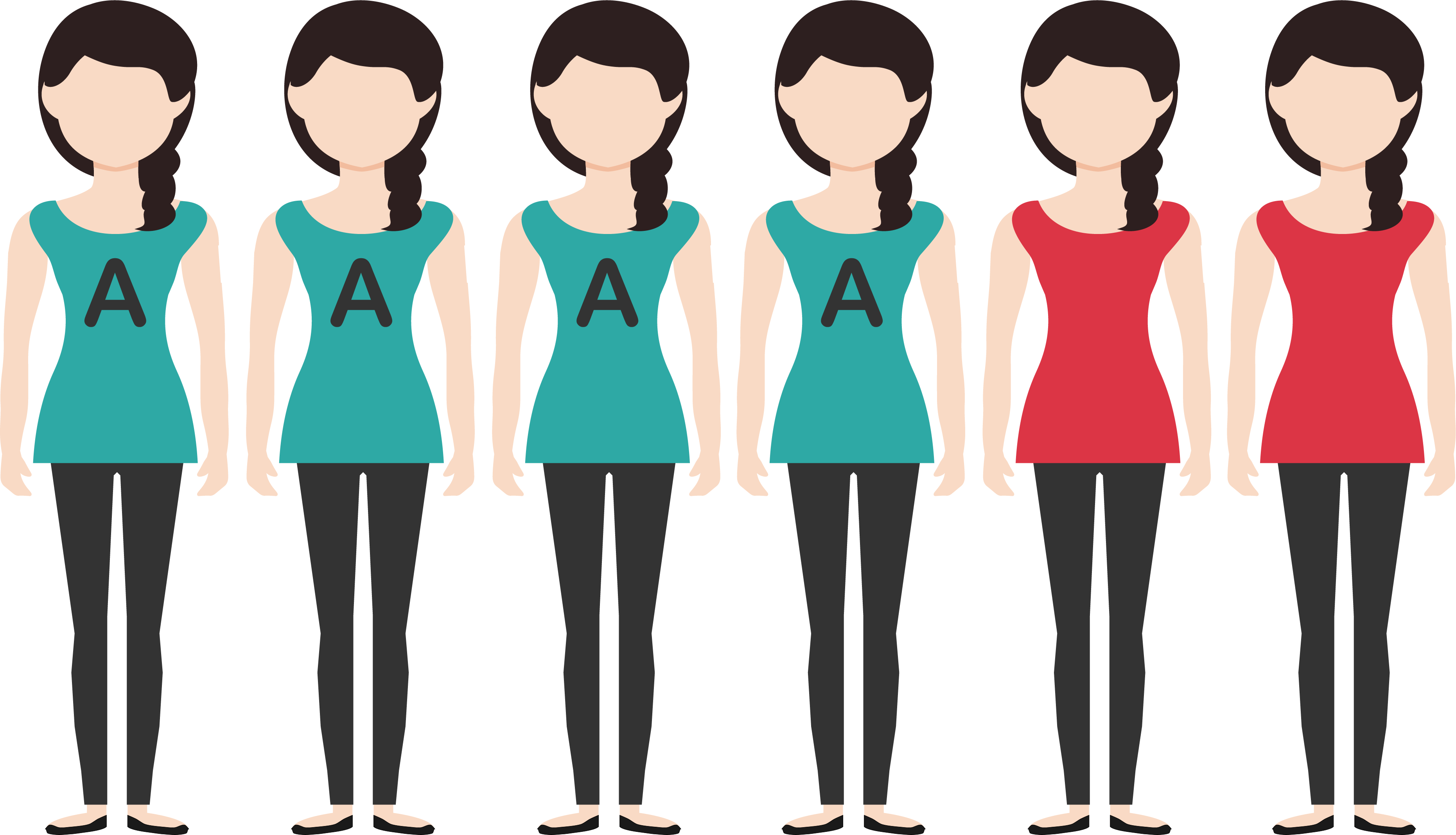}

\begin{enumerate}
    \setcounter{enumi}{8}
    \item Alternative scenario 1:
    
    \vspace{10pt}
    \includegraphics[height=1in]{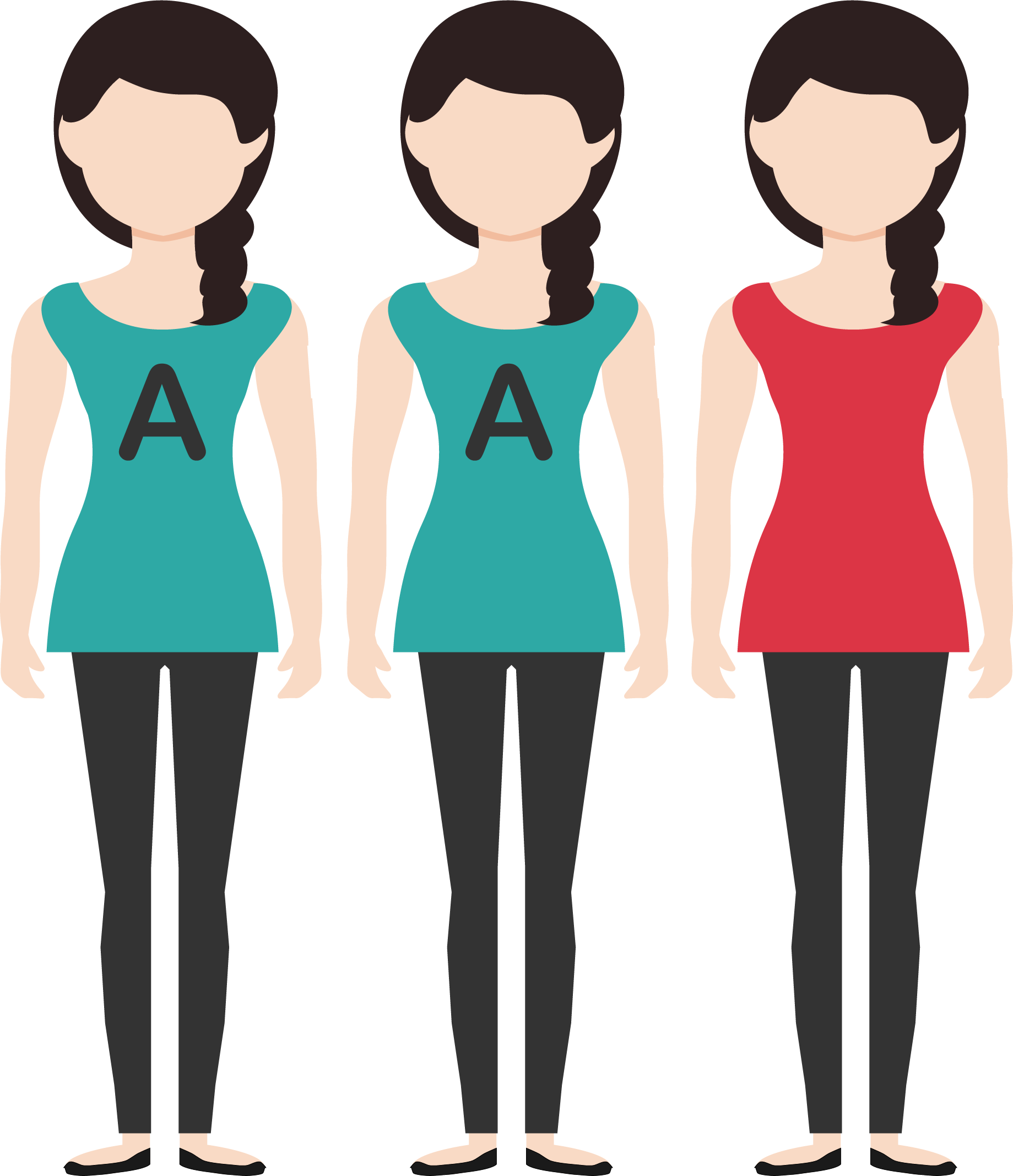}
    
    Does this distribution of awards obey the \textbf{award rule}? \correct{Yes}
    \item Alternative scenario 2:
    
    \vspace{10pt}
    \includegraphics[height=1in]{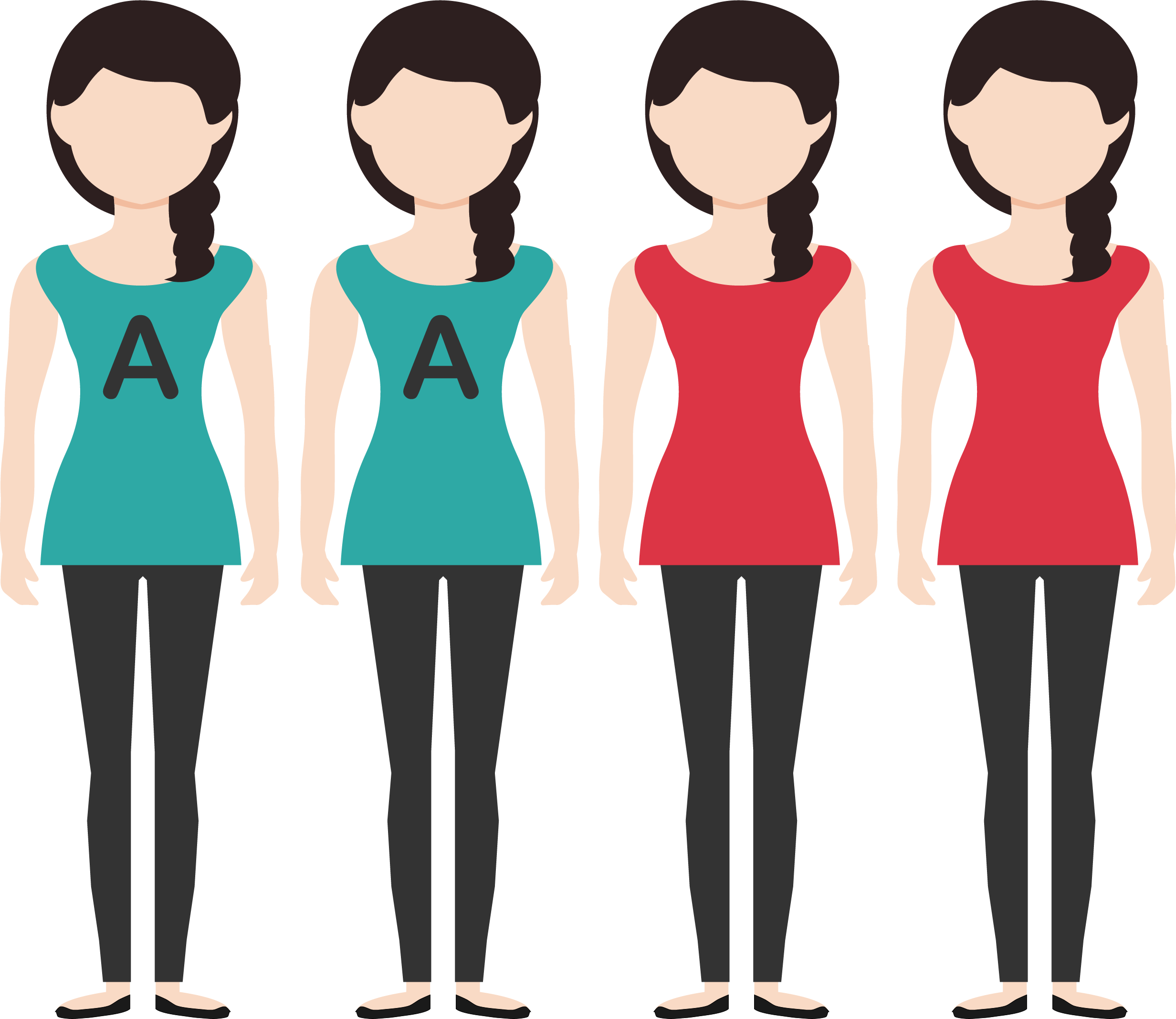}
    
    Does this distribution of awards obey the \textbf{award rule}? \correct{No}
    
    \item Alternative scenario 3:
    
    \vspace{10pt}
    \includegraphics[height=1in]{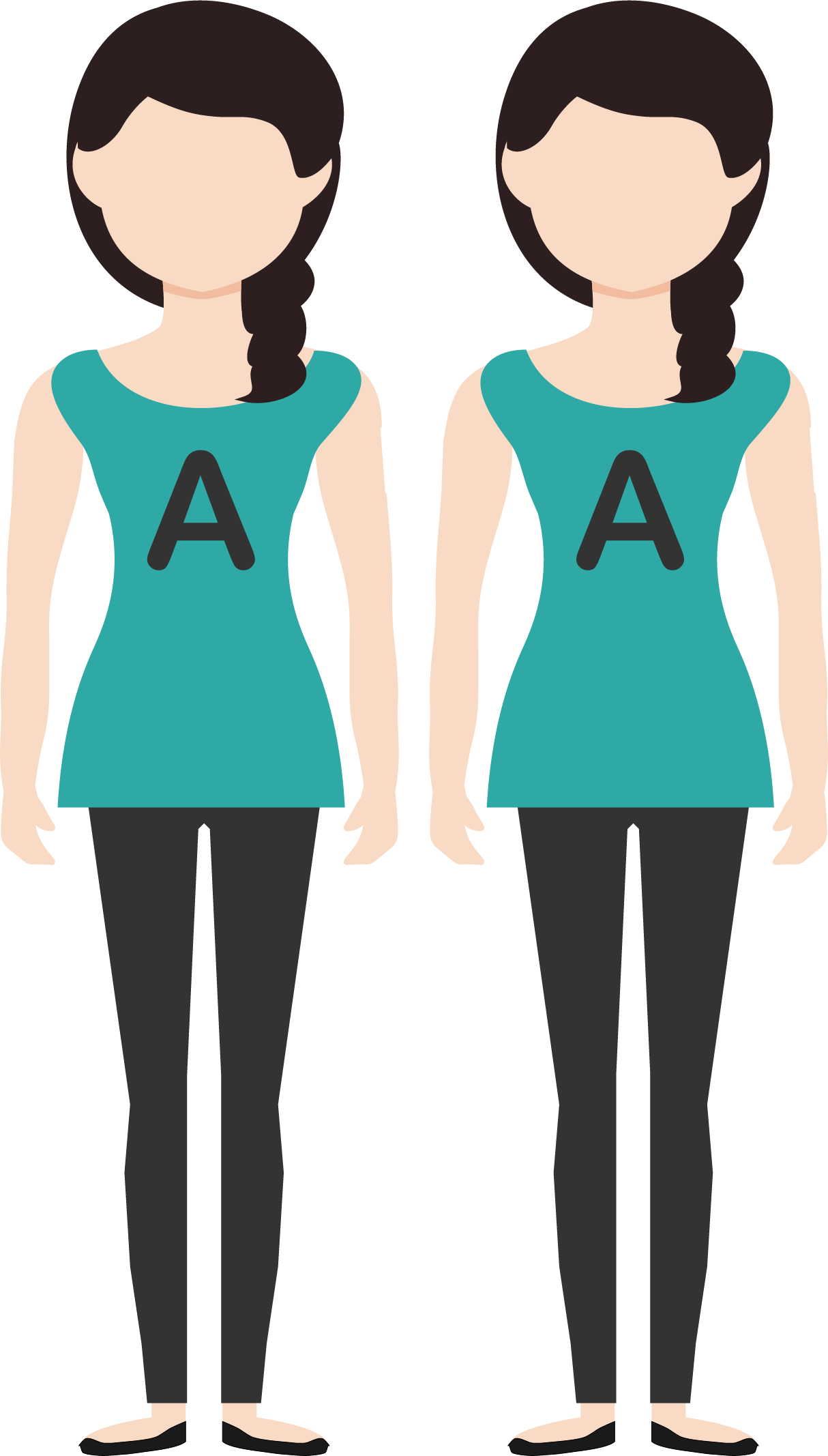}
    
    Does this distribution of awards obey the \textbf{award rule}? \correct{No}
\end{enumerate}

\begin{enumerate}
    \setcounter{enumi}{11}
    \item In your own words, explain the \textbf{award rule}. [short answer] (The rule is not shown above this question)
    \item To what extent do you agree with the following statement: I am confident I know how to \textbf{apply the award rule described above}?
    \begin{itemize}
        \item Strongly agree
        \item Agree
        \item Neither agree nor disagree
        \item Disagree
        \item Strongly Disagree
    \end{itemize}
    \item Please select the choice that best describes your experience: When I answered the previous questions...
    \begin{enumerate}
        \item I applied the provided award rule only.
        \item I used my own ideas of what the correct award decision should be rather than the provided award rule.
        \item I used a combination of the provided award rule and my own ideas of what the correct award decision should be.
    \end{enumerate}
    \item What is your opinion on the award rule? Please explain why. [short answer]
    \item Suppose that you are the teacher whose job it is to distribute lollipops to students based on the criteria listed above (i.e., elaborateness of each project, creativity of each project). How would you ensure that this process is fair? [short answer]
    \item Was there anything about this survey that was hard to understand or answer? [short answer]
\end{enumerate}

\paragraph{Employee awards}
The manager uses the following award rule to distribute awards: \emph{The fraction of employees who receive awards that are female should equal the fraction of employees that are female. Similarly, fraction of employees who receive awards that are male should equal the fraction of employees that are male.}

Example 1: If there are 50 female employees out of 100, then 5 out of the 10 awards should be awarded to female employees (and the remaining 5 would be made to male employees).

Example 2: If there are 30 female employees out of 100, then 3 out of the 10 awards should be awarded to female employees (and the remaining 7 would be made to male employees).

In the next section, we will ask you some questions about the information you have just read. Please note that this is not a test of your abilities. We want to measure the quality of the description you read, not your ability to take tests or answer questions.

\textbf{Please note that we ask you to apply and use ONLY the above award rule when answering the following questions. You will have an opportunity to state your opinions and feelings on the rule later in the survey.}

\begin{enumerate}
    \setcounter{enumi}{2}
    \item Suppose a different manager is considering employees for a different award. There are 100 male employees and 200 female employees, and they decide to give awards to 10 male employees. \textbf{Assuming the manager is required to use the award rule above}, how many female employees do they need to give awards to?
    \begin{enumerate}
        \item 10
        \item \correct{20}
        \item 40
        \item 50
    \end{enumerate}
    \item \textbf{Assuming the manager is required to use the award rule above}, in which of these cases can a manager give more awards to female employees than to male employees?
    \begin{enumerate}
        \item When there are more well-qualified female employees than well-qualified male employees (i.e., more women have better performance reviews, higher mid-year net sales, and more years on the job). 
        \item \correct{When there are more female employees than male employees.}
        \item When female employees receive higher performance reviews than male employees.
        \item This cannot happen under the award rule.
    \end{enumerate}
    \item  \textbf{Assuming the manager is required to use the award rule above}, is the following statement \correct{TRUE} OR FALSE: Even if a male employee’s qualifications look similar to a female employee’s (in terms of performance reviews, mid-year net sales, and years on the job), he can be treated differently (i.e., only one of the employees gets an award).
    \item \textbf{Assuming the manager is required to use the award rule above}, is the following statement TRUE OR \correct{FALSE}: If all female employees are unqualified (i.e., have low performance reviews, low mid-year net sales, and few years on the job), but you give awards to some of them, then awards given to male employees must be made to unqualified male employees.
    \item \textbf{Assuming the manager is required to use the award rule above}, is the following statement \correct{TRUE} OR FALSE: Suppose the manager is distributing 10 awards amongst a pool that includes both male and female employees. Even if all male employees are unqualified for an award (i.e., have low performance reviews, low mid-year net sales, and few years on the job), some of them must still receive awards.
    \item \textbf{Assuming the manager is required to use the award rule above}, is the following statement TRUE OR \correct{FALSE}: This award rule always allows the manager to distribute awards exclusively to the most qualified employees (i.e., employees with better performance reviews, high mid-year net sales, and high number of years on the job).
\end{enumerate}

In the two examples above there are 100 employees. Consider a different scenario, with \textbf{6 employees-- 4 female and 2 male, as illustrated below}. The next three questions each give a potential outcome for all six employees (i.e., which of the 6 employees receive awards). Please indicate which of the outcomes follow \textbf{the award rule above}.

\vspace{10pt}
\includegraphics[height=1in]{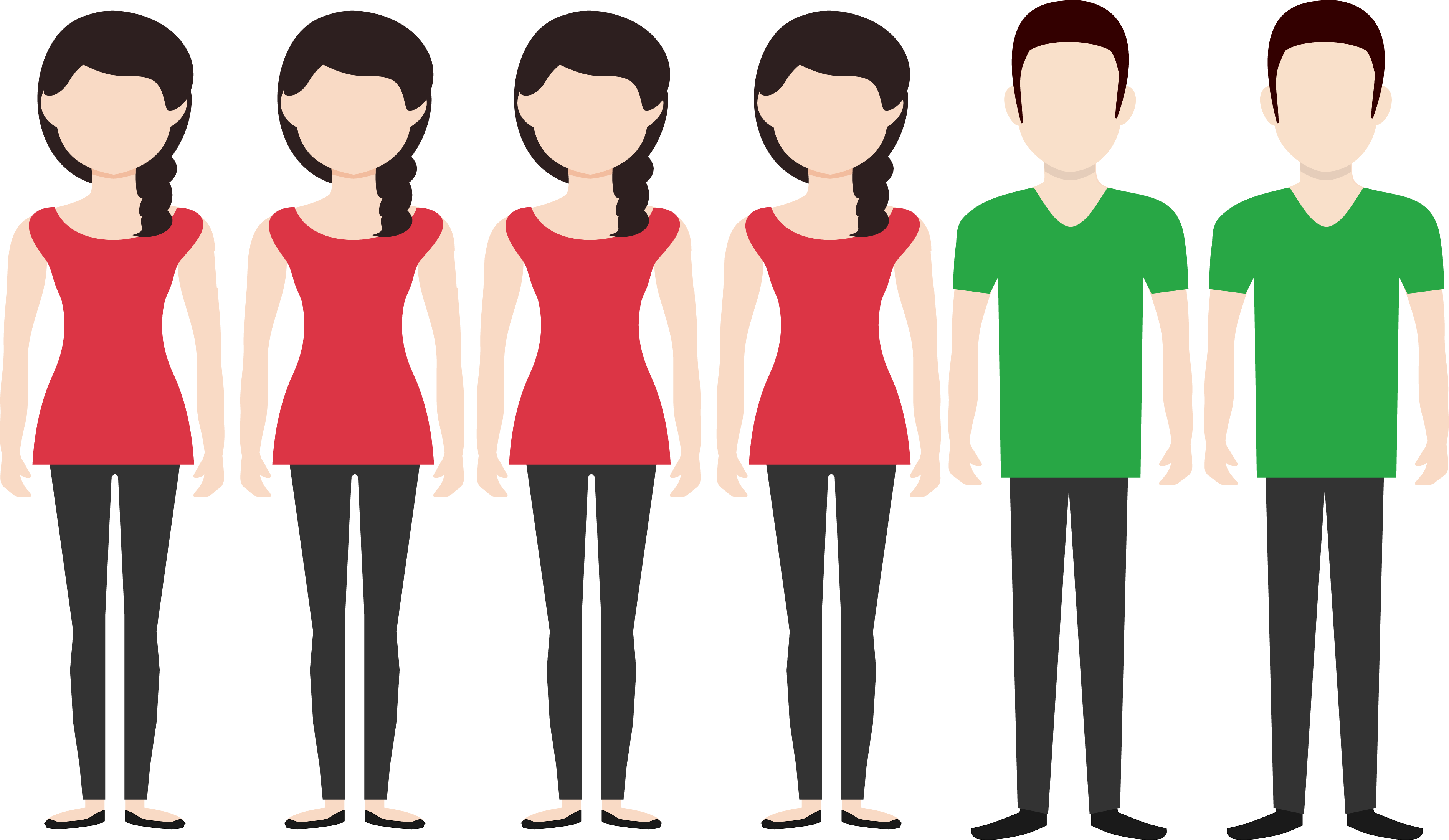}

\begin{enumerate}
    \setcounter{enumi}{8}
    \item Alternative scenario 1:
    
    \vspace{10pt}
    \includegraphics[height=1in]{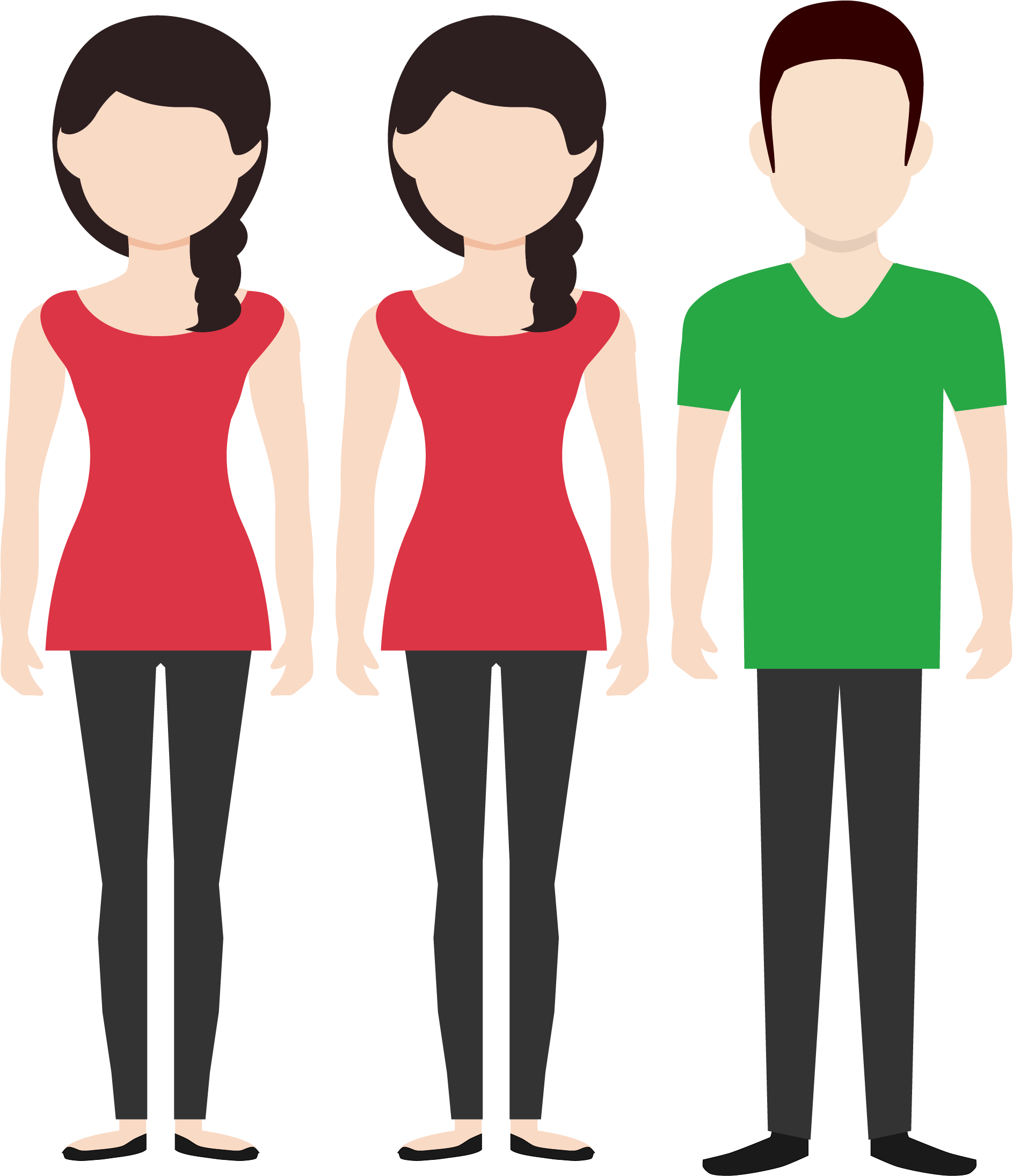}
    
    Does this distribution of awards obey the \textbf{award rule}? \correct{Yes}
    
    \item Alternative scenario 2:
    
    \vspace{10pt}
    \includegraphics[height=1in]{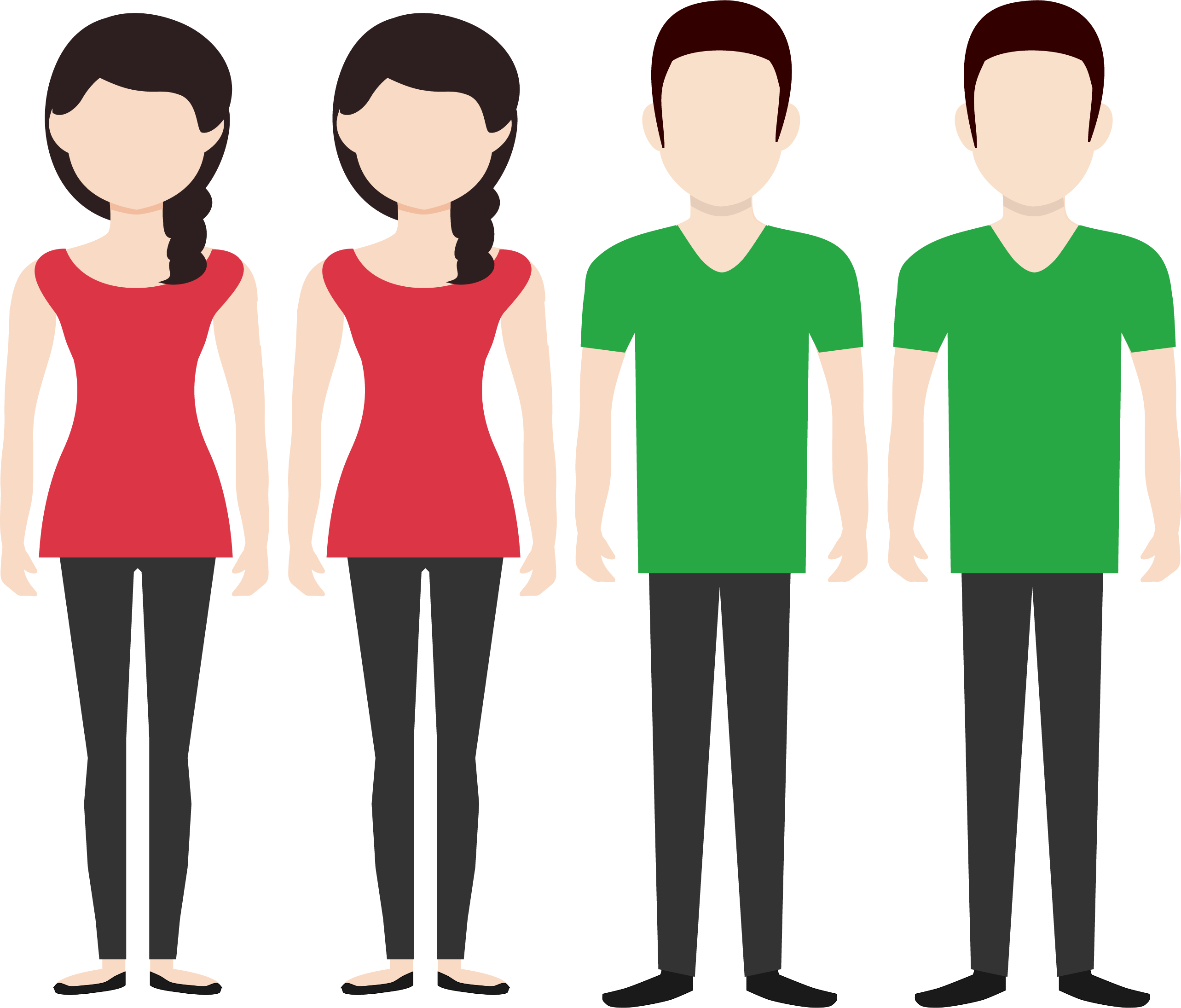}
    
    Does this distribution of awards obey the \textbf{award rule}? \correct{No}
    \item Alternative scenario 3:
    
    \vspace{10pt}
    \includegraphics[height=1in]{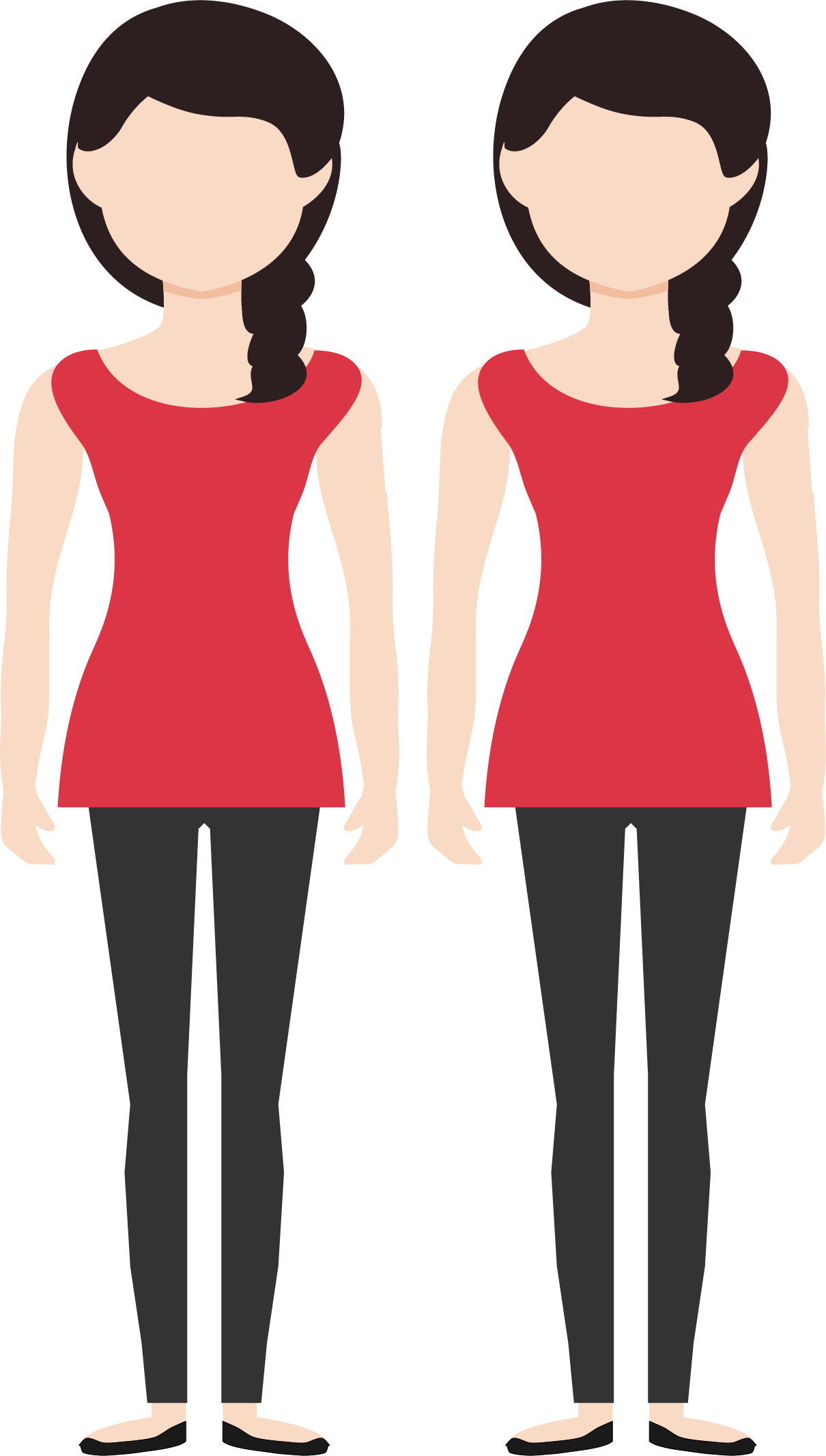}
    
    Does this distribution of awards obey the \textbf{award rule}? \correct{No}
\end{enumerate}

\begin{enumerate}
    \setcounter{enumi}{11}
    \item In your own words, explain the \textbf{award rule}. [short answer] (The rule is not shown above this question)
    \item To what extent do you agree with the following statement: I am confident I know how to \textbf{apply the award rule described above}?
    \begin{itemize}
        \item Strongly agree
        \item Agree
        \item Neither agree nor disagree
        \item Disagree
        \item Strongly Disagree
    \end{itemize}
    \item Please select the choice that best describes your experience: When I answered the previous questions...
    \begin{enumerate}
        \item I applied the provided award rule only.
        \item I used my own ideas of what the correct award decision should be rather than the provided award rule.
        \item I used a combination of the provided award rule and my own ideas of what the correct award decision should be.
    \end{enumerate}
    \item What is your opinion on the award rule? Please explain why. [short answer]
    \item Suppose that you are the manager whose job it is to distribute mid-year awards to employees based on the criteria listed above (i.e., recent performance reviews, mid-year net sales, number of years on the job). How would you ensure that this process is fair? [short answer]
    \item Was there anything about this survey that was hard to understand or answer? [short answer]
\end{enumerate}

\paragraph{Hiring}
The hiring manager uses the following hiring rule to send out offers: \emph{The fraction of applicants who receive job offers that are female should equal the fraction of applicants that are female. Similarly, fraction of applicants who receive job offers that are male should equal the fraction of applicants that are male.}

Example 1: If there are 50 female applicants out of the 100 applicants, then 5 out of the 10 offers would be made to female applicants (and the remaining 5 would be made to male applicants).

Example 2: If there are 30 female applicants out of the 100 applicants, then 3 out of the 10 offers would be made to female applicants (and the remaining 7 would be made to male applicants).

In the next section, we will ask you some questions about the information you have just read. Please note that this is not a test of your abilities. We want to measure the quality of the description you read, not your ability to take tests or answer questions.

\textbf{Please note that we ask you to apply and use ONLY the above hiring rule when answering the following questions. You will have an opportunity to state your opinions and feelings on the rule later in the survey.}

\begin{enumerate}
    \setcounter{enumi}{2}
    \item Suppose a different hiring manager is considering applicants for a different job. There are 100 male applicants and 200 female applicants, and they decide to send offers to 10 male applicants. \textbf{Assuming the hiring manager is required to use the hiring rule above}, how many female applicants do they need to send offers to?
    \begin{enumerate}
        \item 10
        \item \correct{20}
        \item 40
        \item 50
    \end{enumerate}
    \item \textbf{Assuming the hiring manager is required to use the hiring rule above}, in which of these cases can a hiring manager make more job offers to female applicants than to male applicants?
    \begin{enumerate}
        \item When there are more well-qualified female applicants than well-qualified male applicants (i.e., more women have higher interview scores, higher quality recommendation letters, and more years of prior experience in the field). 
        \item \correct{When there are more female applicants than male applicants.}
        \item When female applicants receive better interview scores than male applicants.
        \item This cannot happen under the hiring rule.
    \end{enumerate}
    \item \textbf{Assuming the hiring manager is required to use the hiring rule above}, is the following statement \correct{TRUE} OR FALSE: Even if a male applicant’s qualifications look similar to a female applicant’s (in terms of interview scores, recommendation letters, and years of prior experience in the field), he can be treated differently (i.e., only one of the applicants will receive a job offer).
    \item \textbf{Assuming the hiring manager is required to use the hiring rule above}, is the following statement TRUE OR \correct{FALSE}: If all female applicants are unqualified (i.e., have low interview scores, low-quality recommendation letters, and few years of prior experience in the field), but you send job offers to some of them, then any job offers made to male applicants must be made to unqualified male applicants.
    \item \textbf{Assuming the hiring manager is required to use the hiring rule above}, is the following statement \correct{TRUE} OR FALSE: Suppose the hiring manager is sending out 10 job offers to a pool that includes male and female applicants. Even if all male applicants are unqualified (i.e., have low interview scores, low-quality recommendation letters, and few years of prior experience in the field), some of them must still receive job offers.
    \item \textbf{Assuming the hiring manager is required to use the hiring rule above}, is the following statement TRUE OR \correct{FALSE}: This hiring rule always allows the hiring manager to send offers exclusively to the most qualified applicants (i.e., applicants with high interview scores, high quality recommendation letters, and high number years of prior experience in the field).
\end{enumerate}

In the two examples above there are 100 applicants. Consider a different scenario, with \textbf{6 applicants -- 4 female and 2 male, as illustrated below}. The next three questions each give a potential outcome for all 6 applicants (i.e., which of the 6 applicants receive job offers). Please indicate which of the outcomes follow \textbf{the hiring rule above}.

\vspace{10pt}
\includegraphics[height=1in]{illustrations/total_employees_applicants.png}

\begin{enumerate}
    \setcounter{enumi}{8}
    \item Alternative scenario 1:
    
    \vspace{10pt}
    \includegraphics[height=1in]{illustrations/case1_accept_reject_employees_applicants.png}
    
    Does this distribution of job offers obey the \textbf{hiring rule}? \correct{Yes}
    \item Alternative scenario 2:
    
    \vspace{10pt}
    \includegraphics[height=1in]{illustrations/case2_accept_employees_applicants.png}
    
    Does this distribution of job offers obey the \textbf{hiring rule}? \correct{No}
    \item Alternative scenario 3:
    
    \vspace{10pt}
    \includegraphics[height=1in]{illustrations/case2_reject_case3_accept_employees_applicants.png}
    
    Does this distribution of job offers obey the \textbf{hiring rule}? \correct{No}
\end{enumerate}

\begin{enumerate}
    \setcounter{enumi}{11}
    \item In your own words, explain the \textbf{hiring rule}. [short answer] (The rule is not shown above this question)
    \item To what extent do you agree with the following statement: I am confident I know how to \textbf{apply the hiring rule described above}?
    \begin{itemize}
        \item Strongly agree
        \item Agree
        \item Neither agree nor disagree
        \item Disagree
        \item Strongly Disagree
    \end{itemize}
    \item Please select the choice that best describes your experience: When I answered the previous questions...
    \begin{enumerate}
        \item I applied the provided hiring rule only.
        \item I used my own ideas of what the correct hiring decision should be rather than the provided hiring rule.
        \item I used a combination of the provided hiring rule and my own ideas of what the correct hiring decision should be.
    \end{enumerate}
    \item What is your opinion on the hiring rule? Please explain why. [short answer]
    \item Suppose that you are the hiring manager whose job it is to send job offers to applicants based on the criteria listed above (i.e., interview scores, quality of recommendation letters, number of years of prior experience in the field). How would you ensure that this process is fair? [short answer]
    \vspace{-5pt}
    \item Was there anything about this survey that was hard to understand or answer? [short answer]
    \vspace{-5pt}
\end{enumerate}

\subsection{\studyB{}: Survey}  \label{app:surveyB}

Each of the surveys are split into four main sections. The first section is the consent form which can be found in Appendix~\ref{app:consent}. The second section describes the hiring scenario and asks questions about it (\S\ref{app:st2_scenarios}). The third section describes the fairness metric, defined as the rule, used (in this case it is demographic parity) and asks specific questions about the metric (\S\ref{app:st2_fairness}). Finally the last section asks for demographic information (\S\ref{app:demographics}).

\subsubsection{Scenario description and questions}\label{app:st2_scenarios}
The following is shown to each participant (note that Step 3 is not shown to participants with the DP definition):

It is very important that you read each question carefully and think about your answers. The success of our research relies on our respondents being thoughtful and taking this task seriously.
\begin{itemize}
    \vspace{-8pt}
    \item[\checkbox] I have read the above instructions carefully.
    \vspace{-8pt}
\end{itemize}

A company, Sales-a-lot, is reviewing their hiring process. They want to hire applicants who are high performing, and they also want to make sure that their hiring process is fair to their applicants, no matter their gender. To do this, Sales-a-lot employs an external firm, Recruit-a-matic, which keeps track of all applicants. This review will take place over one year.
 
For clarity at each stage of the hiring process we use images to represent the hiring pool.

\paragraph{Step 1: Applicant Pool.} At the beginning of the year, Sales-a-lot reviews all job applicants, and sends job offers to some of them. The initial applicant pool is shown with a gray background. For example, the following image shows an applicant pool with 15 female applicants and 25 male applicants:

\includegraphics[height=1in]{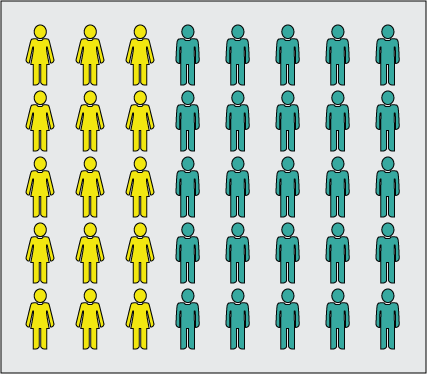}

\paragraph{Step 2: Sending Job Offers.} Next, Sales-a-lot sends job offers \space to some of these \space applicants, using the \space following criteria:
\begin{itemize} \itemsep=0pt
    \vspace{-10pt}
    \item Interview scores
    \item Quality of recommendation letters
    \item Number of years of prior experience in the field
\end{itemize}
Suppose that Sales-a-lot sends offers to 5 female applicants and 8 male applicants (so 10 female and 17 male applicants didn’t receive offers). In the following image, applicants who received a job offer are shown on the left (with a green background) and applicants who didn’t receive a job offer are shown on the right, with a red background):

\includegraphics[height=1in]{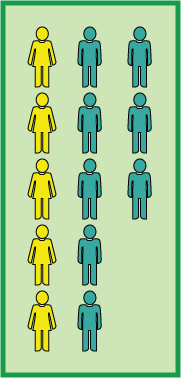}
\space\space\space
\includegraphics[height=1in]{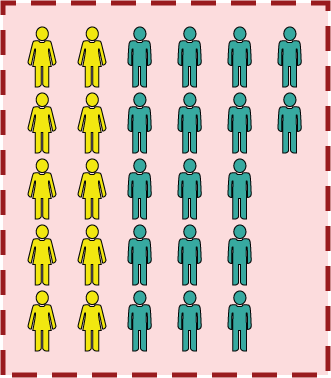}

\paragraph{Step 3: Applicant Evaluation.} For the rest of the year, Recruit-a-matic (the external firm) keeps track of all applicants in the initial pool, whether they received job offers or not. At the end of the year, Rectruit-a-matic finds out which applicants were high performers, i.e. qualified (shown in dark), and which applicants were low performers, i.e. unqualified (shown in light). For example, the following image shows that most of the high performers received job offers, but some did not.

\includegraphics[height=1in]{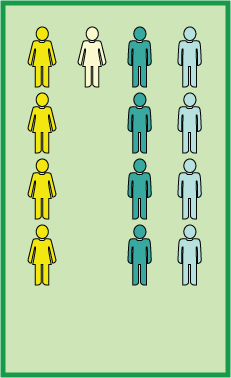}
\space\space\space
\includegraphics[height=1in]{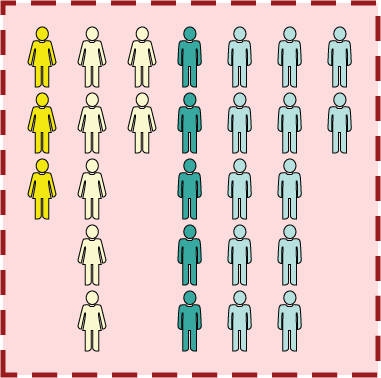}

\begin{tabular}{r|c|c}
     & female & male  \\
    \midrule
    qualified & \includegraphics[height=0.2in]{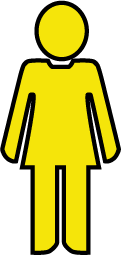} & \includegraphics[height=0.2in]{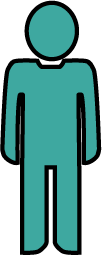} \\
    unqualified & \includegraphics[height=0.2in]{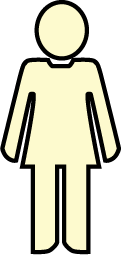} & \includegraphics[height=0.2in]{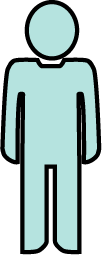} \\
\end{tabular}

\paragraph{Questions}
\begin{enumerate}
    \vspace{-5pt}
    \item To what extent do you agree with the following statement: a scenario similar to the one described above might occur in real life.
    \begin{itemize}
        \item Strongly agree
        \item Agree
        \item Neither agree nor disagree
        \item Disagree
        \item Strongly disagree
    \end{itemize}

    \item How much effort, in hours, should Sales-a-lot put in to make sure these decisions were fair? [short answer - number of hours]
\end{enumerate}

\subsubsection{Rule descriptions and questions}\label{app:st2_fairness}
The following sections provide fairness definitions (presented to participants as \emph{rules}) for Demographic Parity, Equal Opportunity (FNR and FPR), and Equalized Odds. Unless otherwise noted the rule description is shown above each of the questions for reference. Correct answers are noted in \correct{red}.

\paragraph{Demographic Parity.}

Recruit-a-matic uses the following rule to determine whether Sales-a-lot’s hiring decisions were fair:
 
\emph{The fraction of male candidates who receive job offers should equal the fraction of female candidates who receive job offers.}

Example 1: Suppose that over the past year, Recruit-a-matic finds that Sales-a-lot received the following applicants (10 female and 12 male).

\includegraphics[height=1in]{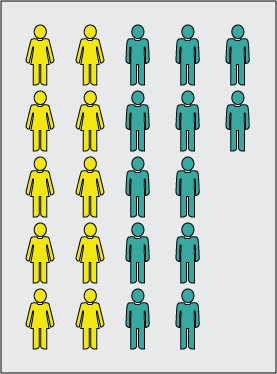}

If Sales-a-lot sent job offers to the following number of applicants (5 female and 6 male), then this would be fair according to the hiring rule (note that there are other possible outcomes that are fair according to the hiring rule).

\includegraphics[height=1in]{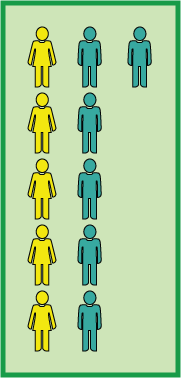}

Example 2: Suppose that over the past year, Recruit-a-matic finds that Sales-a-lot reviewed a total of 100 applicants as follows (40 female and 60 male).

\includegraphics[height=2in]{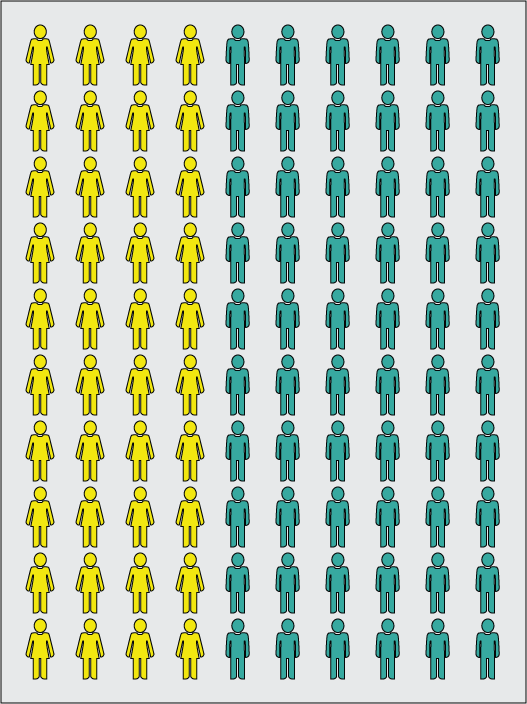}

If Sales-a-lot sent job offers to the following number of applicants (10 female and 15 male), then this would be fair according to the hiring rule (note that there are other possible outcomes that are fair according to the hiring rule).

\includegraphics[height=1in]{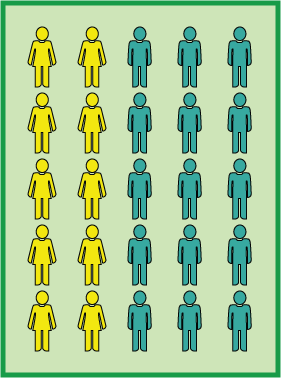}

In the next section, we will ask you some questions about the information you have just read. Please note that this is not a test of your abilities. We want to measure the quality of the description you read, not your ability to take tests or answer questions.

\textbf{Please note that we ask you to apply and use ONLY the above hiring rule when answering the following questions. You will have an opportunity to state your opinions and feelings on the rule later in the survey.}

\begin{enumerate}
\setcounter{enumi}{2}
    \item Suppose a different company considered applicants for a different job. There were 200 female applicants and 100 male applicants,

\includegraphics[height=1.3in]{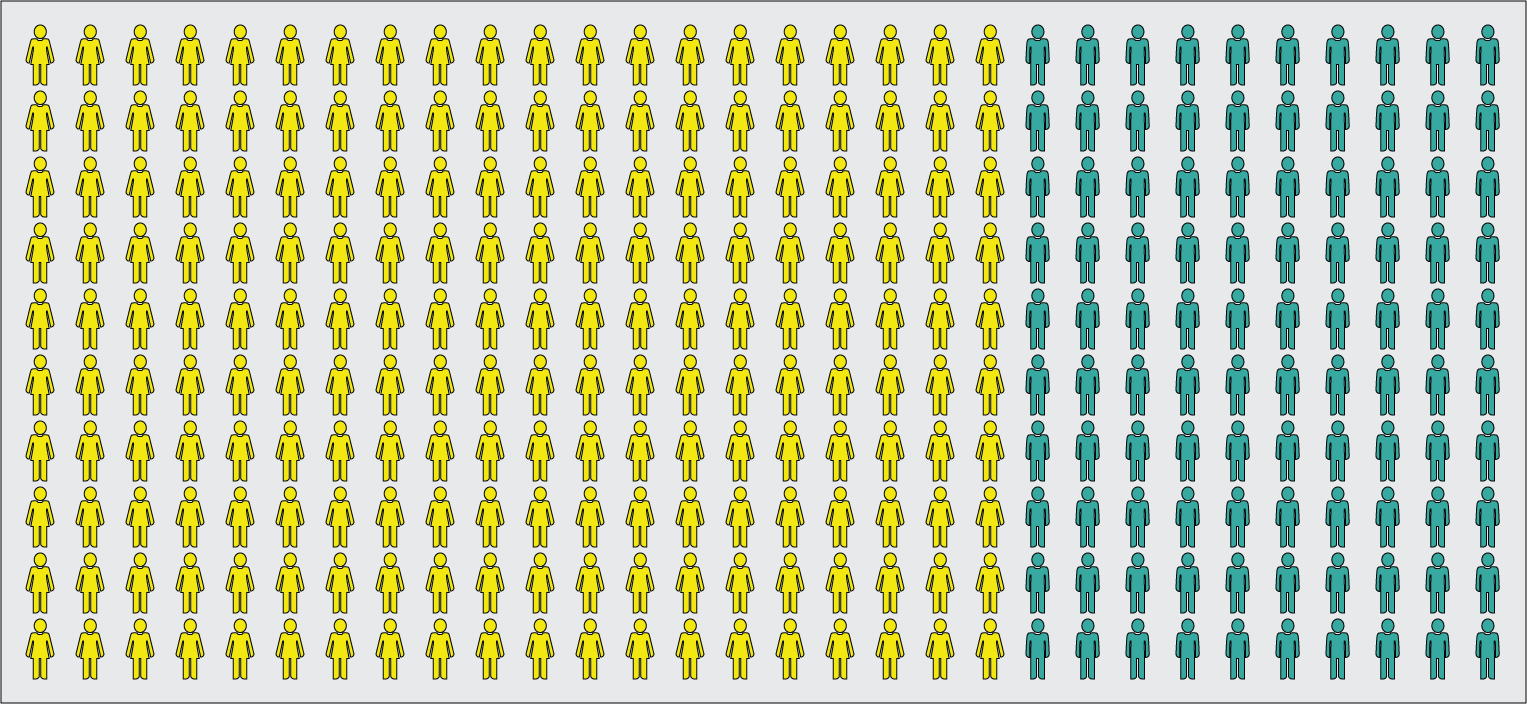}

and they did send job offers to 90 male applicants.

\includegraphics[height=1.3in]{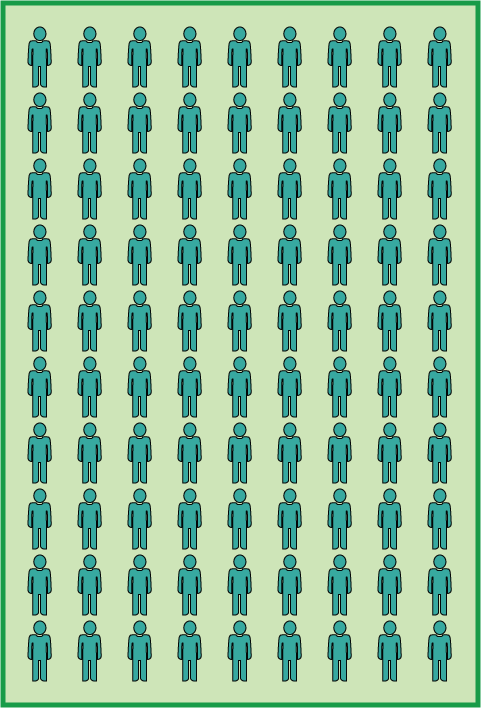}

Assuming that Recruit-a-matic reviews their decisions using the hiring rule above, how many female applicants should have received job offers?
\begin{enumerate}
    \item 190
    \includegraphics[height=1.3in]{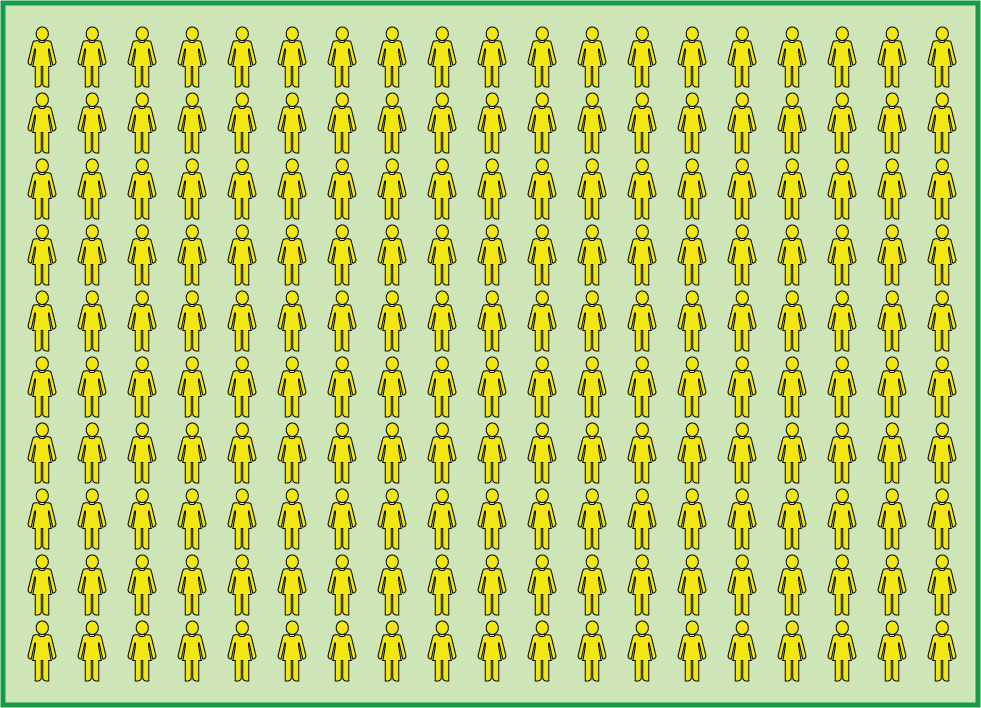}
    \item \correct{180}
    \includegraphics[height=1.3in]{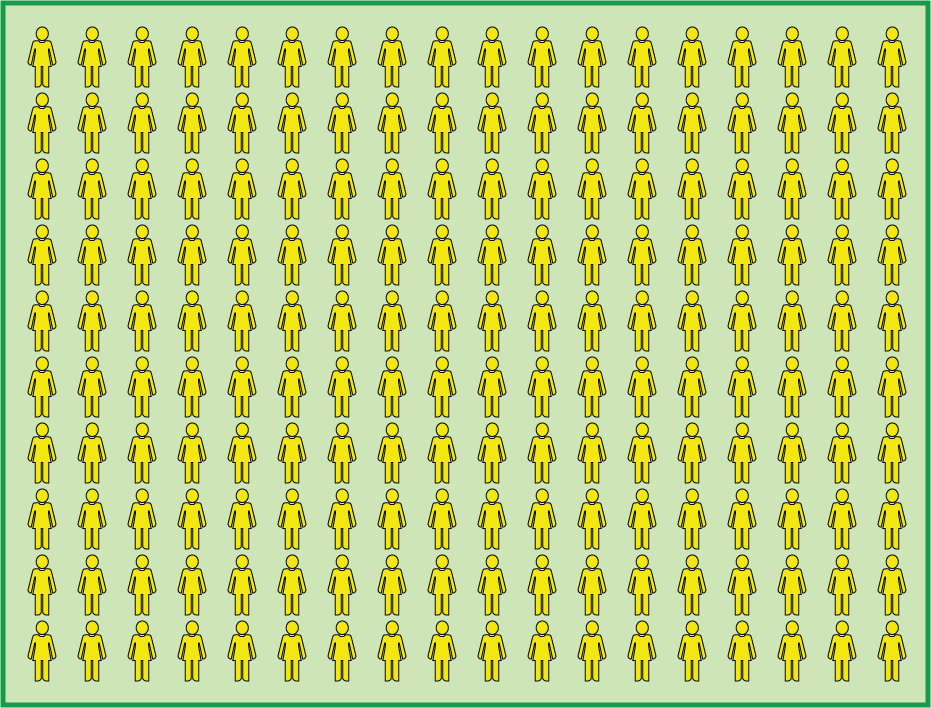}
    \item 160
    \includegraphics[height=1.3in]{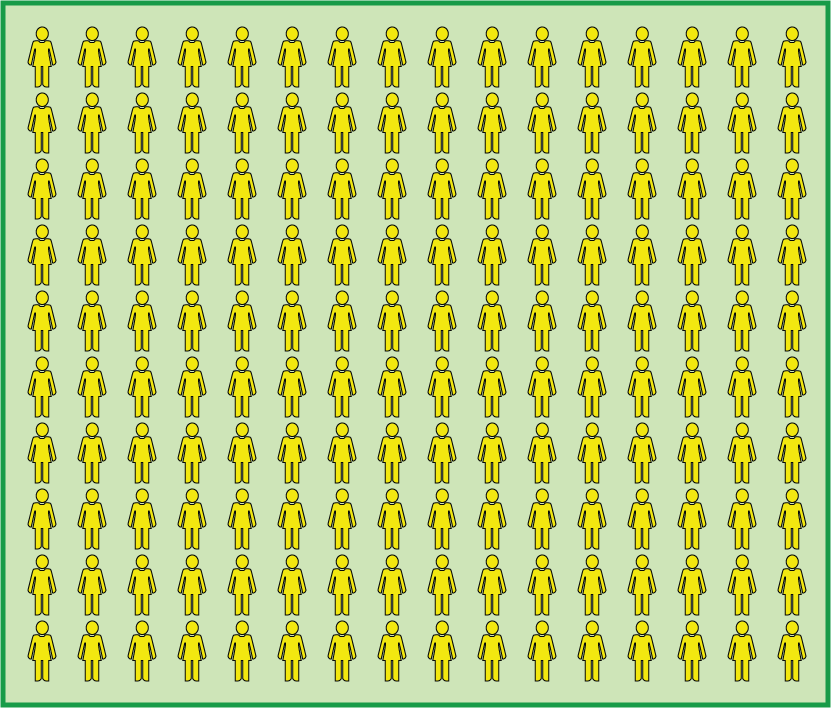}
    \item 150
    \includegraphics[height=1.3in]{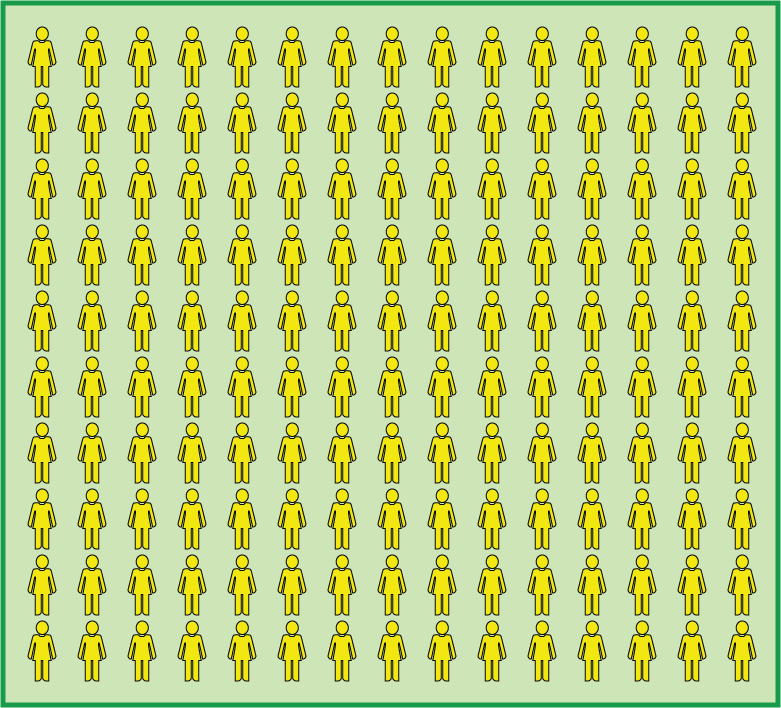}
\end{enumerate}

\item Assuming Recruit-a-matic reviews decisions using the hiring rule above, in which of these cases could Sales-a-lot have accepted more qualified female applicants than qualified male applicants?

\begin{enumerate}
    \item When there are more qualified female applicants than qualified male applicants (i.e., more women had low net sales at the end of the year).
    \item \correct{When there are more female applicants than male applicants.}
    \item When female applicants receive worse interview scores than male applicants.
    \item This cannot happen under the hiring rule.
\end{enumerate}

\item Consider one male applicant and one female applicant, both of whom are similarly qualified for the job (they achieve about the same net sales at the end of their first year). Is the following statement \correct{TRUE} OR FALSE: The hiring rule above allows Sales-a-lot to make a job offer to one of these applicants and not the other.

\item Consider a situation where all female applicants were unqualified (they all achieve low net sales at the end of their first year), but some of them received job offers. Is the following statement TRUE OR \correct{FALSE}: The hiring rule above requires that some job offers made to male applicants must have been made to unqualified male applicants.

\item Suppose Sales-a-lot received 100 male and 100 female applicants, and eventually made 10 job offers. Is the following statement \correct{TRUE} OR FALSE: The hiring rule above requires that even if all male applicants were unqualified (they all achieve low net sales at the end of their first year), some of the unqualified males must have received job offers.

\item Is the following statement TRUE OR \correct{FALSE}: The hiring rule above always allows Sales-a-lot to send job offers only to the most qualified applicants (those who achieve high net sales at the end of their first year).

\end{enumerate}

Consider a different scenario than the two examples above, with 6 applicants -- 4 female and 2 male, as illustrated below. The next three questions each give a different potential outcome for all 6 applicants (i.e., which of the 6 applicants do receive job offers). Please indicate which of the outcomes follow the hiring rule above.

\includegraphics[height=0.4in]{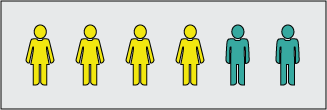}
\begin{enumerate}
\setcounter{enumi}{8}
    \item Sales-a-lot makes the following hiring decisions.

    \includegraphics[height=0.4in]{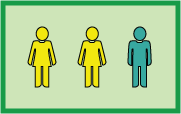}
    \space\space\space
    \includegraphics[height=0.4in]{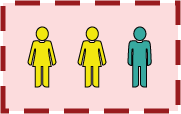}
    
    Do these decisions obey the hiring rule? \correct{Yes}

\item Sales-a-lot makes the following hiring decisions.

    \includegraphics[height=0.4in]{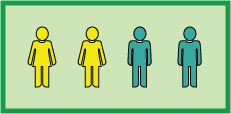}
    \space\space\space
    \includegraphics[height=0.4in]{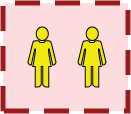}
    
    Do these decisions obey the hiring rule? \correct{No}
    
\item Sales-a-lot makes the following hiring decisions.

    \includegraphics[height=0.4in]{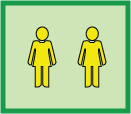}
    \space\space\space
    \includegraphics[height=0.4in]{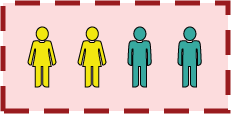}
    
    Do these decisions obey the hiring rule? \correct{No}

\item In your own words, explain the hiring rule. [short answer] [The rule is not shown above this question]

\item To what extent do you agree with the following statement: I am confident I know how to apply the hiring rule described above?
\begin{itemize}
    \item Strongly agree
    \item Agree
    \item Neither agree nor disagree
    \item Disagree
    \item Strongly Disagree
\end{itemize}

\item Please select the choice that best describes your experience: When I answered the previous questions...
\begin{enumerate}
    \item I applied the provided hiring rule only.
    \item I used a combination of the provided hiring rule and my own ideas of what the correct hiring rule should be.
    \item I used only my own ideas of what the correct hiring decision should be rather than the provided hiring rule.
\end{enumerate}

\item To what extent do you agree with the following statement: I like the hiring rule?
\begin{itemize}
    \item Strongly agree
    \item Agree
    \item Neither agree nor disagree
    \item Disagree
    \item Strongly Disagree
\end{itemize}

\item To what extent do you agree with the following statement: I agree with the hiring rule?
\begin{itemize}
    \item Strongly agree
    \item Agree
    \item Neither agree nor disagree
    \item Disagree
    \item Strongly Disagree
\end{itemize}

\item Please explain your opinion on the hiring rule. [short answer]

\item Was there anything about this survey that was hard to understand or answer? [short answer]

\end{enumerate}

\paragraph{Equal Opportunity - FNR.}

Recruit-a-matic uses the following rule to determine whether Sales-a-lot’s hiring decisions were fair:
 
\emph{The fraction of qualified male candidates who do not receive job offers should equal the fraction of qualified female candidates who do not receive job offers.}

Example 1: Suppose that over the past year, Recruit-a-matic finds that Sales-a-lot received the following qualified applicants (10 female and 12 male).

\includegraphics[height=1in]{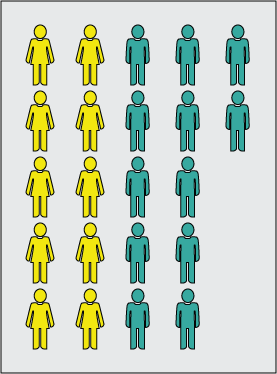}

If Sales-a-lot did not send job offers to the following number of qualified applicants (5 female and 6 male), then this would be fair according to the hiring rule (note that there are other possible outcomes that are fair according to the hiring rule).

\includegraphics[height=1in]{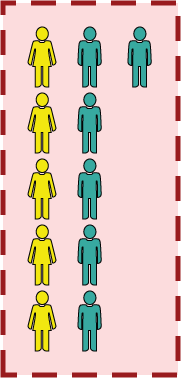}

Example 2: Suppose that over the past year, Recruit-a-matic finds that Sales-a-lot reviewed a total of 100 qualified applicants as follows (40 female and 60 male).

\includegraphics[height=2in]{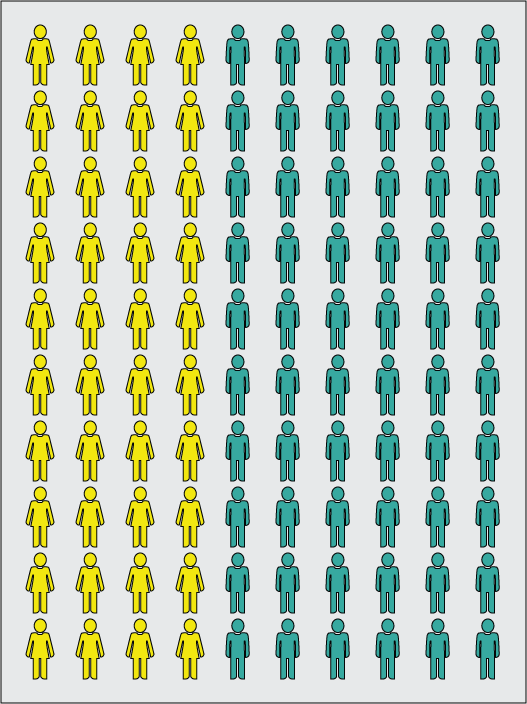}

If Sales-a-lot did not send job offers to the following number of qualified applicants (10 female and 15 male), then this would be fair according to the hiring rule (note that there are other possible outcomes that are fair according to the hiring rule).

\includegraphics[height=1in]{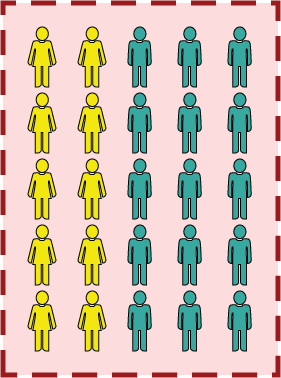}

Note that in the above examples the remaining qualified applicants received job offers, but are not displayed here.

In the next section, we will ask you some questions about the information you have just read. Please note that this is not a test of your abilities. We want to measure the quality of the description you read, not your ability to take tests or answer questions.

\textbf{Please note that we ask you to apply and use ONLY the above hiring rule when answering the following questions. You will have an opportunity to state your opinions and feelings on the rule later in the survey.}

\begin{enumerate}
\setcounter{enumi}{2}
    \item Suppose a different company considered applicants for a different job. There were 200 qualified female applicants and 100 qualified male applicants,

\includegraphics[height=1.2in]{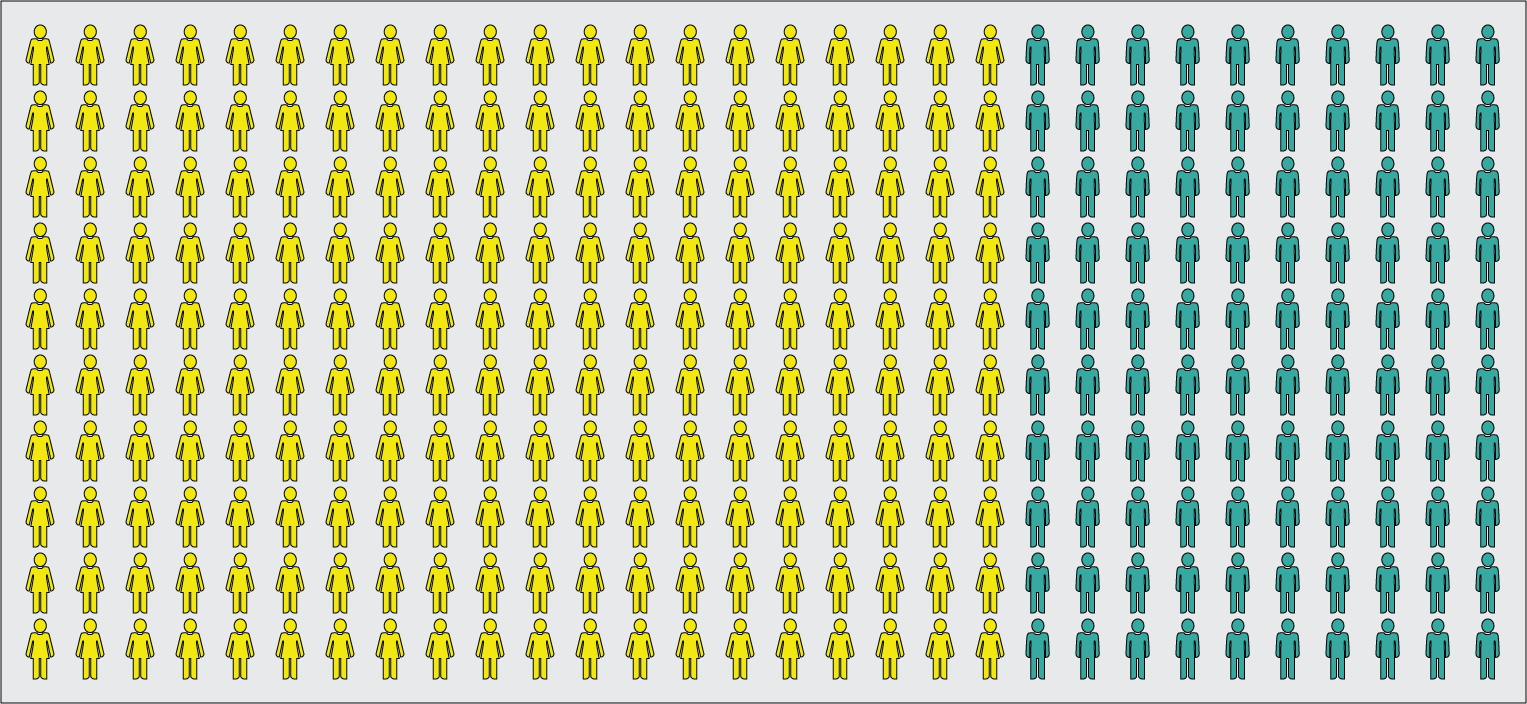}

and they did not send job offers to 90 qualified male applicants.

\includegraphics[height=1.3in]{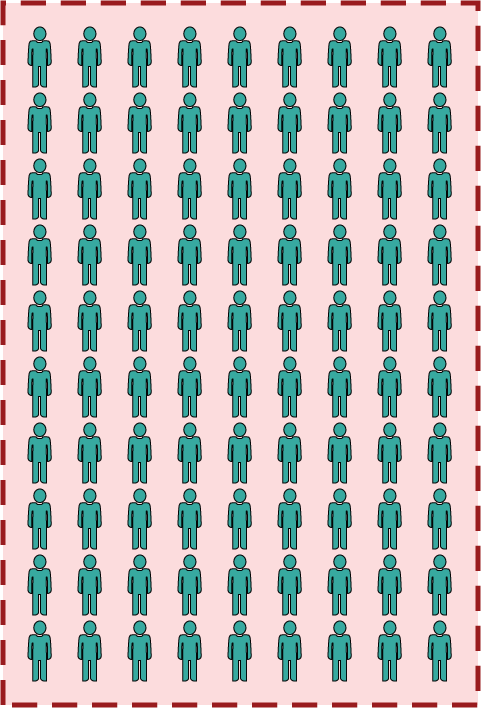}

Assuming that Recruit-a-matic reviews their decisions using the hiring rule above, how many qualified female applicants should not have received job offers?
\begin{enumerate}
    \item 190
    \includegraphics[height=1.3in]{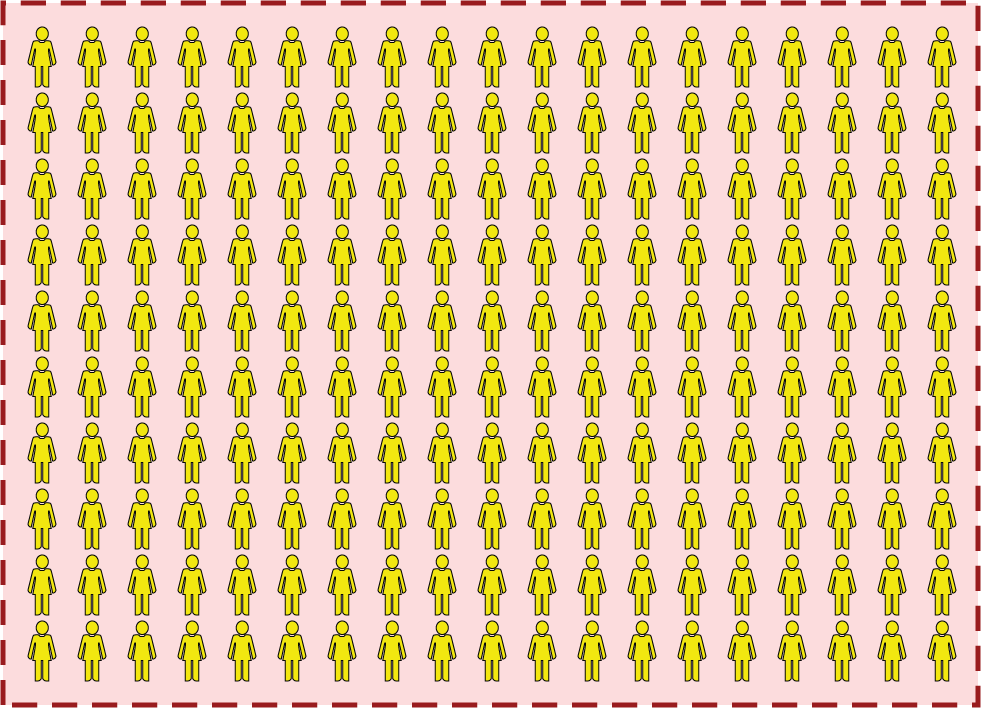}
    \item \correct{180}
    \includegraphics[height=1.3in]{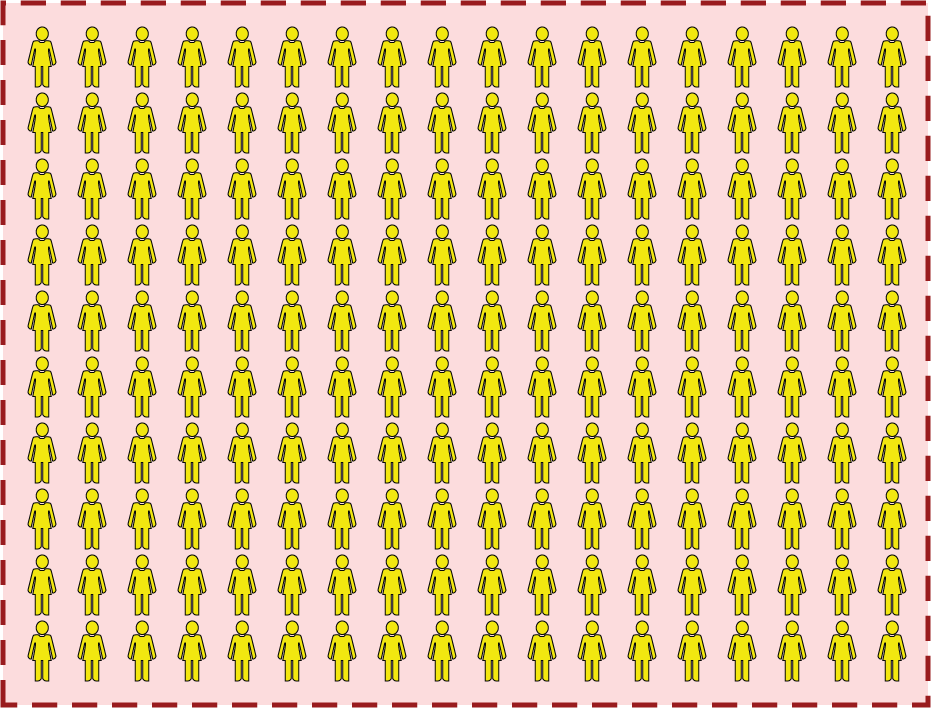}
    \item 160
    \includegraphics[height=1.3in]{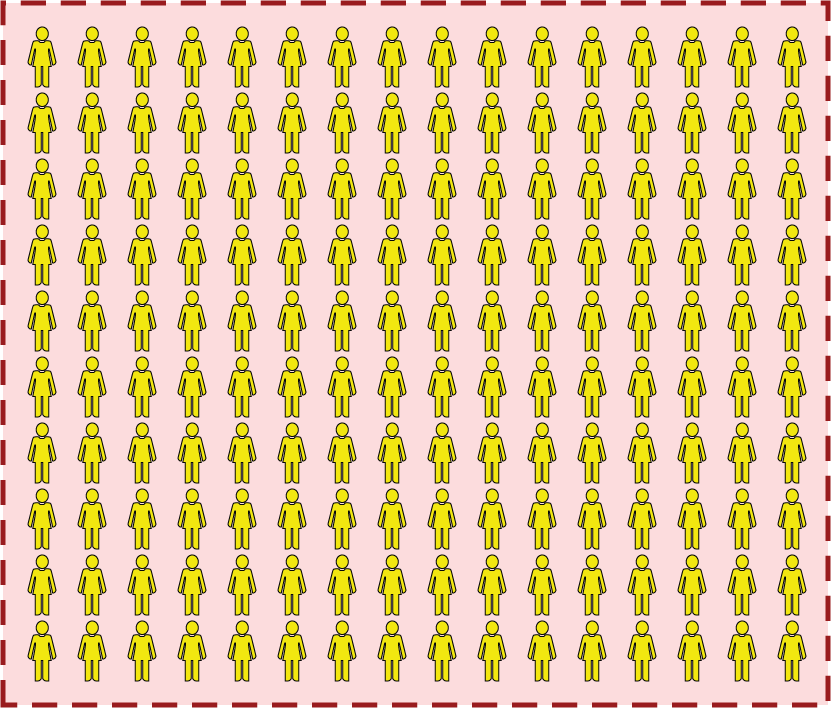}
    \item 150
    \includegraphics[height=1.3in]{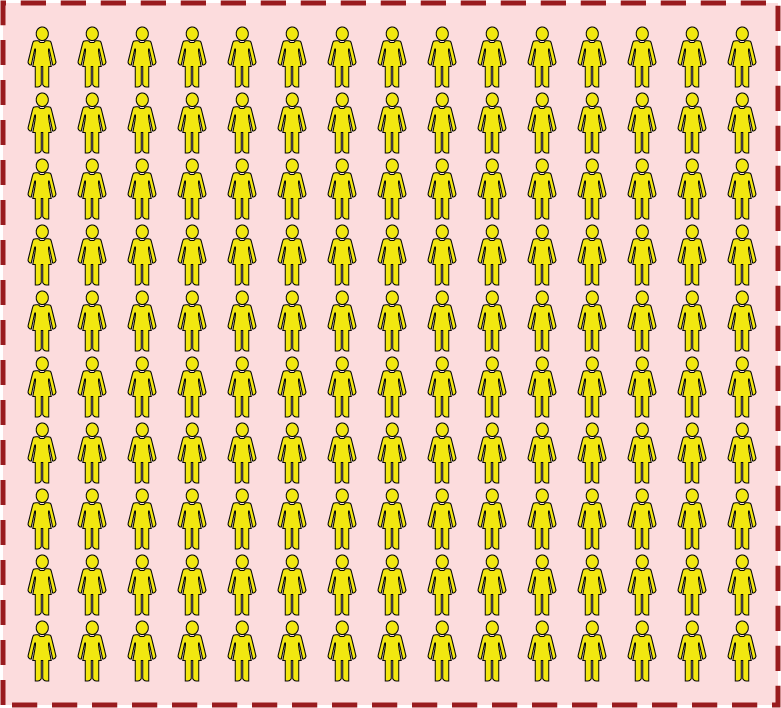}
\end{enumerate}

\item Assuming Recruit-a-matic reviews decisions using the hiring rule above, in which of these cases could Sales-a-lot have rejected more qualified female applicants than qualified male applicants?

\begin{enumerate}
    \item \correct{When there are more qualified female applicants than qualified male applicants (i.e., more women had low net sales at the end of the year).}
    \item When there are more female applicants than male applicants.
    \item When female applicants receive worse interview scores than male applicants.
    \item This cannot happen under the hiring rule.
\end{enumerate}

\item Consider one male applicant and one female applicant, both of whom are similarly qualified for the job (they achieve about the same net sales at the end of their first year). Is the following statement \correct{TRUE} OR FALSE: The hiring rule above allows Sales-a-lot to make a job offer to one of these applicants and not the other.

\item Consider a situation where all female applicants were unqualified (they all achieve low net sales at the end of their first year), but some of them received job offers. Is the following statement TRUE OR \correct{FALSE}: The hiring rule above requires that some job offers made to male applicants must have been made to unqualified male applicants.

\item Suppose Sales-a-lot received 100 male and 100 female applicants, and eventually made 10 job offers. Is the following statement TRUE OR \correct{FALSE}: The hiring rule above requires that even if all male applicants were unqualified (they all achieve low net sales at the end of their first year), some of the unqualified males must have received job offers.

\item Is the following statement \correct{TRUE} OR FALSE: The hiring rule above always allows Sales-a-lot to send job offers only to the most qualified applicants (those who achieve high net sales at the end of their first year).

\end{enumerate}

Consider a different scenario than the two examples above, with 6 qualified applicants -- 4 female and 2 male, as illustrated below. The next three questions each give a different potential outcome for all 6 qualified applicants (i.e., which of the 6 applicants do not receive job offers). Please indicate which of the outcomes follow the hiring rule above.

\includegraphics[height=0.4in]{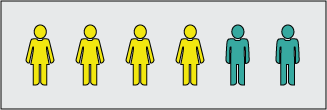}
\begin{enumerate}
\setcounter{enumi}{8}
    \item Sales-a-lot makes the following hiring decisions.

    \includegraphics[height=0.4in]{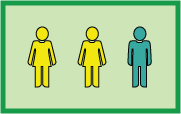}
    \space\space\space
    \includegraphics[height=0.4in]{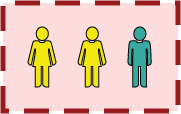}
    
    Do these decisions obey the hiring rule? \correct{Yes}

\item Sales-a-lot makes the following hiring decisions.

    \includegraphics[height=0.4in]{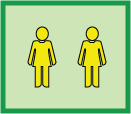}
    \space\space\space
    \includegraphics[height=0.4in]{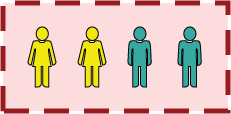}
    
    Do these decisions obey the hiring rule? \correct{No}
    
\item Sales-a-lot makes the following hiring decisions.

    \includegraphics[height=0.4in]{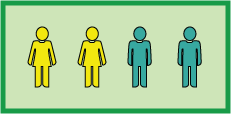}
    \space\space\space
    \includegraphics[height=0.4in]{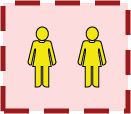}
    
    Do these decisions obey the hiring rule? \correct{No}

\item In your own words, explain the hiring rule. [short answer] [The rule is not shown above this question]

\item To what extent do you agree with the following statement: I am confident I know how to apply the hiring rule described above?
\begin{itemize}
    \item Strongly agree
    \item Agree
    \item Neither agree nor disagree
    \item Disagree
    \item Strongly Disagree
\end{itemize}

\item Please select the choice that best describes your experience: When I answered the previous questions...
\begin{enumerate}
    \item I applied the provided hiring rule only.
    \item I used a combination of the provided hiring rule and my own ideas of what the correct hiring rule should be.
    \item I used only my own ideas of what the correct hiring decision should be rather than the provided hiring rule.
\end{enumerate}

\item To what extent do you agree with the following statement: I like the hiring rule?
\begin{itemize}
    \item Strongly agree
    \item Agree
    \item Neither agree nor disagree
    \item Disagree
    \item Strongly Disagree
\end{itemize}

\item To what extent do you agree with the following statement: I agree with the hiring rule?
\begin{itemize}
    \item Strongly agree
    \item Agree
    \item Neither agree nor disagree
    \item Disagree
    \item Strongly Disagree
\end{itemize}

\item Please explain your opinion on the hiring rule. [short answer]

\item Was there anything about this survey that was hard to understand or answer? [short answer]
\end{enumerate}

\paragraph{Equal Opportunity - FPR.}
Recruit-a-matic uses the following rule to determine whether Sales-a-lot’s hiring decisions were fair:
 
\emph{The fraction of unqualified male candidates who receive job offers should equal the fraction of unqualified female candidates who receive job offers.}

Example 1: Suppose that over the past year, Recruit-a-matic finds that Sales-a-lot received the following unqualified applicants (10 female and 12 male).

\includegraphics[height=1in]{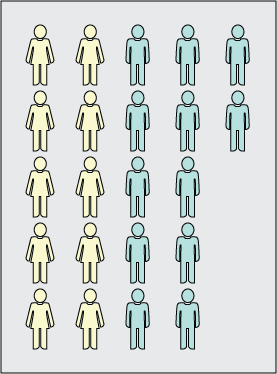}

If Sales-a-lot sent job offers to the following number of unqualified applicants (5 female and 6 male), then this would be fair according to the hiring rule (note that there are other possible outcomes that are fair according to the hiring rule).

\includegraphics[height=1in]{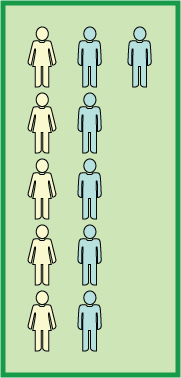}

Example 2: Suppose that over the past year, Recruit-a-matic finds that Sales-a-lot reviewed a total of 100 unqualified applicants as follows (40 female and 60 male).

\includegraphics[height=2in]{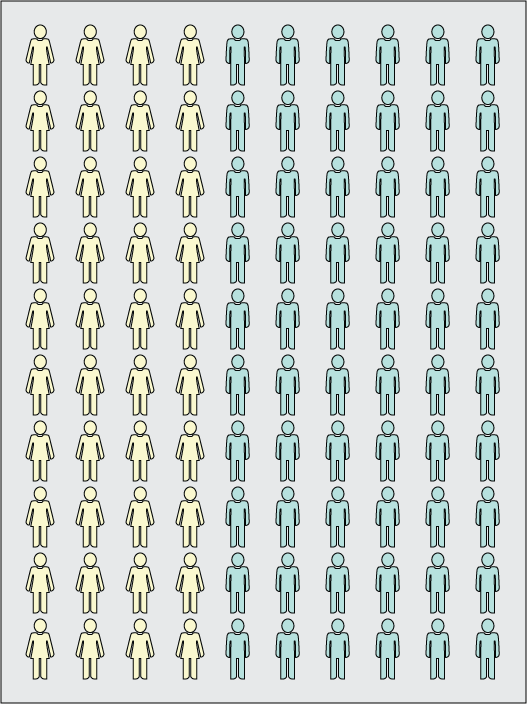}

If Sales-a-lot sent job offers to the following number of unqualified applicants (10 female and 15 male), then this would be fair according to the hiring rule (note that there are other possible outcomes that are fair according to the hiring rule).

\includegraphics[height=1in]{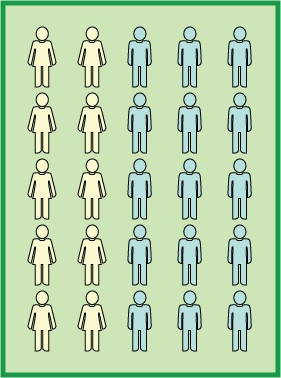}

Note that in the above examples the remaining unqualified applicants did not receive job offers, but are not displayed here.

In the next section, we will ask you some questions about the information you have just read. Please note that this is not a test of your abilities. We want to measure the quality of the description you read, not your ability to take tests or answer questions.

\textbf{Please note that we ask you to apply and use ONLY the above hiring rule when answering the following questions. You will have an opportunity to state your opinions and feelings on the rule later in the survey.}

\begin{enumerate}
\setcounter{enumi}{2}
    \item Suppose a different company considered applicants for a different job. There were 200 unqualified female applicants and 100 unqualified male applicants,

\includegraphics[height=1.3in]{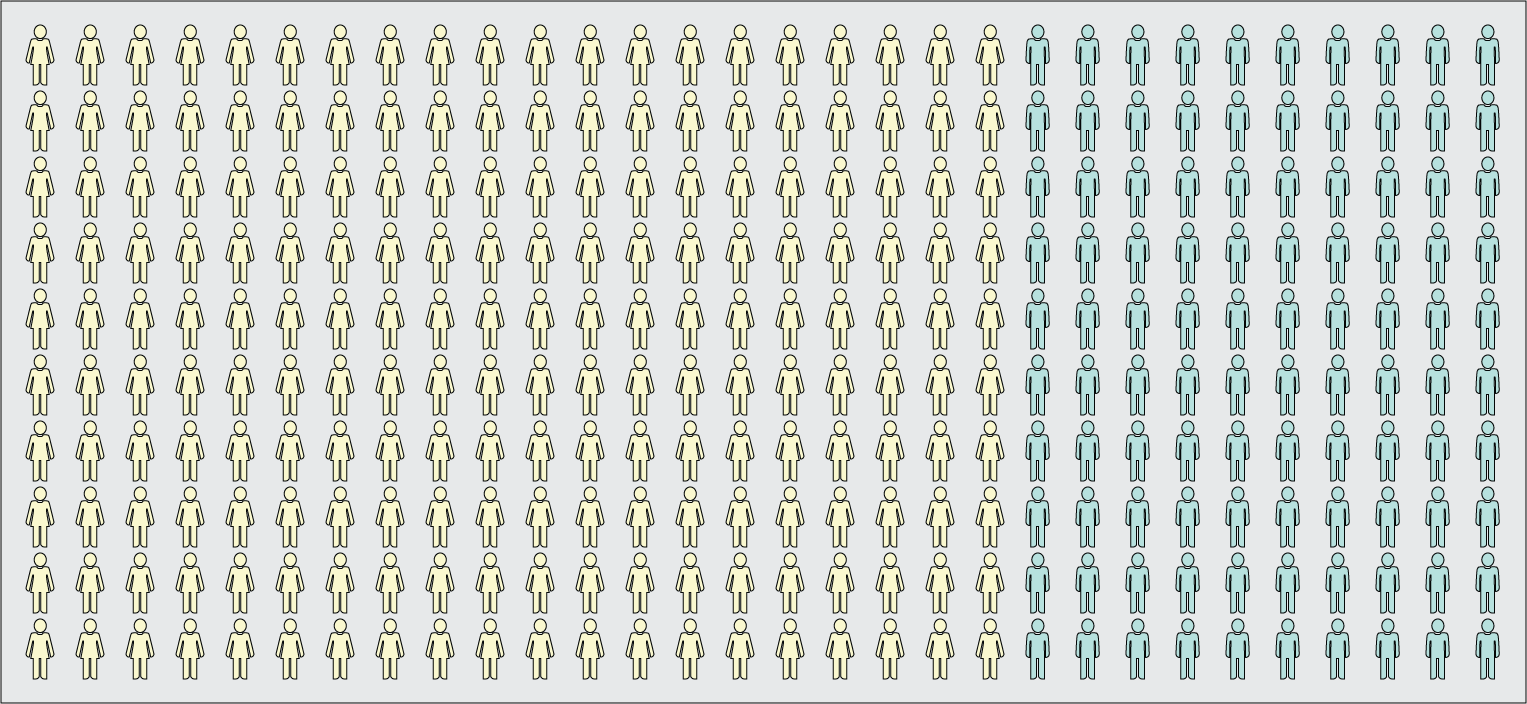}

and they did send job offers to 10 unqualified male applicants.

\includegraphics[height=1.3in]{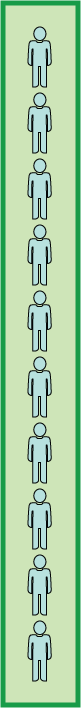}

Assuming that Recruit-a-matic reviews their decisions using the hiring rule above, how many unqualified female applicants should have received job offers?
\begin{enumerate}
    \item 10
    \includegraphics[height=1.3in]{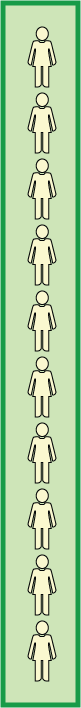}
    \item \correct{20}
    \includegraphics[height=1.3in]{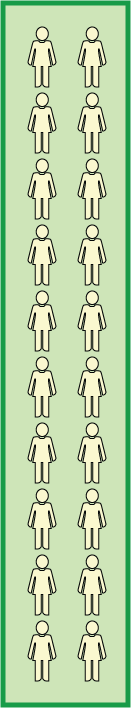}
    \item 40
    \includegraphics[height=1.3in]{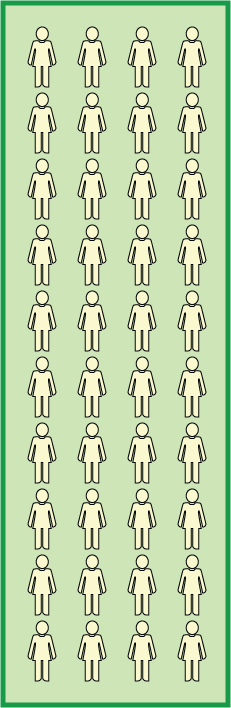}
    \item 50
    \includegraphics[height=1.3in]{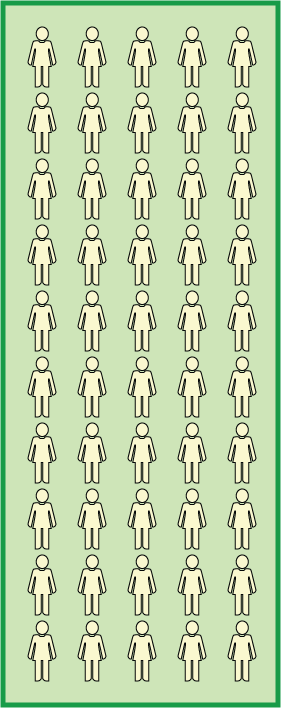}
\end{enumerate}

\item Assuming Recruit-a-matic reviews decisions using the hiring rule above, in which of these cases could Sales-a-lot have accepted more unqualified female applicants than unqualified male applicants?

\begin{enumerate}
    \item \correct{When there are more unqualified female applicants than unqualified male applicants (i.e., more women had low net sales at the end of the year).}
    \item When there are more female applicants than male applicants.
    \item When female applicants receive worse interview scores than male applicants.
    \item This cannot happen under the hiring rule.
\end{enumerate}

\item Consider one male applicant and one female applicant, both of whom are similarly qualified for the job (they achieve about the same net sales at the end of their first year). Is the following statement \correct{TRUE} OR FALSE: The hiring rule above allows Sales-a-lot to make a job offer to one of these applicants and not the other.

\item Consider a situation where all female applicants were unqualified (they all achieve low net sales at the end of their first year), but some of them received job offers. Is the following statement \correct{TRUE} OR FALSE: The hiring rule above requires that some job offers made to male applicants must have been made to unqualified male applicants.

\item Suppose Sales-a-lot received 100 male and 100 female applicants, and eventually made 10 job offers. Is the following statement TRUE OR \correct{FALSE}: The hiring rule above requires that even if all male applicants were unqualified (they all achieve low net sales at the end of their first year), some of the unqualified males must have received job offers.

\item Is the following statement \correct{TRUE} OR FALSE: The hiring rule above always allows Sales-a-lot to send job offers only to the most qualified applicants (those who achieve high net sales at the end of their first year).

\end{enumerate}

Consider a different scenario than the two examples above, with 6 unqualified applicants -- 4 female and 2 male, as illustrated below. The next three questions each give a different potential outcome for all 6 applicants (i.e., which of the 6 applicants receive job offers). Please indicate which of the outcomes follow the hiring rule above.

\includegraphics[height=0.4in]{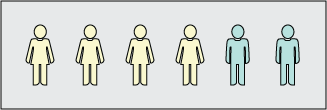}
\begin{enumerate}
\setcounter{enumi}{8}
    \item Sales-a-lot makes the following hiring decisions.

    \includegraphics[height=0.4in]{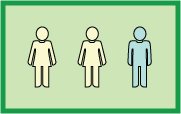}
    
    Do these decisions obey the hiring rule? \correct{Yes}

\item Sales-a-lot makes the following hiring decisions.

    \includegraphics[height=0.4in]{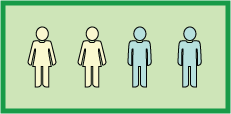}
    
    Do these decisions obey the hiring rule? \correct{No}
    
\item Sales-a-lot makes the following hiring decisions.

    \includegraphics[height=0.4in]{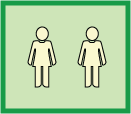}
    
    Do these decisions obey the hiring rule? \correct{No}

\item In your own words, explain the hiring rule. [short answer] [The rule is not shown above this question]

\item To what extent do you agree with the following statement: I am confident I know how to apply the hiring rule described above?
\begin{itemize}
    \item Strongly agree
    \item Agree
    \item Neither agree nor disagree
    \item Disagree
    \item Strongly Disagree
\end{itemize}

\item Please select the choice that best describes your experience: When I answered the previous questions...
\begin{enumerate}
    \item I applied the provided hiring rule only.
    \item I used a combination of the provided hiring rule and my own ideas of what the correct hiring rule should be.
    \item I used only my own ideas of what the correct hiring decision should be rather than the provided hiring rule.
\end{enumerate}

\item To what extent do you agree with the following statement: I like the hiring rule?
\begin{itemize}
    \item Strongly agree
    \item Agree
    \item Neither agree nor disagree
    \item Disagree
    \item Strongly Disagree
\end{itemize}

\item To what extent do you agree with the following statement: I agree with the hiring rule?
\begin{itemize}
    \item Strongly agree
    \item Agree
    \item Neither agree nor disagree
    \item Disagree
    \item Strongly Disagree
\end{itemize}

\item Please explain your opinion on the hiring rule. [short answer]

\item Was there anything about this survey that was hard to understand or answer? [short answer]

\end{enumerate}

\paragraph{Equalized Odds.}
Recruit-a-matic uses the following rule to determine whether Sales-a-lot’s hiring decisions were fair:
 
\emph{The fraction of qualified male candidates who do not receive job offers should equal the fraction of qualified female candidates who do not receive job offers. Similarly, the fraction of unqualified male candidates who receive job offers should equal the fraction of unqualified female candidates who receive job offers.}

Example 1: Suppose that over the past year, Recruit-a-matic finds that Sales-a-lot received the following qualified applicants (10 female and 12 male) and unqualified applicants (10 female and 12 male).

\includegraphics[height=1in]{illustrations/eo/example_1_POOL.png}

If Sales-a-lot did send offers to the following number of unqualified applicants (left, 5 female and 6 male), and did not send job offers to the following number of qualified applicants (right, 5 female and 6 male), then this would be fair according to the hiring rule (note that there are other possible outcomes that are fair according to the hiring rule).

\includegraphics[height=1in]{illustrations/eo/example_1b_offer.png}
\space\space\space
\includegraphics[height=1in]{illustrations/eo/example_1b_no_offer.png}

Example 2: Suppose that over the past year, Recruit-a-lot finds that Sales-a-lot reviewed a total of 100 qualified applicants (40 female and 60 male) and 100 unqualified applicants (40 female and 60 male).

\includegraphics[height=2in]{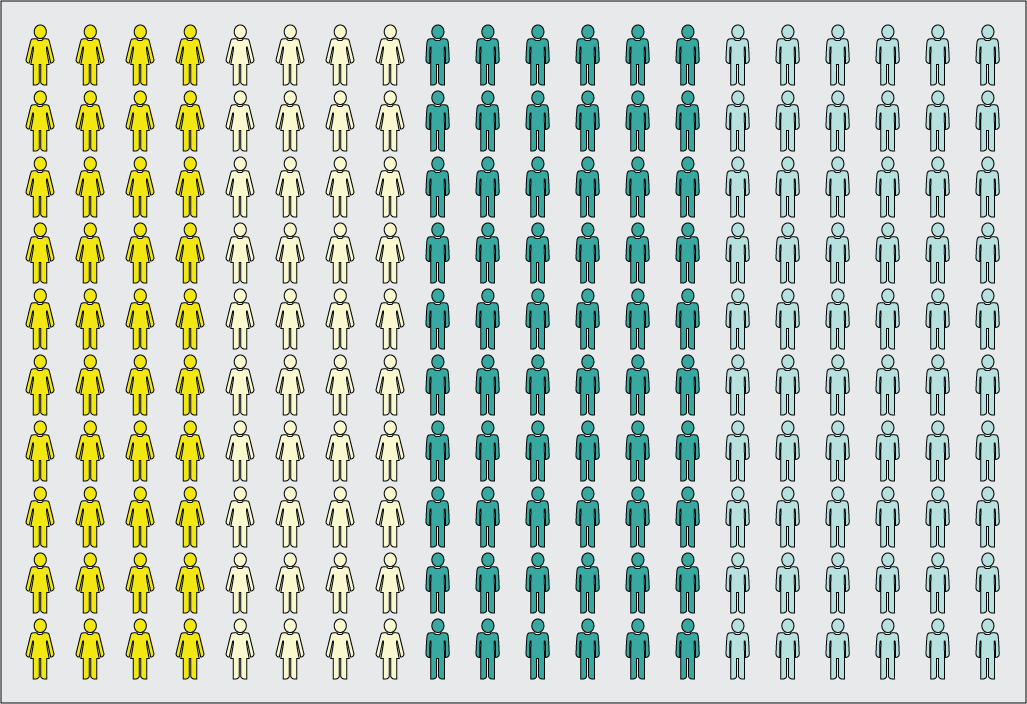}

If Sales-a-lot did send offers to the following number of unqualified applicants (left, 10 female and 15 male), and did not send job offers to the following number of qualified applicants (right, 10 female and 15 male), then this would be fair according to the hiring rule (note that there are other possible outcomes that are fair according to the hiring rule).

\includegraphics[height=1in]{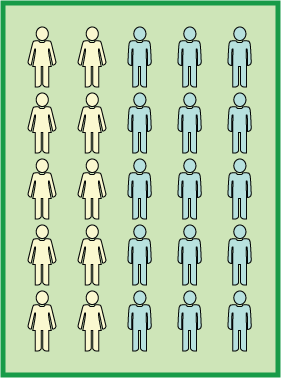}
\space\space\space
\includegraphics[height=1in]{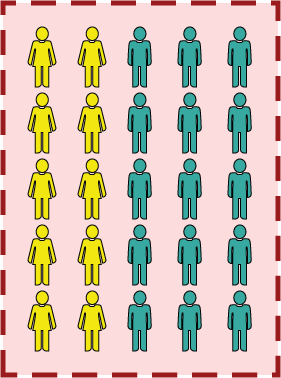}

Note that in the above examples the remaining unqualified applicants did not receive job offers, but are not displayed here. Similarly, the remaining qualified applicants received job offers, but are not displayed here.

In the next section, we will ask you some questions about the information you have just read. Please note that this is not a test of your abilities. We want to measure the quality of the description you read, not your ability to take tests or answer questions.

\textbf{Please note that we ask you to apply and use ONLY the above hiring rule when answering the following questions. You will have an opportunity to state your opinions and feelings on the rule later in the survey.}

\begin{enumerate}
\setcounter{enumi}{2}
    \item Suppose a different company considered applicants for a different job. There were 200 qualified female applicants and 100 qualified male applicants,

\includegraphics[height=1.2in]{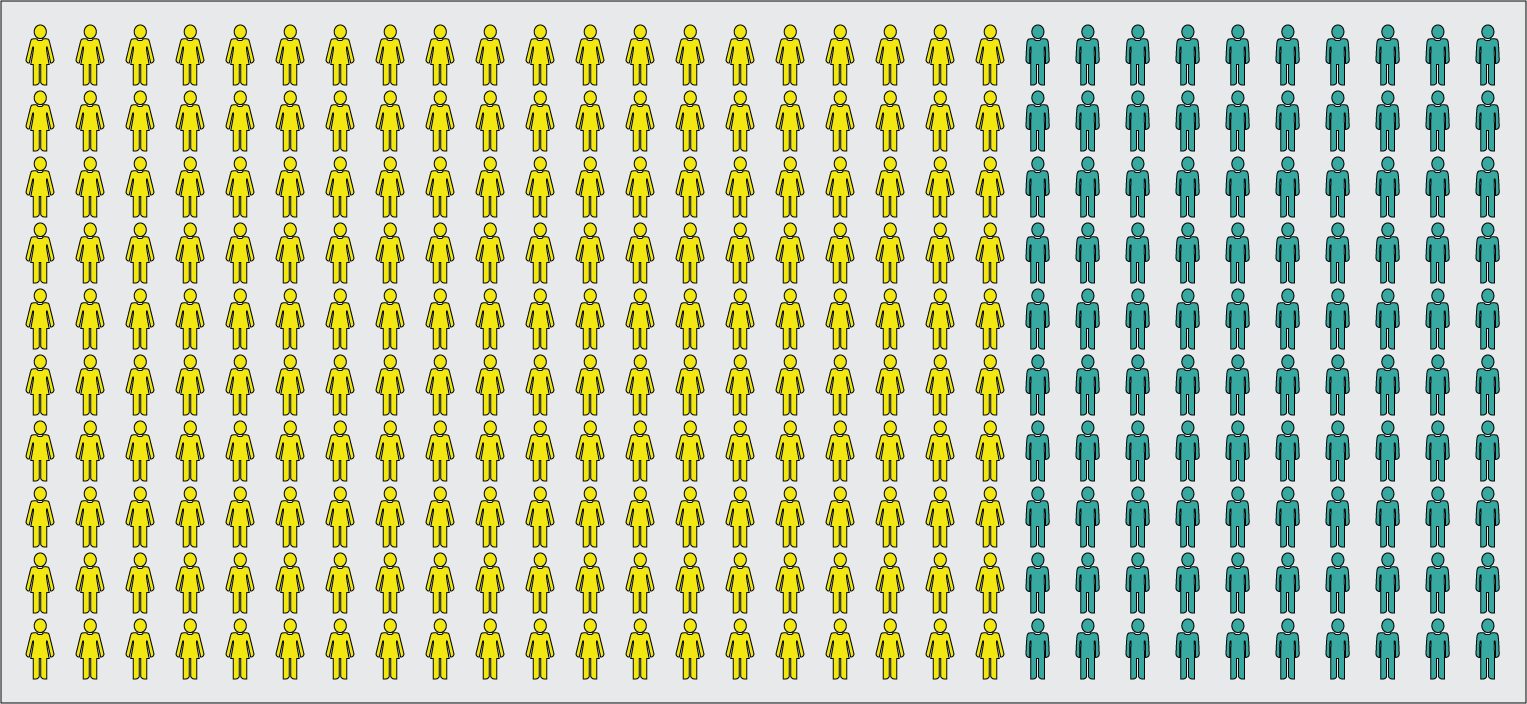}

and they did not send job offers to 90 qualified male applicants.

\includegraphics[height=1.3in]{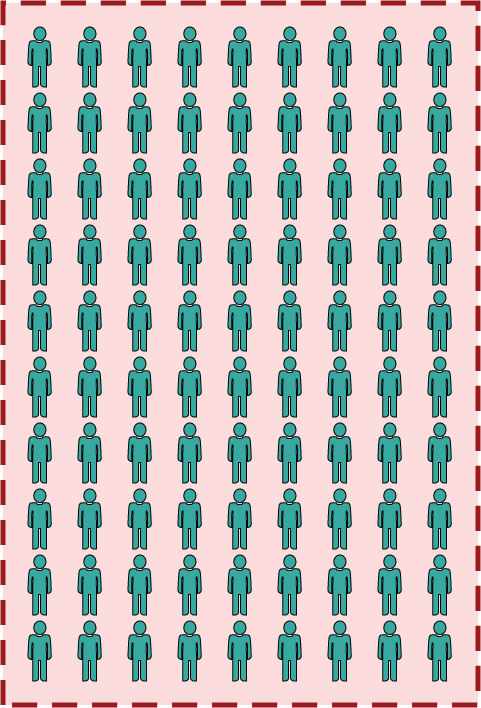}

Assuming that Recruit-a-matic reviews their decisions using the hiring rule above, how many qualified female applicants should not have received job offers?
\begin{enumerate}
    \item 190
    \includegraphics[height=1.3in]{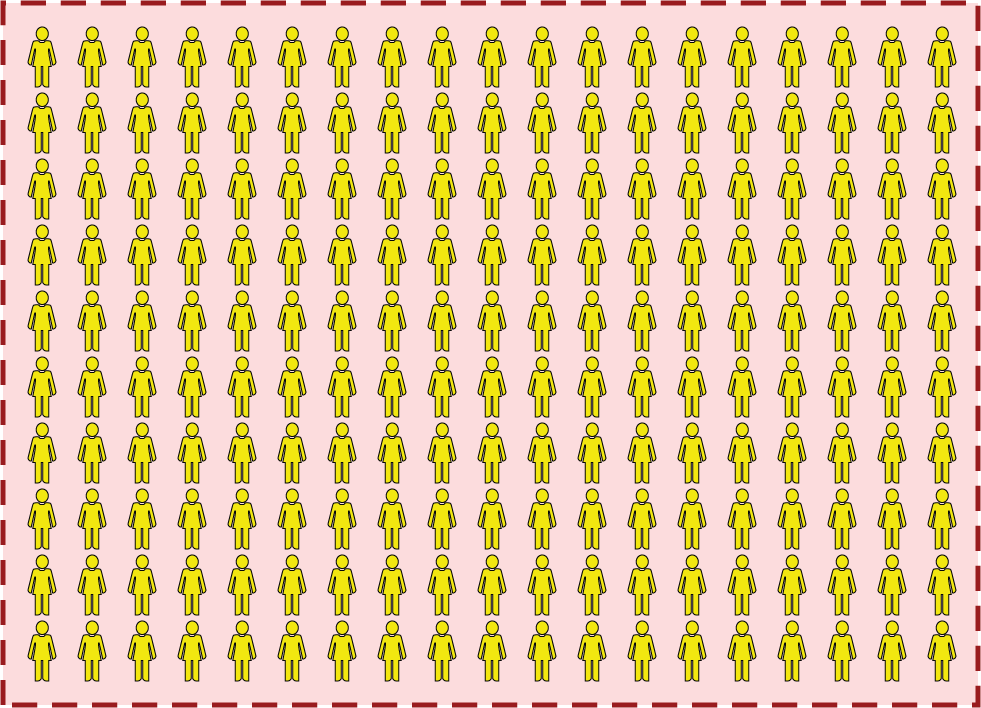}
    \item \correct{180}
    \includegraphics[height=1.3in]{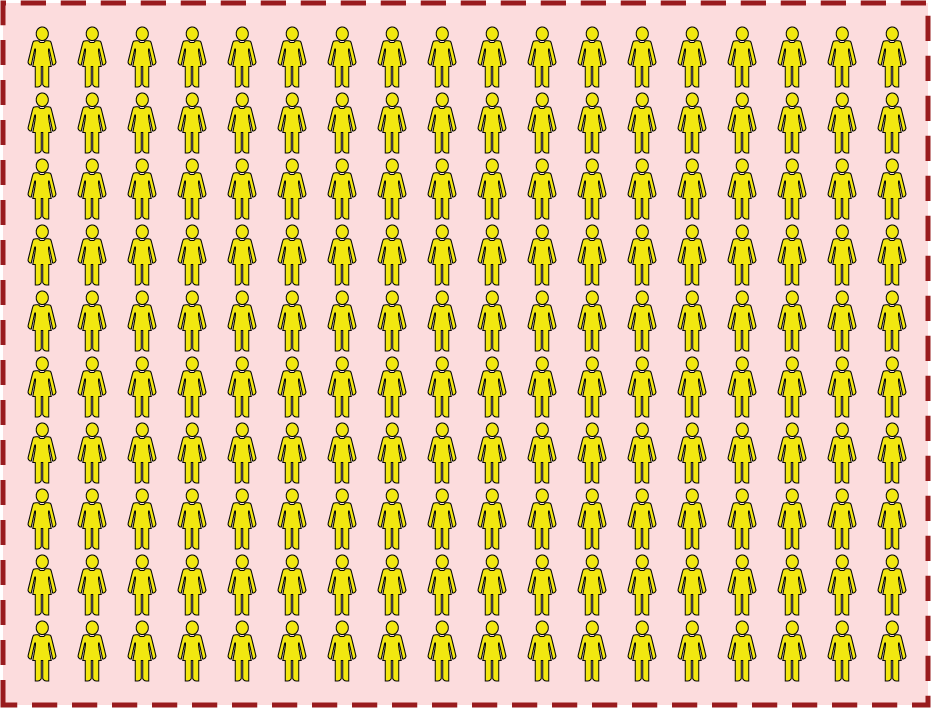}
    \item 160
    \includegraphics[height=1.3in]{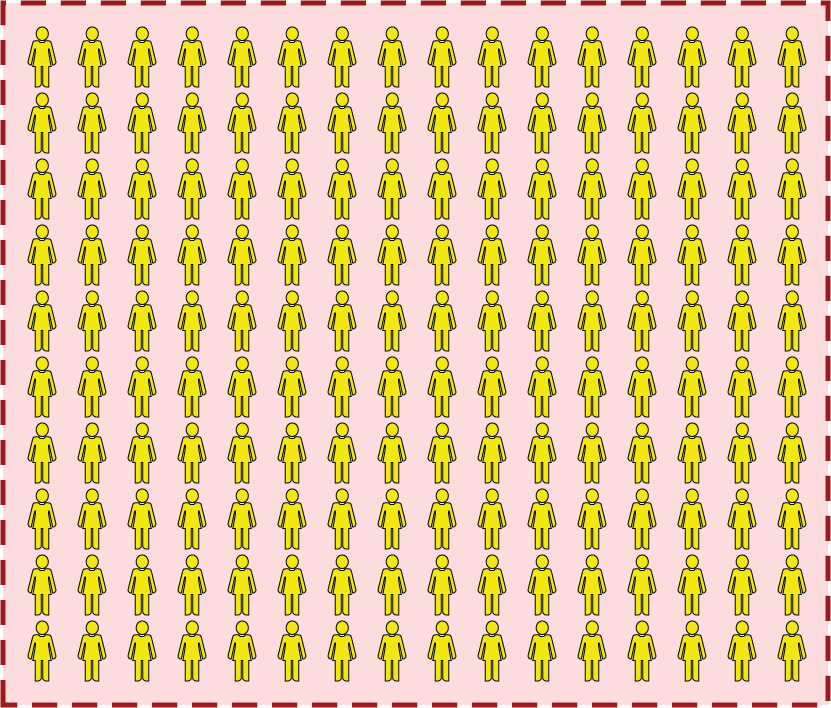}
    \item 150
    \includegraphics[height=1.3in]{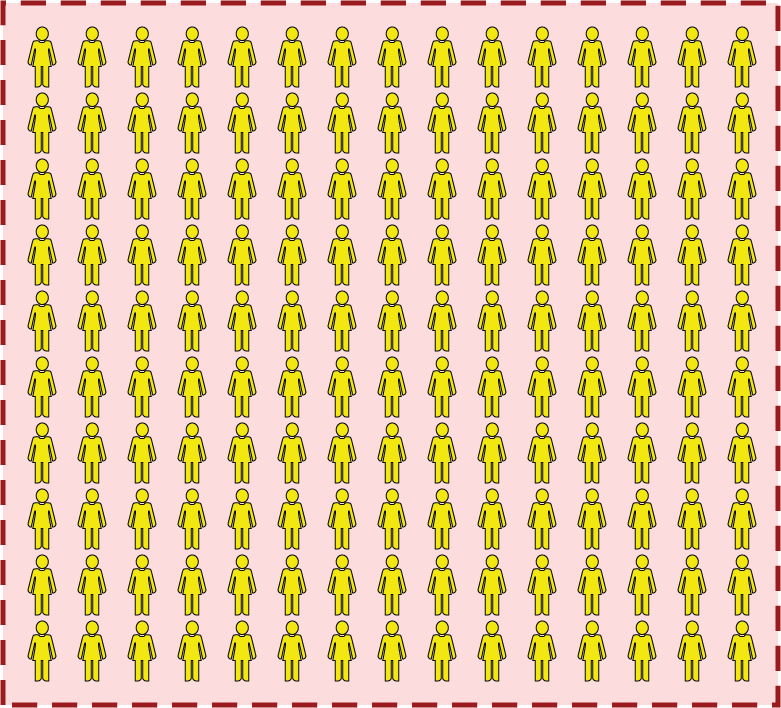}
\end{enumerate}

\item Assuming Recruit-a-matic reviews decisions using the hiring rule above, in which of these cases could Sales-a-lot have accepted more unqualified female applicants than unqualified male applicants?

\begin{enumerate}
    \item \correct{When there are more unqualified female applicants than unqualified male applicants (i.e., more women had low net sales at the end of the year).}
    \item When there are more female applicants than male applicants.
    \item When female applicants receive worse interview scores than male applicants.
    \item This cannot happen under the hiring rule.
\end{enumerate}

\item Consider one male applicant and one female applicant, both of whom are similarly qualified for the job (they achieve about the same net sales at the end of their first year). Is the following statement \correct{TRUE} OR FALSE: The hiring rule above allows Sales-a-lot to make a job offer to one of these applicants and not the other.

\item Consider a situation where all female applicants were unqualified (they all achieve low net sales at the end of their first year), but some of them received job offers. Is the following statement \correct{TRUE} OR FALSE: The hiring rule above requires that some job offers made to male applicants must have been made to unqualified male applicants.

\item Suppose Sales-a-lot received 100 male and 100 female applicants, and eventually made 10 job offers. Is the following statement TRUE OR \correct{FALSE}: The hiring rule above requires that even if all male applicants were unqualified (they all achieve low net sales at the end of their first year), some of the unqualified males must have received job offers.

\item Is the following statement \correct{TRUE} OR FALSE: The hiring rule above always allows Sales-a-lot to send job offers only to the most qualified applicants (those who achieve high net sales at the end of their first year).

\end{enumerate}

Consider a different scenario than the two examples above, with 6 qualified applicants -- 4 female and 2 male; and 6 unqualified applicants -- 4 female and 2 male. The next three questions each give a different potential outcome for the applicants (i.e., which of the applicants did or did not receive job offers). Please indicate which of the outcomes follow the hiring rule above.

\includegraphics[height=0.4in]{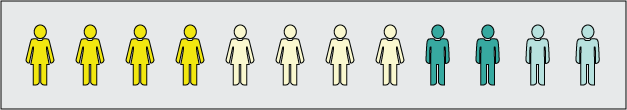}
\begin{enumerate}
\setcounter{enumi}{8}
    \item Sales-a-lot makes the following hiring decisions.

    \includegraphics[height=0.4in]{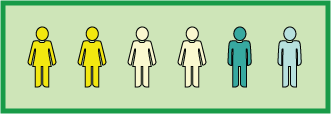}
    \space\space\space
    \includegraphics[height=0.4in]{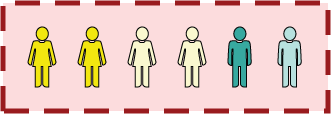}
    
    Do these decisions obey the hiring rule? \correct{Yes}

\item Sales-a-lot makes the following hiring decisions.

    \includegraphics[height=0.4in]{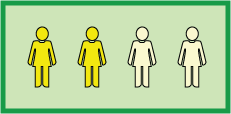}
    \space\space\space
    \includegraphics[height=0.4in]{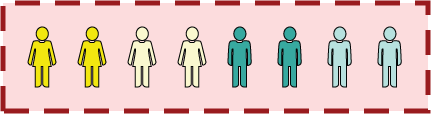}
    
    Do these decisions obey the hiring rule? \correct{No}
    
\item Sales-a-lot makes the following hiring decisions.

    \includegraphics[height=0.4in]{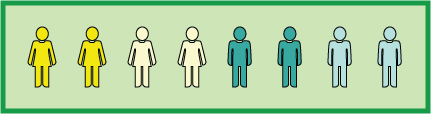}
    \space\space\space
    \includegraphics[height=0.4in]{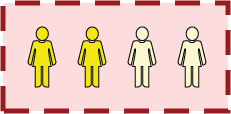}
    
    Do these decisions obey the hiring rule? \correct{No}

\item In your own words, explain the hiring rule. [short answer] [The rule is not shown above this question]

\item To what extent do you agree with the following statement: I am confident I know how to apply the hiring rule described above?
\begin{itemize}
    \item Strongly agree
    \item Agree
    \item Neither agree nor disagree
    \item Disagree
    \item Strongly Disagree
\end{itemize}

\item Please select the choice that best describes your experience: When I answered the previous questions...
\begin{enumerate}
    \item I applied the provided hiring rule only.
    \item I used a combination of the provided hiring rule and my own ideas of what the correct hiring rule should be.
    \item I used only my own ideas of what the correct hiring decision should be rather than the provided hiring rule.
\end{enumerate}

\item To what extent do you agree with the following statement: I like the hiring rule?
\begin{itemize}
    \item Strongly agree
    \item Agree
    \item Neither agree nor disagree
    \item Disagree
    \item Strongly Disagree
\end{itemize}

\item To what extent do you agree with the following statement: I agree with the hiring rule?
\begin{itemize}
    \item Strongly agree
    \item Agree
    \item Neither agree nor disagree
    \item Disagree
    \item Strongly Disagree
\end{itemize}

\item Please explain your opinion on the hiring rule. [short answer]

\item Was there anything about this survey that was hard to understand or answer? [short answer]
\end{enumerate}

\subsection{Demographic Information}\label{app:demographics}

\begin{enumerate}
    \item Please specify the gender with which you most closely identify:
    \begin{itemize}
        \item Male
        \item Female
        \item Other
        \item Prefer not to answer
    \end{itemize}
    \item Please specify your year of birth
    \item Please specify your ethnicity (you may select more than one):
    \begin{itemize}
        \item White
        \item Hispanic or Latinx	
        \item Black or African American
        \item American Indian or Alaska Native
        \item Asian, Native Hawaiian, or Pacific Islander
        \item Other
    \end{itemize}
    \item Please specify the highest degree or level of school you have completed:
    \begin{itemize}
        \item Some high school credit, no diploma or equivalent
        \item High school graduate, diploma or the equivalent (for example: GED)
        \item Some college credit, no degree
        \item Trade/technical/vocational training
        \item Associate’s degree
        \item Bachelor’s degree
        \item Master’s degree
        \item Professional or doctoral degree (JD, MD, PhD)
    \end{itemize}
    \item How much experience do you have in each of the following areas? (1 - no experience, 2 - limited experience, 3 - significant experience, 4 - expert)
    \begin{enumerate}
        \item Human resources (making hiring decisions)
        \item Management (of employees)
        \item Education (teaching)
        \item IT infrastructure/systems administration
        \item Computer science/programming
        \item Machine learning/data science
    \end{enumerate}
\end{enumerate}

\textbf{We will maintain privacy of the information you have provided here. Your information will only be used for data analysis purposes.}

\section{Consent} \label{app:consent}

\subsection{Online Survey Consent Form} \label{app:survey_consent}

\subsubsection{Project Title}
Fairness Evaluation and Comprehension

\subsubsection{Purpose of the Study}
This research is being conducted by Michelle Mazurek at the University of Maryland, College Park. We are inviting you to participate in this research project because you are above 18. The purpose of this research project is to understand lay comprehension of different fairness metrics.  

\subsubsection{Procedures}
The procedures will start with reading a brief description of a decision-making scenario. You will then be asked to answer some comprehension questions about the scenario. The questions will look like the following: What are the pros and cons of the notion of fairness described above?

Finally, you will be asked some demographics questions. The entire survey will take approximately 20 minutes or less.

\subsubsection{Potential Risks and Discomforts}
There are several questions to answer over the course of this study, so you may find yourself growing tired towards the end. Outside of this, there are minimal risks to participating in this research study. All data collected in this study will be maintained securely (see Confidentiality section) and will be deleted at the conclusion of the study.

However, if at any time you feel that you wish to terminate your participation for any reason, you are permitted to do so.

\subsubsection{Potential Benefits}
There are no direct benefits from participating in this research. We hope that, in the future, other people might benefit from this study through improved understanding of fairness metrics and their applications. 

\subsubsection{Confidentiality}
Any potential loss of confidentiality will be minimized by storing all data (including information such as MTurk IDs and demographics) will be stored securely (a) in a password-protected computer located at the University of Maryland, College Park or (b) using a trusted third party (Qualtrics). Personally identifiable information that is collected (MTurk IDs, IP addresses, cookies) will be deleted upon study completion. All other data gathered will be stored for three years post study completion, after which it will be erased.
The only persons that will have access to the data are the Principle Investigator and the Co-Investigators.

If we write a report or article about this research project, your identity will be protected to the maximum extent possible.  Your information may be shared with representatives of the University of Maryland, College Park or governmental authorities if you or someone else is in danger or if we are required to do so by law. 

\subsubsection{Compensation}
You will receive \$3. You will be responsible for any taxes assessed on the compensation.  

If you will earn \$100 or more as a research participant in this study, you must provide your name, address and SSN to receive compensation.

If you do not earn over \$100 only your name and address will be collected to receive compensation.

\subsubsection{Right to Withdraw and Questions}
Your participation in this research is completely voluntary.  You may choose not to take part at all.  If you decide to participate in this research, you may stop participating at any time.  If you decide not to participate in this study or if you stop participating at any time, you will not be penalized or lose any benefits to which you otherwise qualify.

If you decide to stop taking part in the study, if you have questions, concerns, or complaints, or if you need to report an injury related to the research, please contact the investigator:

{\centering
Michelle Mazurek \\
5236 Iribe Center, \\University of Maryland, College Park 20742\\
mmazurek@cs.umd.edu\\
(301) 405-6463\\}

\subsubsection{Participant Rights}
If you have questions about your rights as a research participant or wish to report a research-related injury, please contact: 

{\centering
University of Maryland College Park \\
Institutional Review Board Office\\
1204 Marie Mount Hall \\
College Park, Maryland, 20742 \\
E-mail: irb@umd.edu \\
Telephone: 301-405-0678 \\}

\vspace{5pt}
For more information regarding participant rights, please visit:

\url{https://research.umd.edu/irb-research-participants}

This research has been reviewed according to the University of Maryland, College Park IRB procedures for research involving human subjects.

\subsubsection{Statement of Consent}
By agreeing below you indicate that you are at least 18 years of age; you have read this consent form or have had it read to you; your questions have been answered to your satisfaction and you voluntarily agree to participate in this research study. 
Please ensure you have made a copy of the above consent form for your records.

Pease ensure you have made a copy of the above consent form for your records. A copy of this consent form can be found here [link to digital copy].

\begin{itemize}
    \item[\checkbox] I am age 18 or older
    \item[\checkbox] I have read this consent form
    \item[\checkbox] I voluntarily agree to participate in this research study
\end{itemize}

\subsection{Cognitive Interview Consent Form} \label{app:cognitive_consent}

\subsubsection{Project Title}
Fairness Cognitive Interview

\subsubsection{Purpose of the Study}
This research is being conducted by Michelle Mazurek at the University of Maryland, College Park. We are inviting you to participate in this research project because you are above the age of 18, and fluent in English. The purpose of this research project is to understand lay comprehension of different fairness metrics.  

\subsubsection{Procedures}
The procedure involves completing an interview. The full procedure will be approximately 1 hour in duration.

During the interview you will be audio recorded, if you agree to be recorded. You will be asked to first read a brief description of a decision-making scenario. You will then be asked to fill out a survey about the scenario. While answering questions you will be asked verbal questions related to how you reached your answer in the survey.

Sample survey question:
Is the following statement true or false? This hiring rule allows the hiring manager to send offers exclusively to the most qualified applicants.

Sample interview question:
How did you reach your answer to that survey question?

\subsubsection{Potential Risks and Discomforts}
There are several questions to answer over the course of this study, so you may find yourself growing tired towards the end. Outside of this, there are minimal risks to participating in this research study. All data collected in this study will be maintained securely (see Confidentiality section) and will be deleted at the conclusion of the study.

However, if at any time you feel that you wish to terminate your participation for any reason, you are permitted to do so.

\subsubsection{Potential Benefits}
There are no direct benefits from participating in this research. We hope that, in the future, other people might benefit from this study through improved understanding of fairness metrics and their applications. 

\subsubsection{Confidentiality}
Any potential loss of confidentiality will be minimized by storing all data (including information such as demographics) securely (a) in a password protected computer located at the University of Maryland, College Park or (b) using a trusted third party (Qualtrics). Personally identifiable information that is collected will be deleted upon study completion. All other data gathered will be stored for three years post study completion, after which it will be erased. The only persons that will have access to the data are the principle Investigator and the Co-Investigators.

If we write a report or article about this research project, your identity will be protected to the maximum extent possible.  Your information may be shared with representatives of the University of Maryland, College Park or governmental authorities if you or someone else is in danger or if we are required to do so by law. 

\subsubsection{Compensation}
You will receive \$30.  You will be responsible for any taxes assessed on the compensation.  

If you will earn \$100 or more as a research participant in this study, you must provide your name, address and SSN to receive compensation.

If you do not earn over \$100 only your name and address will be collected to receive compensation.

\subsubsection{Right to Withdraw and Questions}
Your participation in this research is completely voluntary.  You may choose not to take part at all.  If you decide to participate in this research, you may stop participating at any time.  If you decide not to participate in this study or if you stop participating at any time, you will not be penalized or lose any benefits to which you otherwise qualify.

If you decide to stop taking part in the study, if you have questions, concerns, or complaints, or if you need to report an injury related to the research, please contact the investigator:

{\centering
Michelle Mazurek \\
5236 Iribe Center, \\University of Maryland, College Park 20742\\
mmazurek@cs.umd.edu\\
(301) 405-6463\\}

\subsubsection{Participant Rights}
If you have questions about your rights as a research participant or wish to report a research-related injury, please contact: 

{\centering
University of Maryland College Park \\
Institutional Review Board Office\\
1204 Marie Mount Hall \\
College Park, Maryland, 20742 \\
E-mail: irb@umd.edu \\
Telephone: 301-405-0678 \\}
\vspace{5pt}
For more information regarding participant rights, please visit:

\url{https://research.umd.edu/irb-research-participants}

This research has been reviewed according to the University of Maryland, College Park IRB procedures for research involving human subjects.

\subsubsection{Statement of Consent}
Your signature indicates that you are at least 18 years of age; you have read this consent form or have had it read to you; your questions have been answered to your satisfaction and you voluntarily agree to participate in this research study. You will receive a copy of this signed consent form.

Please initial all that apply (you may choose any number of these statements):
\begin{itemize}
    \item[\checkbox] I agree to be audio recorded
    \item[\checkbox] I agree to allow researchers to use my audio recording in research publications and presentations.
\end{itemize}

\begin{itemize}
    \item[\checkbox] I do not agree to be audio recorded
\end{itemize}

If you agree to participate, please sign your name below.

\end{document}